\documentclass[twocolumn]{aastex701}
\usepackage{xcolor}
\newcommand {\msun}{\,M$_\odot \ $}

\newcommand {\kms}{$\rm{km\,s^{-1}}$}

\begin{document}

\title{Exploring the central region of SNR 0540-69.3 with JWST I: 3D morphology}

\correspondingauthor{J. Larsson}

\author[0000-0003-0065-2933]{J. Larsson}
\affiliation{Department of Physics, KTH Royal Institute of Technology, The Oskar Klein Centre, AlbaNova, SE-106 91 Stockholm, Sweden}
\affiliation{Stellenbosch Institute for Advanced Study (STIAS), Wallenberg Research Centre at Stellenbosch University, Stellenbosch 7600, South Africa}
\email{josla@kth.se}

\author[0009-0007-3359-5767]{C. Tegkelidis}
\affiliation{Department of Physics, KTH Royal Institute of Technology, The Oskar Klein Centre, AlbaNova, SE-106 91 Stockholm, Sweden}
\email{cteg@kth.se}

\author[0000-0001-8532-3594]{C.\ Fransson}
\affiliation{Department of Astronomy, Stockholm University, The Oskar Klein Centre, AlbaNova, SE-106 91 Stockholm, Sweden}
\email{author email}

\author[0000-0002-3664-8082]{P. Lundqvist}
\affiliation{Department of Astronomy, Stockholm University, The Oskar Klein Centre, AlbaNova, SE-106 91 Stockholm, Sweden}
\email{author email}

\author[0000-0003-1546-6615]{J. Sollerman}
\affiliation{Department of Astronomy, Stockholm University, The Oskar Klein Centre, AlbaNova, SE-106 91 Stockholm, Sweden}
\email{author email}

\author[0000-0001-6815-4055]{J. Spyromilio}
\affiliation{European Southern Observatory, Karl-Schwarzschild-Strasse 2, D-85748 Garching, Germany}
\email{author email}


\begin{abstract}
 
The young supernova remnant SNR~0540-69.3 in the Large Magellanic Cloud offers a detailed view of an energetic pulsar-wind nebula interacting with the surrounding ejecta. We present infrared observations of the central region of SNR~0540-69.3 obtained with the JWST NIRSpec and MRS integral field units. From the observations we reconstruct the 3D morphology of the strongest emission lines in the inner ejecta ($\lesssim 1000$~\kms), which reveals the distribution of H I, He I, [\ion{Ne}{2}], [\ion{Ne}{3}], [\ion{S}{3}], [\ion{S}{4}], [\ion{Fe}{2}], and [Ni II].  The 3D morphology of most lines is dominated by two highly fragmented lobes of approximately similar size. Based on the assumption that the lobes are symmetric around the pulsar, we infer a pulsar kick velocity of $\sim 300$~\kms\ away from the observer. There are differences in the 3D morphologies of individual emission lines due to a combination of varying physical conditions and abundances. The detection of \ion{H}{1}~1.8756~$\mu$m in the inner ejecta confirms the classification of the SN as a Type II and shows that hydrogen was mixed down to low velocities of $< 400$~\kms\ in the explosion. We compare the results to the Crab nebula and conclude that asymmetries originating in the explosion most likely play a major role in shaping the PWNe. 

\end{abstract}


\section{Introduction}

One of the possible outcomes of a core-collapse supernova (SN) is that the core of the progenitor star collapses to a rapidly spinning pulsar. The particle wind from the pulsar can inflate a bubble that interacts with the expanding ejecta, referred to as a pulsar wind nebula (PWN, \citealt{Buhler2014,Olmi2023}).  Highly energetic pulsars/PWNe are prime candidates for powering some of the most luminous SN types \citep{Woosley2010,Metzger2015,Rodriguez2024}, while less extreme versions of PWNe are commonly observed in SN remnants (SNRs) in the Milky Way and Magellanic Clouds \citep[e.g.][]{Gaensler2006,Kargaltsev2008}. A major advantage of studying nearby PWNe is that they provide a spatially resolved view of the complex processes that drive their hydrodynamical evolution and multiwavelength emission. 

While PWNe are most commonly observed through bright X-ray synchrotron emission, some objects also show optical and infrared (IR) emission lines from the ejecta, as in the well known case of the Crab nebula \citep{Hester2008,Temim2024}. Such observations are of great interest as the emission lines also provide information about the explosions and progenitors. A particularly interesting diagnostic is offered by spatially resolved spectroscopy of the expanding ejecta, which makes it possible to reconstruct their 3D distribution, which in turn probes both the asymmetries in the explosion and the impact of the pulsar wind. The small number of PWNe for which 3D information is available includes the Crab \citep{Martin2021}, the Galactic remnant 3C~58 \citep{Fesen2008} and the object studied in this paper: SNR~0540-69.3 (SNR~0540 from here on, \citealt{Larsson2021}). 

SNR~0540 is located in the Large Magellanic Cloud (LMC) and has an age of $\sim 1,100$~years \citep{Larsson2021}. It has a composite morphology, showing both a pulsar surrounded by a PWN \citep[e.g.,][]{Petre2007,Lundqvist2011,Mignani2012,Tenhu2024} and a large-scale shell where the SN blast wave interacts with the surrounding medium \citep{Hwang2001, Brantseg2014, Lundqvist2020,Tenhu2025}. The first detection of optical emission lines from the ejecta was presented by \cite{Mathewson1980}, which was followed by more detailed studies in the optical \citep{Kirshner1989,Serafimovich2005,Morse2006,Sandin2013,Larsson2021,Lundqvist2022}, while \cite{Williams2008} presented the first IR spectrum based on Spitzer observations. These observations show that most of the line emission is confined to a small region of $\sim 4\arcsec$ radius, surrounded by an [\ion{O}{3}] halo extending to $\sim$10--14$\arcsec$. For reference, $1\arcsec$ corresponds to 0.24~pc in the LMC (assuming a distance of 49.6~kpc, \citealt{Pietrzynski2019}), which is equivalent to 214~\kms\ for freely expanding ejecta for an age of 1,100~years. While the presence of Balmer lines in the spectrum was debated for the early observations, it was later confirmed \citep{Serafimovich2005,Morse2006,Larsson2021}, establishing the explosion as a Type II SN. 

The 3D morphology of the optical emission lines has been presented in \cite{Sandin2013} and \cite{Larsson2021}, based on observations from the Very Large Telescope (VLT) integral field units (IFU) VIMOS and MUSE, respectively.  The MUSE observations had a larger field of view (FOV), covering the whole remnant, which showed that the [\ion{O}{3}] halo in fact forms a torus of ejecta with a radius of $\sim 1600$~\kms\ (extending to $\sim 3000$ \kms). The MUSE observations also revealed a large blob of H$\alpha$ and H$\beta$ emission located south-east of the pulsar at velocities $\sim 1500$--3000~\kms, which may be ejecta from a highly asymmetric explosion or the blown-off envelope of a binary companion \citep{Larsson2021}.  The 3D morphology of the emission interior to the [\ion{O}{3}] torus was mapped in H$\beta$, [\ion{O}{3}], [\ion{S}{2}], [\ion{S}{3}], and [\ion{Ar}{3}], which revealed clumpy ring-like structures with velocities $\lesssim 1000$~\kms. 

Here we present a more detailed investigation of the 3D morphology of the innermost region using JWST IFU observations with NIRSpec \citep{Jakobsen2022} and the MIRI Medium-Resolution Spectrometer (MRS; \citealt{Wells2015}). These observations cover the full wavelength interval from 1--28~$\mu$m, which includes strong emission lines from elements not probed by the optical IFU observations (such as Fe and Ne), thus providing a more complete view of the inner ejecta. The JWST observations also offer better signal-to-noise ratio (S/N) and contrast compared to the MUSE data, for which the use of adaptive optics gives a point spread function (PSF) with a narrow core and very broad wings. This paper is the first in a series about the JWST observations of SNR~0540. Forthcoming papers will include a detailed analysis of the full emission line spectrum, including many weaker lines not suitable for 3D mapping, spectral modeling with shock and photoionization models, as well as studies of the continuum emission from the pulsar and PWN. 

We describe the observations and data reduction in Section~\ref{sec:obs} and then present the methods for the 3D reconstructions in Section~\ref{sec-mapmaking}. The results are presented in Section~\ref{sec:results}, followed by a discussion in Section~\ref{sec:discussion} and conclusions in Section~\ref{sec:summary}. We refer to spectral lines by their vacuum wavelengths for the JWST data in the full IR range, while we use air wavelengths when discussing emission lines observed with MUSE in the optical range.

\section{Observations and data reduction}
\label{sec:obs}

\begin{figure}[t]
\centering
\includegraphics[width=\hsize]{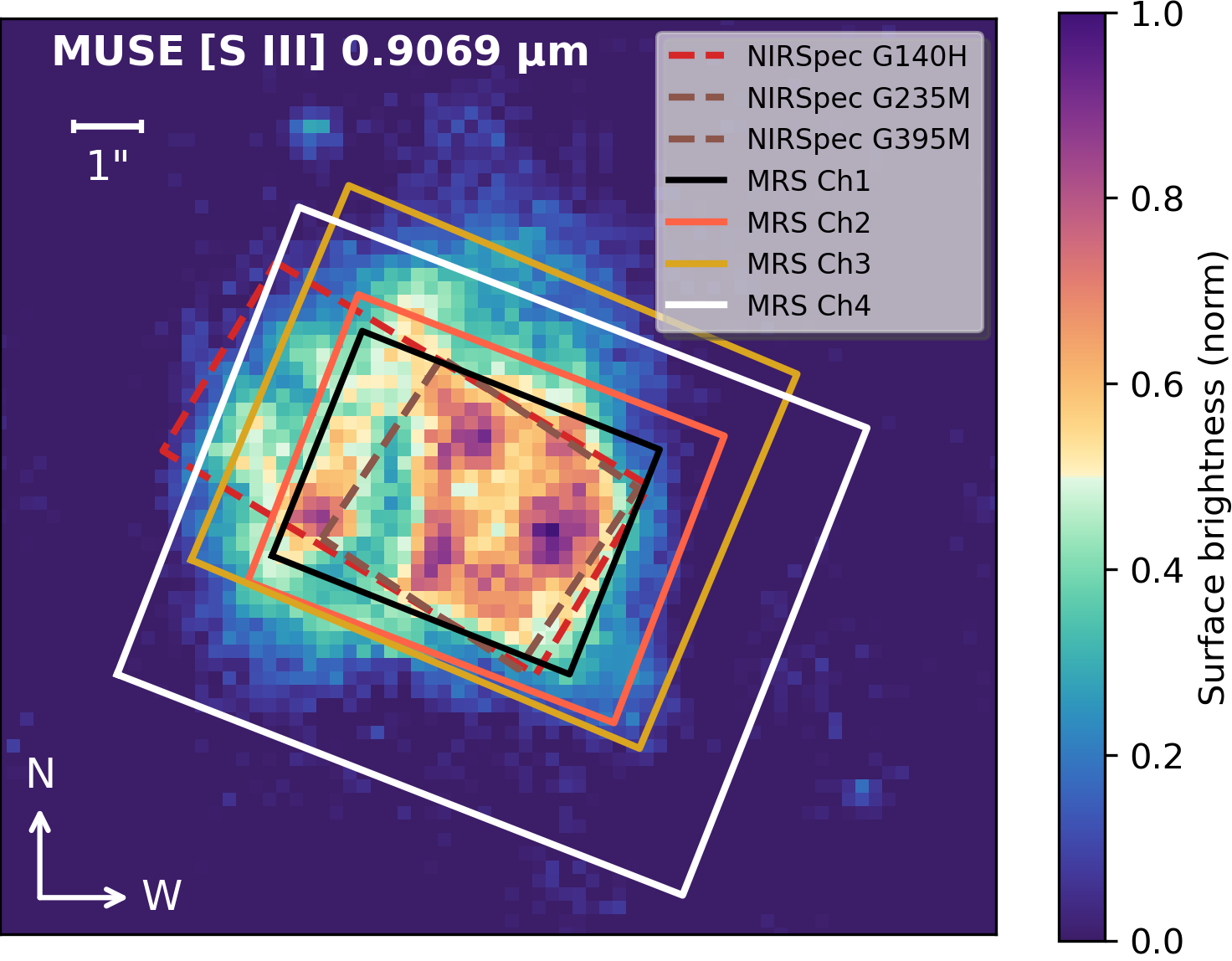}
\caption{The footprints of the JWST NIRSpec and MIRI/MRS observations of SNR~0540, shown superposed on a line map of [\ion{S}{3}]~0.9069~$\mu$m from VLT/MUSE.}
\label{fig:fov}
\end{figure}

JWST Cycle 3 observations of SNR~0540 were carried out with MIRI/MRS on 2025 February 14 and with the NIRSpec IFU on 2025 March 25--26 (Program ID 4712, PI J. Larsson, see Table~\ref{tab:obs}). The data from the two instruments together provide spatially resolved spectroscopy over the 1--28~$\mu$m wavelength range. The observations cover the main part of the inner line-emitting ejecta, with some differences in the size and orientation of the FOV of the different NIRSpec pointings and MRS channels, illustrated in Figure~\ref{fig:fov}. Below we first describe the details of the NIRSpec and MRS observations, and then describe the correction of the world coordinate solution (WCS) and the determination of the systemic velocity. We end the Section with a brief overview of the previous VLT/MUSE observations, which are used for comparison.

\subsection{NIRSpec observations}
\label{sec:obs-nirspec}

 The details of the NIRSpec observations of SNR~0540 are summarized in Table~\ref{tab:obs}. There were two pointings with the G140H/F100LP grating/filter combination, and one pointing each with G235M/F170LP and G395M/F290LP at the same position as the first G140H/F100LP pointing.  The NIRSpec IFU has a $3^{\prime \prime} \times 3^{\prime \prime}$ FOV with $0\farcs{1} \times 0\farcs{1}$ spatial sampling. The footprint of these observations is illustrated in Figure~\ref{fig:fov}. All the observations were performed using a small four-point dither pattern and the NRSIRS2RAPID readout mode. No dedicated background observations were carried out with NIRSpec, as the background at these wavelengths is low compared to the signal from the remnant. 

The high-resolution grating G140H provides a spectral resolving power ($R=\lambda / \Delta \lambda$) that ranges from $R \sim 1900$--$3600$ over 0.97--1.89$\ \mu$m, with a small gap between 1.41--1.49~$\mu$m due to the physical gap between the two NIRSpec detectors \citep{Jakobsen2022}. The medium-resolution gratings G235M and G395M only use one detector and therefore do not have wavelength gaps. They cover the wavelength ranges 1.66--3.17~$\mu$m and 2.87--5.27~$\mu$m, respectively, where $R$ increases from $\sim$700--1300 from short to long wavelengths in each grating. The use of the high-resolution grating at the shortest wavelengths is motivated by the high prevalence of spectral lines, while the longer NIR wavelength range is dominated by continuum emission.  In this paper we only analyze the strongest emission lines and therefore only use the G140H/F100LP data, while the data from other gratings will be included in future publications. For completeness, we provide a description of the full NIRSpec data set below. 

The data were downloaded from the Mikulski Archive for Space Telescopes (MAST) and processed with version 1.19.1 of the JWST Calibration Pipeline \citep{Bushouse2023}, using the default version “jwst\_1413.pmap” of the Calibration Reference Data System (CRDS) context. The data were processed with default parameters, except for a some adjustments of the stage 1 \texttt{jump} step, where we set the expand\_factor to 3, and the stage 3 \texttt{outlier detection} step, where we used a threshold of 50\% and a kernel size of 7 pixels. These adjustments allow for more effective rejection of cosmic rays and other artifacts, such as bad pixels. 

It is known that NIRSpec IFU observations may be contaminated by light from outside the FOV that leaks through the microshutter array. While the use of dithered observations should mitigate this, there may be remaining issues in crowded fields, such as that surrounding SNR~0540. Inspection of the observations revealed two positions in G235M/F170LP and G395M/F290LP where bright stars coincided with the location of stuck open shutters, creating parasitic signals in the data cubes. The affected pixels of the “cal" files were flagged manually to be excluded from the cube building process. This introduces uncertainties at the corresponding locations of the cubes, but the affected regions are at the edges of the FOV where there is very little signal from SNR~0540, implying that the impact on the science is small. 

Another common issue with NIRSpec IFU observations is that the  undersampling of the PSF results in sinusoidal-like “wiggles" of spectra from single spatial pixels (spaxels). This primarily affects bright point sources and was detected for the pulsar in the observations of SNR~0540. The observations were corrected for this using the algorithm provided by \cite{Dumont2025}. The line-emitting ejecta studied in this paper are not affected by wiggles.

\begin{deluxetable*}{llccc}[t]
\tablecaption{JWST observations of SNR~0540 \label{tab:obs}}
\tablecolumns{5}
\tablenum{1}
\tablewidth{0pt}
\tablehead{
\colhead{Instrument\tablenotemark{a}} &
\colhead{Grating/filter} &
\colhead{Date} & 
\colhead{$t_{\rm exp}$} &
\colhead{$\lambda$ range} \\
\colhead{} &
\colhead{} &
\colhead{(YYYY-mm-dd)} &
\colhead{(s)} &
\colhead{($\mu$m)} 
}
\startdata
NIRSpec & G140H/F100LP & 2025-03-26 & 8753 & 0.97--1.89\tablenotemark{b} \\
NIRSpec & G140H/F100LP & 2025-03-26 & 8753 & 0.97--1.89\tablenotemark{b} \\
NIRSpec & G235M/F170LP & 2025-03-25 & 2918 & 1.66--3.17 \\
NIRSpec & G395M/F290LP & 2025-03-26 & 2918 & 2.87--5.27 \\
MRS & \nodata & 2025-02-14 & 5483\tablenotemark{c} & 4.9--28 \\
MRS\tablenotemark{d} & \nodata & 2025-02-14 & 2187\tablenotemark{c} & 4.9--28 \\
\enddata
\tablenotetext{a}{The footprints of the observations are shown in Figure~\ref{fig:fov}. The two G140H/F100LP pointings form a small mosaic, covering the total region shown by the dashed red line.}
\tablenotetext{b}{Wavelength gap between 1.41--1.49~$\mu$m.}
\tablenotetext{c}{Per sub-band.}
\tablenotetext{d}{Background field.}
\end{deluxetable*}

\subsection{MRS observations}
\label{sec:obs-mrs}

SNR~0540 was observed  with all four MRS channels (1--4) and the three dispersers, SHORT (A), MEDIUM (B), and LONG (C), providing complete spectral coverage from 4.9 to 28\,$\mu\rm{m}$ across twelve bands. The spectral resolving power decreases from $R \sim$~3500--2000 between channels 1--4, though it increases with wavelength within each band \citep{Pontoppidan2024}. The FOV increases from $3\farcs{2} \times 3\farcs{7}$ in channel 1 to $6\farcs{6} \times 7\farcs{7}$ in channel 4 \citep{Wells2015,Argyriou2023} as illustrated in  Figure~\ref{fig:fov}. 

For the target field observations, the four--point extended-source negative dither pattern was used, with five integrations of 98 groups per exposure in the FASTR1 readout mode, for a total exposure time of 5483.5~s per MIRI MRS band. An additional dedicated background observation was acquired with the same dither pattern and readout mode, with two integrations of 98 groups per exposure, for a total exposure time of 2186.7~s per MIRI MRS band. The background field was selected as a source-free region $\sim 33^{\prime \prime}$ south of SNR~0540.

All data were downloaded from MAST and processed with version 1.18.0 of the JWST Science Calibration Pipeline \citep{Bushouse2023}, using CRDS version 12.1.8 and CRDS context \texttt{jwst\_1364.pmap}. All level~1b ramp files (\texttt{\*\/\_uncal.fits}) were processed through the \texttt{calwebb\_detector1} pipeline to produce level~2a uncalibrated rate images (\texttt{\*\/\_rate.fits}). These rate images were then processed with \texttt{calwebb\_spec2} to produce level~2b calibrated rate images (\texttt{\*\/\_cal.fits}). In \texttt{calwebb\_spec2}, we applied the image-from-image background subtraction using the dedicated background exposures. In addition, three non-default corrections were applied during this stage: (1) the “clean\_showers" option in the \texttt{straylight} step, which removes residual cosmic-ray showers and is newly available in pipeline version~1.18.0; (2) the \texttt{residual\_fringe} step, which applies a 2D correction based on known fringe frequencies; and (3) the \texttt{badpix\_selfcal} step, with “flagfrac\_lower" and “flagfrac\_upper" set to 0.005, which flags persistent hot and cold pixels via self-calibration. We processed the calibrated rate images through \texttt{calwebb\_spec3} to produce the spectral cubes for each of the twelve MRS bands by setting the \texttt{cube\_build} parameter “output\_type" to “band". Furthermore, we set the “threshold\_percent" parameter of the \texttt{outlier\_detection} step to 99 to remove outliers in the background subtracted images that made it through both the \texttt{calwebb\_detector1} and \texttt{calwebb\_spec2} pipelines.

\subsection{WCS correction}
\label{sec:wcs}

All MRS data products have an absolute WCS uncertainty of $0\farcs{3}$ due to the guide star catalog and spacecraft roll uncertainty \citep{Patapis2024}. To improve the astrometry, we aligned the astrometric solution with the Gaia~DR3 frame \citep{GaiaCollaboration2016,GaiaCollaboration2023} using the bright pulsar PSR~J0540--6919, which is visible in all MRS channels. We fitted 2D Gaussian functions to estimate the pulsar centroid in channels 1 and 2, integrating over continuum regions and masking strong emission-line features. We were unable to accurately determine the pulsar position in channels 3 and 4 due to their low resolution. Despite this, we used the morphology of the extended emission to confirm that the MRS IFU cubes from different bands are internally well aligned. 

We used the channel 1 fits for the final results as the fits in channel 2 were more uncertain due to contamination by a nearby star. The mean pulsar position from the three channel 1 sub-bands was cross-matched with the Gaia~DR3 coordinate (${\alpha = 5^{\mathrm{h}}~ 40^{\rm{m}}~ 11^{\rm{s}}.2109}$, ${\delta = -69^{\circ}~ 19'~ 54''.1190}$), which revealed an offset of approximately $0\farcs{15}$ (J2016 epoch). The offset is larger if the epoch is set to the observational epoch ($\sim$J2025), but the relative proper motion uncertainties are $\sim 60\%$. We applied the same translational shift to the \texttt{CRVAL} keyword values in all headers to match the data to the absolute Gaia~DR3 frame.

For consistency with the MRS analysis above, we aligned the NIRSpec data with the Gaia~DR3 frame using the same procedure. We estimated the pulsar position in each of the three gratings and found an offset relative to the Gaia~DR3 frame of $\sim 0\farcs{13}$.

The fitting positional errors of each MRS band in channel 1 were $\sim 15$~mas, while the standard deviation of the three bands was 8~mas. The fitting positional errors of each NIRSpec grating were $<4$~mas, while the standard deviation of the three gratings was 4~mas. The standard deviation of all NIRSpec gratings and channel 1 bands was $\sim 18$~mas.

\subsection{Systemic velocity}
\label{sec:sysvel}

We determined the systemic velocity of SNR~0540 from narrow emission lines that originate from the interstellar medium (ISM) surrounding the remnant. We extracted spectra from different peripheral regions of the remnant to reduce contamination from continuum and broad emission. We selected the following lines from the NIRSpec G140H/F100LP grating/filter and MRS channels 2, 3, and 4: \ion{He}{1}~1.0833~$\mu$m, \ion{H}{1}~1.0941~$\mu$m, \ion{H}{1}~1.2822~$\mu$m, \ion{H}{1}~1.8756~$\mu$m, [\ion{Ar}{3}]~8.9914~$\mu$m, [\ion{S}{4}]~10.5105~$\mu$m, [\ion{Ne}{2}]~12.8135~$\mu$m, and [\ion{S}{3}]~18.7130~$\mu$m. These lines have high S/N ratios and widths comparable to the instrumental resolution at the respective wavelengths. We estimated an average systemic velocity of 281.6~\kms\ with a standard deviation about the average of 3.9~\kms. This is consistent with the systemic velocity of $277.5 \pm 6.5$~\kms\ estimated from the previous MUSE observations, where the uncertainty represents the standard deviation obtained by fitting different lines \citep{Larsson2021}. All spectra presented below have been corrected for the systemic velocity of 281.6~\kms.

\subsection{MUSE observations}

SNR~0540 was observed with MUSE in 2019 Jan--March as part of ESO program 0102.D-0769. The observations were taken in the wide field mode, which provides a $1\arcmin \times 1 \arcmin$ FOV with $0\farcs{2}$ sampling over the 4650--9300~$\AA$ wavelength interval. The spectral resolving power increases with wavelength from $R \sim$ 1750--3750. The observations used AO, which results in a PSF with a narrow core and broad wings (see also \citealt{Tenhu2024}). If approximated by a single Gaussian, the FWHM decreases from $\sim 1\farcs{0} - 0\farcs{55}$ over the observed wavelength interval. 

A 3D reconstruction of the brightest emission lines from these observations has been presented in \cite{Larsson2021}. Here we focus our comparison on the [\ion{S}{3}]~0.9069~$\mu$m line, which offers the best resolution, and the deblended [\ion{O}{3}]~0.4959, 0.5007~$\mu$m doublet (referred to as [\ion{O}{3}]~0.5007~$\mu$m from here on), which exhibits a somewhat different morphology. For consistency with the JWST observations, we corrected the MUSE data for the new systemic velocity estimated in Section~\ref{sec:sysvel}, though we note that the small difference is not noticeable in the 3D maps.  

\section{Construction of 3D emissivity maps}
\label{sec-mapmaking}

\begin{deluxetable*}{lrcccc}[t]
\tablecaption{Strong emission lines used for 3D emissivity maps \label{tab:lines}}
\tablecolumns{6}
\tablenum{2}
\tablewidth{0pt}
\tablehead{
\colhead{Line ID} &
\colhead{$\lambda$} & 
\colhead{Instrument} &
\colhead{Channel/grating\tablenotemark{a}} &
\colhead{$1/R$\tablenotemark{b}} &
\colhead{$\theta$\tablenotemark{c}} \\
\colhead{} &
\colhead{($\mu$m)} &
\colhead{} &
\colhead{} &
\colhead{(\kms)} &
\colhead{(\kms)} 
}
\startdata
\ion{He}{1} & 1.0833 & NIRSpec & G140H & 157 & 39 \\
$[$\ion{Fe}{2}$]$ & 1.2570 & NIRSpec & G140H & 125 & 39\\
$[$\ion{Fe}{2}$]$ & 1.6440 & NIRSpec & G140H  & 93 & 39 \\
\ion{H}{1} & 1.8756 & NIRSpec & G140H & 80 & 39 \\
$[$\ion{Fe}{2}$]$ & 5.3402  & MRS & Ch1 & 99 & 60 \\
$[$\ion{Ni}{2}$]$ & 6.6360 & MRS & Ch1 & 87 & 69 \\
$[$\ion{Ar}{2}$]$ & 6.9853 & MRS & Ch1 & 82 & 72 \\
$[$\ion{Ar}{3}$]$ & 8.9914 & MRS & Ch2 & 92 & 86 \\
$[$\ion{S}{4}$]$ & 10.5105 & MRS & Ch2 & 118 & 97 \\
$[$\ion{Ne}{2}$]$ & 12.8135 & MRS & Ch3 & 100 & 113 \\
$[$\ion{Ne}{3}$]$ & 15.5550 & MRS & Ch3 & 125 & 132 \\
$[$\ion{Fe}{2}$]$ & 17.9360  & MRS & Ch4 & 161 & 159 \\
$[$\ion{S}{3}$]$ & 18.7130 & MRS & Ch4 & 147 &  155 \\
\enddata
\tablenotetext{a}{See Figure~\ref{fig:fov} for the corresponding FOV.}
\tablenotetext{b}{Spectral resolution at the line wavelength, translated to  FWHM in velocity space. $R$ taken from \citet{Pontoppidan2024} for MRS and from the JWST user documentation (\url{https://jwst-docs.stsci.edu/\#gsc.tab=0}, JDox) for NIRSpec.}
\tablenotetext{c}{Spatial resolution at the line wavelength, expressed as a FWHM for the velocity of freely expanding ejecta in SNR~0540 (see text for details). The resolution was measured from the pulsar in NIRSpec and taken from \citet[][their equation 1]{Law2023} for the MRS.
}
\end{deluxetable*}

The premise for obtaining 3D emissivity maps from the observations is that the line-emitting ejecta are expanding freely. This implies that $v_{\rm ej} = d/t_{\rm exp}$, where  $v_{\rm ej}$ is the velocity of ejecta at a distance $d$ from the center of the explosion, and $t_{\rm exp}$ is the time since explosion. The assumption of free expansion should be reasonable even in the likely scenario that the ejecta are ionized by shocks, as the shock velocity into the ejecta is expected to be low. Specifically, \cite{Williams2008} find $\sim 20$~\kms\ for shocks in the dense clumps, which is small compared to the typical free-expansion velocities of $~\sim$400--1000~\kms\ in the observed region. 

As the center of explosion, we take the position of the pulsar on the sky (Section~\ref{sec:wcs}) and the systemic velocity of 282~\kms\ measured for the local ISM (Section~\ref{sec:sysvel}). The assumption that the pulsar is close to the center of explosion in the sky plane is supported by the upper limit on its transverse velocity of $<250$~\kms\ \citep{Mignani2010}. We further assume that $t_{\rm exp}=1100$~years, which was also used for the previous 3D mapping of SNR~0540 with MUSE \citep{Larsson2021}, in line with the age estimate of $1146 \pm 116$~years obtained from the same MUSE observations. The expansion of the remnant in the five years between the MUSE and JWST observations is negligible compared to the resolution of the observations (discussed below).

We perform the reconstruction in velocity space as the velocities of the freely expanding ejecta are directly related to the physical properties of the explosion. To this end, we translate the position of each spaxel in the data cubes to a velocity in the sky plane based on the separation from the pulsar. For a distance to the LMC of 49.6~kpc \citep{Pietrzynski2019} and $t_{\rm exp}=1100$~years, an angular separation of $1^{\prime \prime}$ is equivalent to 214~\kms. We define $v_{\rm x}$ as the velocity in the east-west direction and $v_{\rm y}$ as the velocity in the south-north direction, implying that $v_{\rm x}$ is negative east of the pulsar and $v_{\rm y}$ is negative south of the pulsar. Similarly, the Doppler shifts of emission lines in each spaxel are translated to a velocity along the line of sight, $v_{\rm z}$, where negative (blueshifted) velocities are on the near side towards the observer, following the standard convention. 

We inspected the full wavelength range of the observations from 1--28~$\mu$m for lines that are suitable for 3D mapping. The main criteria are that the lines should be strong (giving high S/N in individual spaxels), isolated from other emission lines, and allow for an accurate continuum subtraction in individual spaxels. The selected lines are listed in Table~\ref{tab:lines}. The only strong line that was disqualified due to the last condition is [\ion{Fe}{3}]~22.9250~$\mu$m, where there is a high noise level in the surrounding continuum, while the strongest lines excluded due to blending are [\ion{O}{4}]~25.8903~$\mu$m and [\ion{Fe}{2}]~25.9884~$\mu$m. These lines and the many weaker emission lines detected in the full spectrum will be discussed in C. Tegkelidis et al. (in preparation). 

Table~\ref{tab:lines} also provides the resolution in velocity space for all lines in both the spectral and spatial directions, expressed as the full width at half maximum (FWHM) for an unresolved line from a point source. The velocity-space spectral resolution is simply based on the Doppler shift, while the angular resolution ($\theta$) was translated to velocity using the conversion $1^{\prime \prime}= 214$~\kms\ noted above. We note that the PSF is slightly elongated in both NIRSpec \citep[e.g.,][]{Bentz2025} and MIRI/MRS (especially in Channels 1 and 4, \citealt{Law2023}), while the values in Table~\ref{tab:lines} represent an average. As is clear from  Table~\ref{tab:lines}, the resolution is strongly wavelength-dependent, which needs to be kept in mind for comparisons between different lines. The decrease in spatial resolution with increasing wavelength implies that the velocity-space resolution is better in the spatial than spectral direction for the lines at the shortest wavelengths observed with NIRSpec, reaching a maximum difference of a factor $\sim 4$ for \ion{He}{1}~1.0833~$\mu$m. The resolutions in the two directions are more comparable for the MRS data, where they increases from $\sim$80~\kms\ in Channel 1 to $\sim$150~\kms\ in Channel 4. 

To obtain the final 3D maps for all the lines in Table~\ref{tab:lines}, we need to subtract the continuum emission and any narrow lines from the ISM. We determine the continuum level in each spaxel by fitting a linear model to 500--1000~\kms\ wide velocity intervals on both sides of the lines, and then subtract the best-fitting model from the data. The narrow lines from the ISM vary significantly across the FOV, resulting in different levels of contamination in each spaxel. For NIRSpec, we correct for this following a similar approach to that used for the MUSE data in \cite{Larsson2021}. We fit narrow wavelength intervals around the systemic velocity in each spaxel with a model comprising a narrow Gaussian and linear component, where the former captures the ISM component and the latter approximates the broad emission from the ejecta over a narrow interval. The width of the narrow Gaussian is constrained to be consistent with the spectral resolution, and the width of the fitted wavelength interval is set to $2.2 \times$ the FWHM of the resolution at the relevant wavelength. The main caveat with this approach is that the fits fail to capture the narrow ISM component if there is a strong peak in the ejecta emission at a similar wavelength. We detect narrow ISM components in all the NIRSpec lines listed in Table~\ref{tab:lines}, but only fit and subtract it for \ion{He}{1}~1.0833~$\mu$m and \ion{H}{1}~1.8756~$\mu$m, as it is weak compared to the ejecta emission in the [\ion{Fe}{2}] lines.

In the case of the MRS data, the ISM components are removed as part of the background subtraction (Section~\ref{sec:obs-mrs}), as the background field contained some diffuse emission. This in fact oversubtracts the ISM component in most cases, which is especially apparent for the Ne lines. We experimented with using the MRS cubes without background subtraction and correcting for the ISM line as for NIRSpec, which solves the problem of oversubtraction, but comes at the expense of increased instrumental background/artifacts. We therefore present results based on the background-subtracted cubes, but inspect the non-background subtracted ones to compare with and verify the main results. In summary, the final continuum- and ISM-corrected cubes in both NIRSpec and MRS are more uncertain close to $v_{\rm z}=0$~\kms, though the uncertainties are different for the two detectors. For the NIRSpec data, we also manually masked out a small number of field stars and obvious artifacts in the final cubes. This masking was typically applied to narrow wavelength intervals, the main exception being a star located just south of the pulsar, which was fully masked out in all NIRSpec cubes. As a final step, we interpolated all cubes onto a uniform grid with a step size of $20$~\kms\ to facilitate 3D visualization.  

\section{Results}
\label{sec:results}

\begin{figure*}[t]
\centering
\includegraphics[width=\hsize]{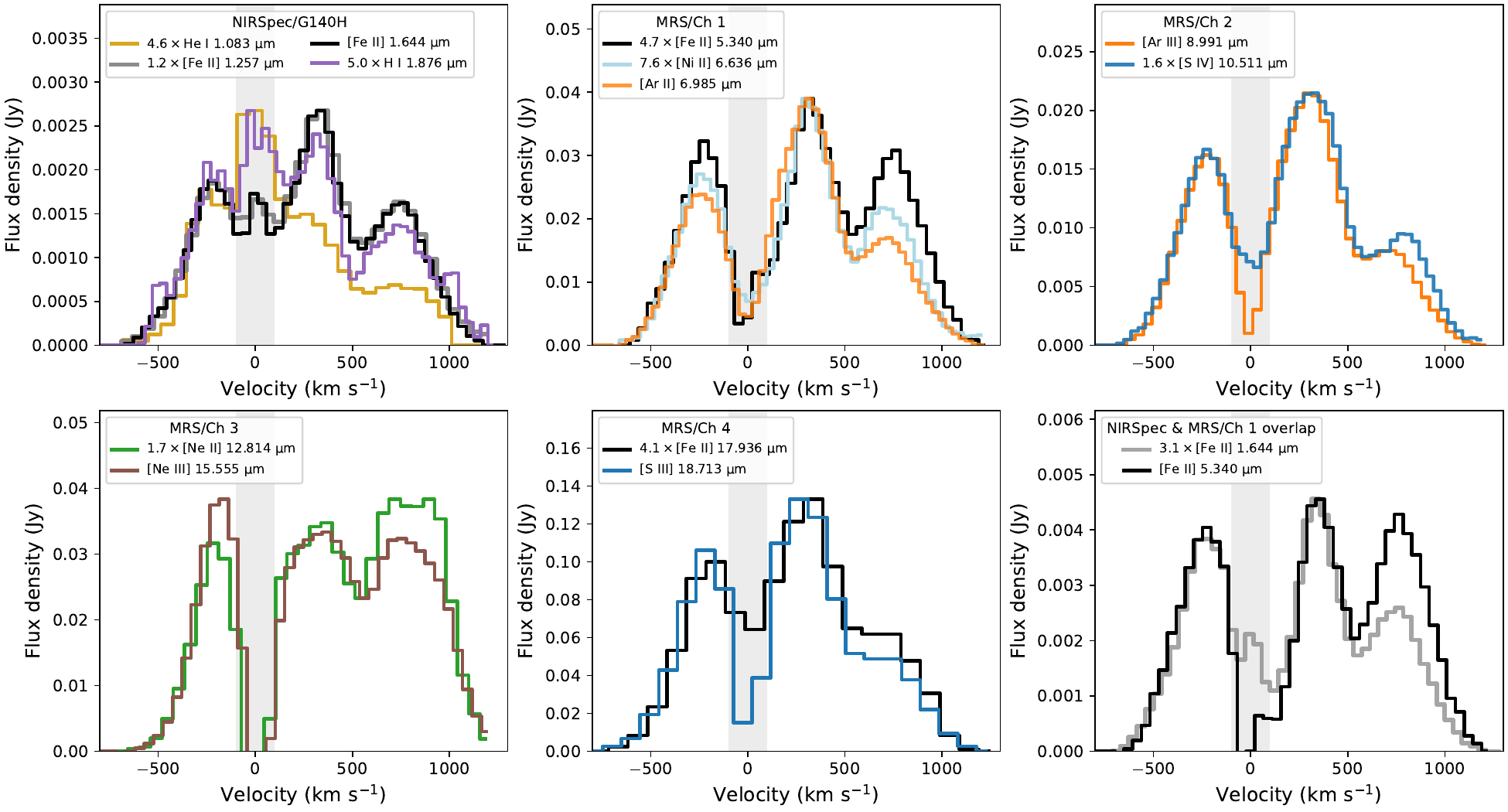}
\caption{Velocity profiles of all the lines listed in Table~\ref{tab:lines}. The profiles in the first five panels were extracted from the full FOV of NIRSpec/G140H and the different MRS channels (shown in Figure~\ref{fig:fov}). The bottom, right panel shows [\ion{Fe}{2}] lines extracted from the overlapping region between NIRSpec/G140H and MRS/Ch1. Note that the velocity interval $\sim \pm 100$~\kms\ (gray shaded region) is uncertain due to lines from the ISM, which are undersubtracted to different degrees for NIRSpec and typically oversubtracted for the MRS (especially the Ne line in the bottom, left panel). } 
\label{fig:profiles}
\end{figure*}
\begin{figure*}[t]
\centering
\includegraphics[width=\hsize]{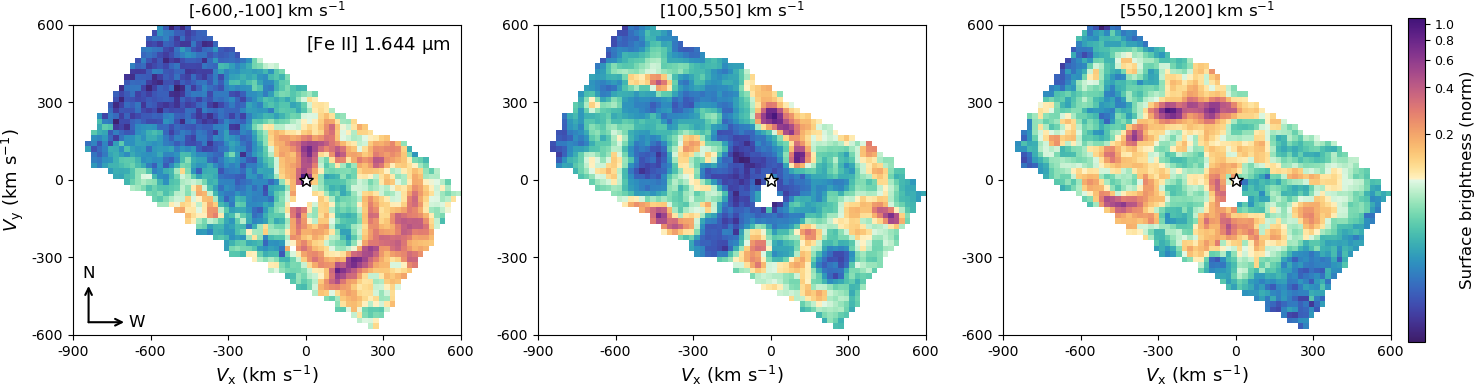}
\caption{Images of the [\ion{Fe}{2}]~1.644~$\mu$m emission in three Doppler shift intervals, indicated at the top of each panel. The three intervals correspond to the three main peaks in the integrated line profile (Figure~\ref{fig:profiles}, top left). The velocities of freely expanding ejecta in the plane of the sky are shown on the x- and y-axes. The white star symbol shows the position of the pulsar, which is taken as the center of explosion at 0~\kms. The white region just south of the pulsar was masked out due to contamination by a star.} 
\label{fig:fe_slices}
\end{figure*}

\begin{figure*}[h!]
\centering
\includegraphics[width=\hsize]{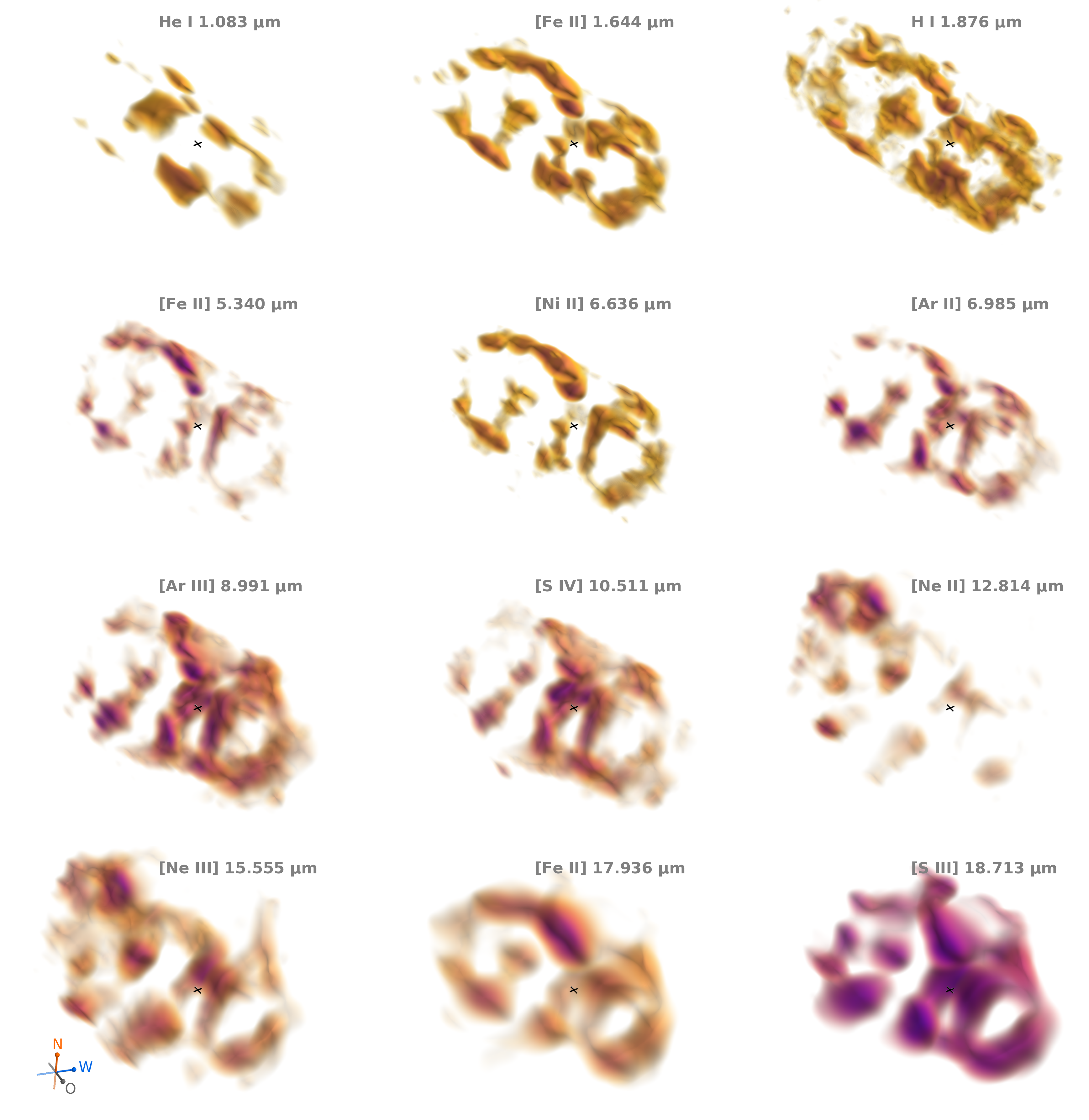}
\caption{3D volume renderings of bright emission lines in SNR~0540. Darker more opaque colors correspond to brighter emission in all 3D maps, while three different opacity transfer functions have been used to highlight the bright large-scale structure in each line, considering resolution and line strength. For the NIRSpec lines (top row) and [\ion{Ni}{2}], the lowest intensities plotted is 30\% of the peak value, and higher opacities are given to medium intensities (30--70\%) compared to the other lines. The lowest intensities plotted for the other lines correspond to 15\% of the peak, except for [\ion{S}{3}] for which the threshold is set to 35\%. The NIRSpec data have been convolved with a Gaussian in the spatial direction to further highlight the large-scale structures (see text for details). The viewing angle is shown by the compass in the bottom left panel, where the direction of the observer is out of the page. The axes of the compass all have a length of 200~\kms. The 3D maps extend to 1200~\kms\ in all directions, though the FOVs do not fully cover this region (with differences between the lines, see Figure~\ref{fig:fov} and Appendix~\ref{sec:app-sideviews}). The assumed center of explosion is marked by a black cross. An animated (rotating) version of this figure is available. The video shows one rotation of all the panels.
}   
\label{fig:3dvol_a}
\end{figure*}
\begin{figure*}[h!]
\centering
\includegraphics[width= \hsize]{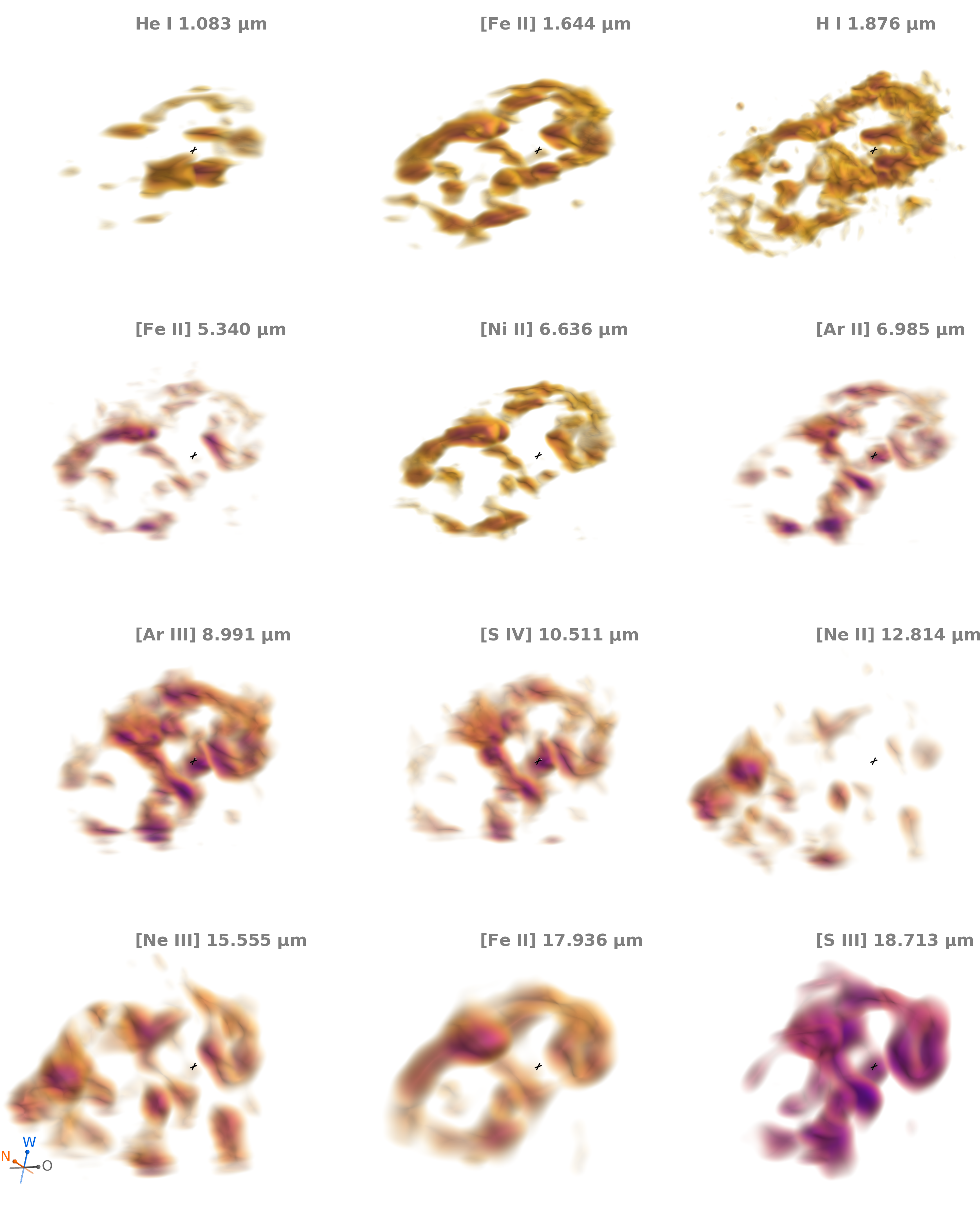}
\caption{3D volume renderings as in Figure~\ref{fig:3dvol_a}, but with a different viewing angle, indicated by the compass in the bottom left panel. The length of each compass axis is 200~\kms\ and the north direction is out of the page. An animated (rotating) version of this figure is available. The video shows one rotation of all the panels.} 
\label{fig:3dvol_b}
\end{figure*}
\begin{figure*}[h!]
\centering
\includegraphics[width= \hsize]{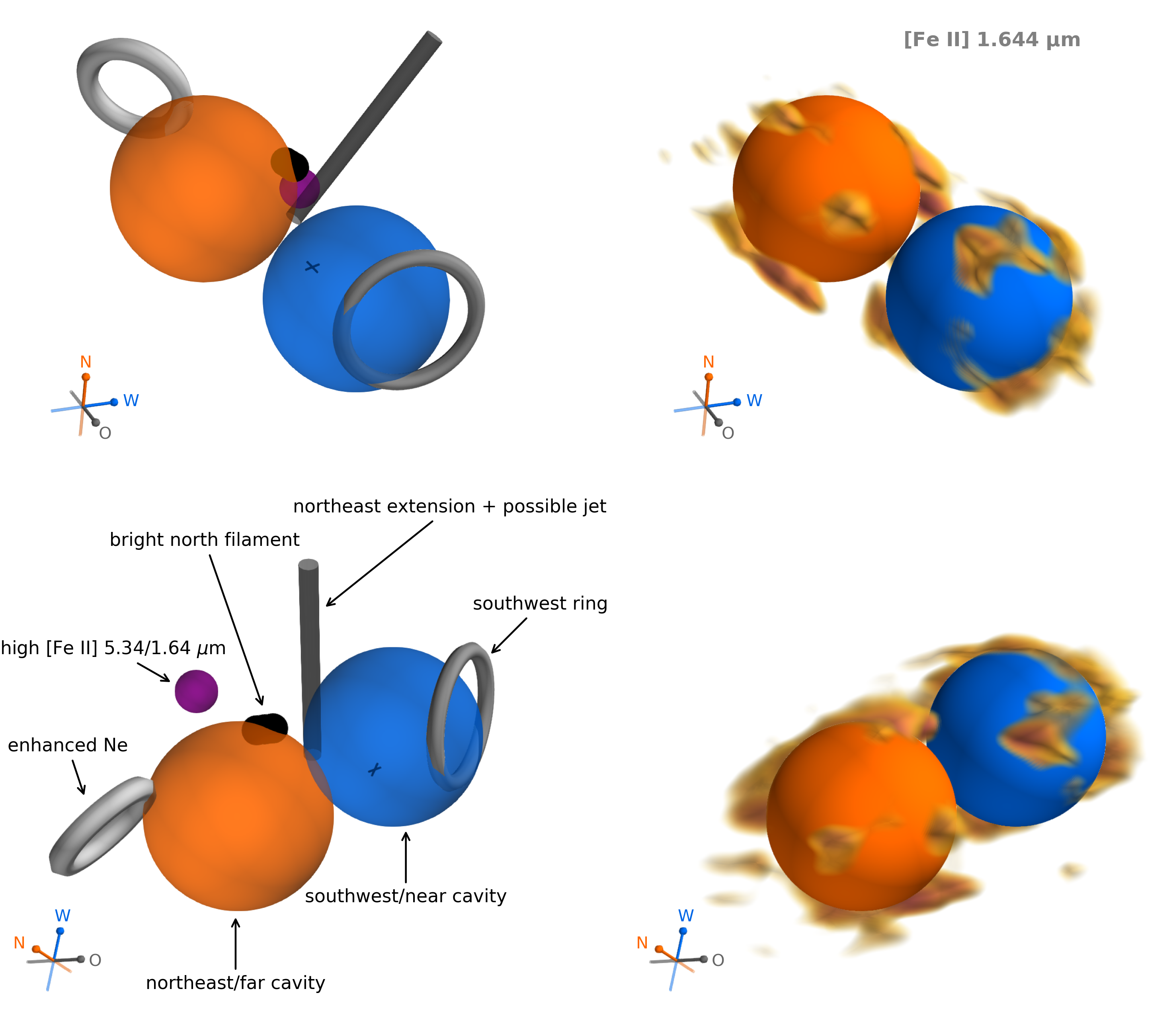}
\caption{Schematic view of the main components of the 3D morphology. The top and bottom rows show the same viewing angles as Figures \ref{fig:3dvol_a} and \ref{fig:3dvol_b}, respectively, as indicated by the compasses in the lower left corners of all panels. The length of each compass axis is 200~\kms\ and the assumed center of explosion is marked by a black cross.  All the morphological components are  labeled in the bottom left panel. We note that much of the northeast extension is outside the FOV of the JWST observations, but captured in the MUSE observations (Figure~\ref{fig:fov} and middle panel of Figure~\ref{fig:3dvol_linecomp}). The narrow tube is only meant to point out the direction of this extension and possible jet, while the observed emission covers a larger area. In the case of the enhanced Ne emission and high [\ion{Fe}{2}] 5.3402/1.6440~$\mu$m ratio, we have only highlighted some prominent features that stand out in the irregular emission regions. The two spheres that approximate the cavity have diameters of 600~\kms. The right panels show the [\ion{Fe}{2}]~1.6440~$\mu$m emission together with the two spheres, where the latter are plotted with full opacity to highlight the 3D structure of the emission.} 
\label{fig:cartoon}
\end{figure*}

To provide a first overview of the line emission, we show the integrated velocity profiles from the entire observed region in Figure~\ref{fig:profiles}. We show the lines from NIRSpec/G140H and all the MRS channels in five separate panels as the FOVs are slightly different (Figure~\ref{fig:fov}), while the sixth panel shows the [\ion{Fe}{2}]~1.6440~$\mu$m and 5.3402~$\mu$m lines extracted from the overlapping region between NIRSpec/G140H and MRS channel 1. The figure shows that all emission lines are more extended on the redshifted side, covering a total velocity interval $\sim [-700, 1200]$~\kms, consistent with previous observations at both optical and IR wavelengths \citep{Kirshner1989,Morse2006, Williams2008,Sandin2013,Larsson2021,Lundqvist2022}. 

The profiles are all characterized by three broad peaks at $\sim -230$, 320, and 750~\kms, albeit with some differences in the relative intensities of the peaks for different lines. It is clear that some of these differences are due to spatial variations of the physical conditions of the ejecta, considering that there are differences between [\ion{Ne}{2}]~12.8135~$\mu$m and [\ion{Ne}{3}]~15.5550~$\mu$m (Figure~\ref{fig:3dvol_linecomp}, bottom left panel) and [\ion{Fe}{2}]~1.6440 and 5.3402~$\mu$m (Figure~\ref{fig:3dvol_linecomp} bottom right panel), where each pair of lines was extracted from the same region. The over/under subtraction of the narrow ISM component at $\sim 0$~\kms\ discussed in Section~\ref{sec-mapmaking} is also evident in many of the line profiles in Figure~\ref{fig:profiles}. 

Figure~\ref{fig:fe_slices} shows three images of the [\ion{Fe}{2}]~1.6440~$\mu$m line, created by integrating the emission over each of the spectral peaks. The [\ion{Fe}{2}]~1.6440~$\mu$m line is the brightest of the lines in the NIRSpec range, where the spatial resolution is highest, which makes it the best suited for investigating the spatial morphology of the emission. Figure~\ref{fig:fe_slices} shows that the emission originates from complex extended filaments and knots, rather than one main emission region per peak. The most coherent structure is a ring of blueshifted emission in the southwest, seen in the left panel of Figure~\ref{fig:fe_slices}.  

Figures~\ref{fig:3dvol_a} and \ref{fig:3dvol_b} show 3D volume renderings of all the emission lines for two different viewing angles. We also provide animations of all the lines as online material, which give the clearest view of the 3D morphology. The only line from Table~\ref{tab:lines} omitted in Figures~\ref{fig:3dvol_a} and \ref{fig:3dvol_b} is [\ion{Fe}{2}]~1.2570~$\mu$m, which shows the same morphology as [\ion{Fe}{2}]~1.6440~$\mu$m and will be discussed in more detail in Section~\ref{sec:ratios}. The opacity transfer functions (i.e., the mapping between the intensity in the cubes and the opacity of the plotted data points) have been set to highlight the strongest emission in each line. The lowest levels plotted correspond to 15--35\% of the peak intensities of the different lines, while the noise levels measured in the adjacent continuum is $\lesssim 1$\%\ (1$\sigma$) for all lines except \ion{H}{1}~1.8756~$\mu$m, for which it is $\sim 3$\%. To highlight the large-scale structure, we have also smoothed the 3D maps from NIRSpec with a Gaussian in the spatial direction to match the spatial resolution of the [\ion{Fe}{2}]~5.3402~$\mu$m line in the short-wavelength end of the MRS. The large differences between the resolution in the spatial and spectral directions for NIRSpec make it harder to see the main 3D structures without this smoothing. 

Most of the 3D maps show that the strongest emission is distributed in a thin layer around an inner cavity, which can be approximated by two adjacent spheres, as illustrated by the schematic in Figure~\ref{fig:cartoon}. The emission does not uniformly surround the cavity, but is highly fragmented into knots, filaments and ring-like features. Figures~\ref{fig:3dvol_a} and \ref{fig:3dvol_b} also show a number of differences in the 3D morphology of the emission lines, some of which are related to the wavelength-dependent resolution (Table~\ref{tab:lines}) and FOVs (Figure~\ref{fig:fov}). The most notable real difference is in the two Ne lines, which show additional extended emission on the far side from the observer, including a small redshifted ring that strongly dominates the [\ion{Ne}{2}]~12.8135~$\mu$m emission (most clearly seen for the viewing angle in Figure~\ref{fig:3dvol_a}, see also schematic in Figure~\ref{fig:cartoon}). 

\begin{figure}[t]
\centering
\includegraphics[width=\hsize]{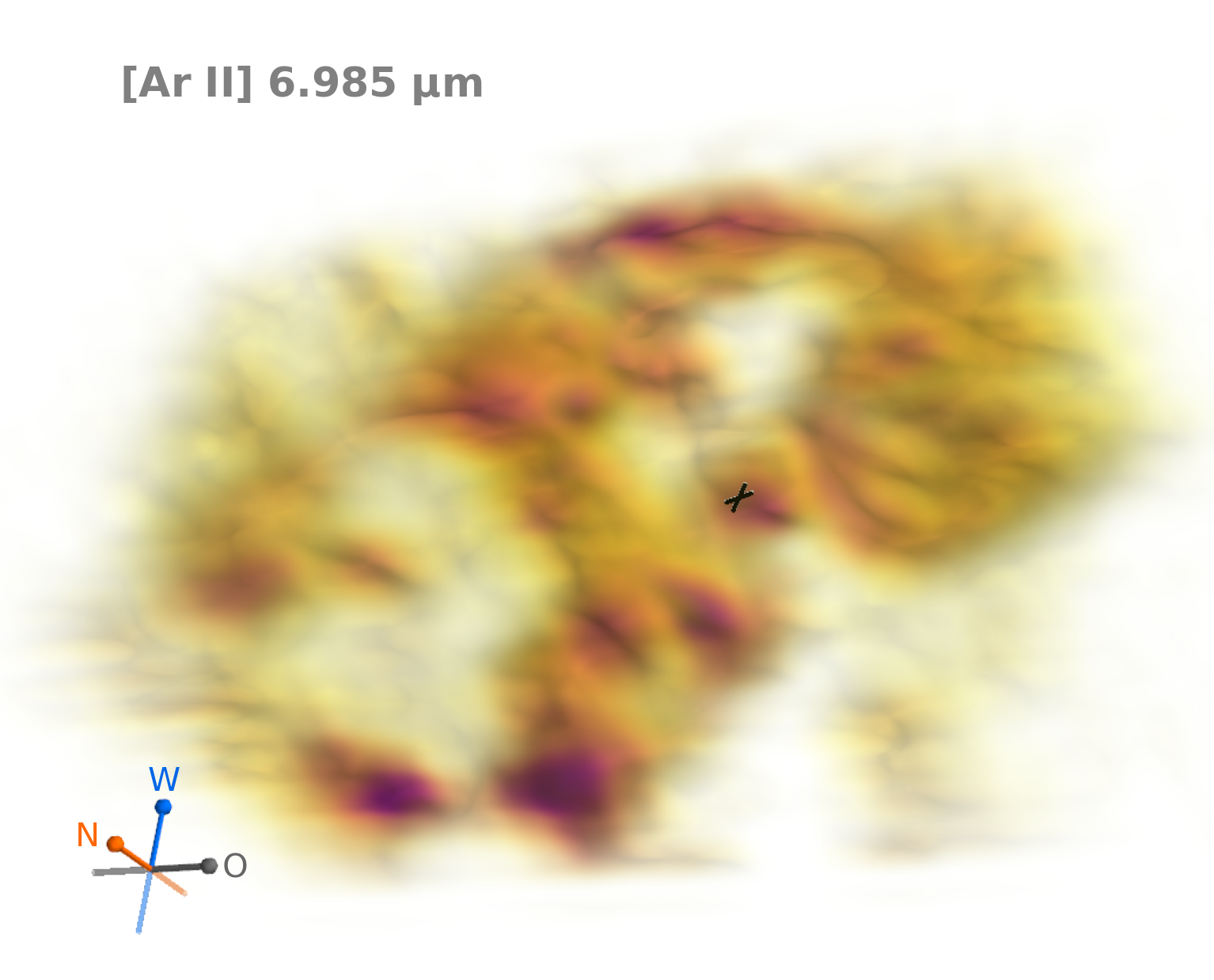}
\caption{3D volume rendering for [\ion{Ar}{2}]~6.9853~$\mu$m. Same as the right panel of the second row in Figure~\ref{fig:3dvol_b}, but with the opacity transfer function set to highlight faint emission (a higher opacity is used for faint emission). The threshold for the faintest emission included corresponds to 10\% of the peak intensity. Darker more opaque colors correspond to brighter emission. An animated (rotating) version is available. The video shows one rotation.} 
\label{fig:3dvol_ar2_neb}
\end{figure}

\begin{figure*}[t]
\centering
\includegraphics[width=\hsize]{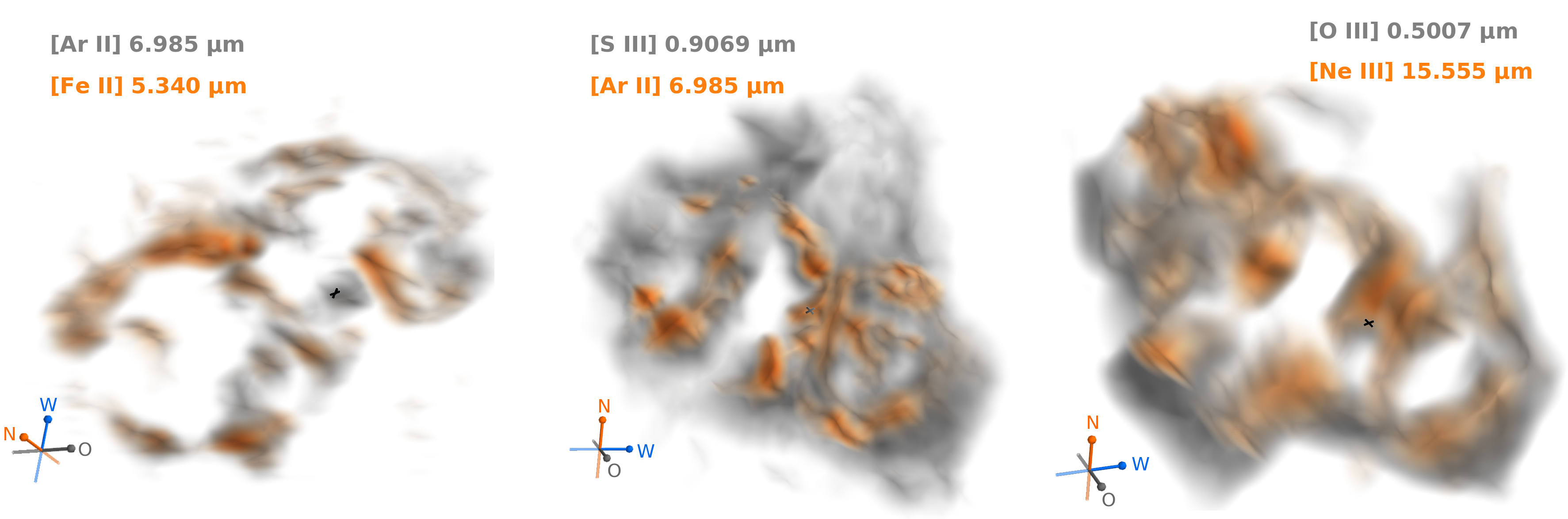}
\caption{Comparison of 3D emissivity maps of different lines, as indicated by the labels in each panel. The [\ion{S}{3}]~0.9069~$\mu$m and [\ion{O}{3}]~0.5007~$\mu$m lines are from MUSE \citep{Larsson2021}, while the others are from the JWST observations presented in Figures~\ref{fig:3dvol_a} and ~\ref{fig:3dvol_b}. The viewing angles indicated by the compass in each panel has been set to best highlight the differences and similarities in each comparison. The left and right panels have the same angle as Figures~\ref{fig:3dvol_b} and \ref{fig:3dvol_a}, respectively, while the middle panel is slightly rotated around the north axis with respect to the right panel. The length of each compass axis is 200~\kms. Darker more opaque colors correspond to brighter emission. The lowest intensities plotted correspond to 15, 20 and 35\% of the peak intensities for the JWST lines, [\ion{S}{3}]~0.9069~$\mu$m and [\ion{O}{3}]~0.5007~$\mu$m, respectively. The 3D maps extend to 1200~\kms\ in all directions, though we note that the FOVs of the JWST observations do not fully cover this region in the spatial direction (with differences between the lines, see Figure~\ref{fig:fov} and Appendix~\ref{sec:app-sideviews}).  An animated (rotating) version of this figure is available. The video shows one rotation of all the panels.} 
\label{fig:3dvol_linecomp}
\end{figure*}

The [\ion{Ar}{2}]~6.9853~$\mu$m is the brightest line in the short-wavelength range of the MRS, where the combined spatial and spectral resolution is the best. This makes it optimal for investigating the 3D morphology down to fainter levels. Such a 3D map is seen in Figure~\ref{fig:3dvol_ar2_neb}, which reveals faint filamentary emission outside the two main lobes. This faint emission can also be seen in collapsed images of the 3D cubes of all the emission lines along different directions, the results of which we include in  Appendix~\ref{sec:app-sideviews}. These collapsed images also facilitate the quantitative comparison of different emission lines, though we caution that there is clearly a loss of information compared to the full 3D maps due to projection effects.  

Finally, we show a comparison between the 3D morphology of different pairs of lines in Figure~\ref{fig:3dvol_linecomp}. The left panel shows [\ion{Fe}{2}]~5.3402~$\mu$m together with [\ion{Ar}{2}]~6.9853~$\mu$m, which are both observed in MRS channel 1. This comparison highlights that both lines have the same overall morphology, but with spatial variation of the relative intensity, such as the filament on the north-far side that is very bright in [\ion{Fe}{2}] (also included in the schematic in Figure~\ref{fig:cartoon}). The middle panel instead shows the [\ion{Ar}{2}]~6.9853~$\mu$m line compared to [\ion{S}{3}]~0.9069~$\mu$m from MUSE, which serves to illustrate the 3D morphology of the emission that falls outside the MRS FOV. This includes a cone-like extension to the north, as well as the eastern end of the lobe on the far side. The resolution in velocity space for the [\ion{S}{3}]~0.9069~$\mu$m line is 92~\kms\ along the line of sight (corresponding to the spectral resolution) and $\sim 130$~\kms\ in the spatial direction, where we have approximated the MUSE PSF by a Gaussian. The better resolution of the [\ion{Ar}{2}]~6.9853~$\mu$m line (Table~\ref{tab:lines}) is evident in Figure~\ref{fig:3dvol_linecomp}, which makes the two lobes and fragmented nature of the emission stand out more clearly. The faint extended wings of the MUSE PSF further reduces the contrast in the 3D map of the [\ion{S}{3}] line.  

Among all the optical emission lines observed with MUSE, [\ion{O}{3}]~0.5007~$\mu$m shows the biggest differences compared to [\ion{S}{3}]~0.9069~$\mu$m, exhibiting more extended emission, in particular on the eastern and far sides. These differences are significant also considering the lower resolution at the [\ion{O}{3}] line (169~\kms\ and $\sim 180$~\kms\ in the spectral and spatial directions, respectively). We compare the morphology of the [\ion{O}{3}] line to [\ion{Ne}{3}]~15.5550~$\mu$m in the right panel of Figure~\ref{fig:3dvol_linecomp}. The two lines show a similar overall extent, though the easternmost side of the [\ion{O}{3}] emission is outside the FOV of the MRS. It is also notable that the small ring of [\ion{Ne}{3}] emission on the far side is bright relative to [\ion{O}{3}], though there is clearly some [\ion{O}{3}] emission from the same location, unlike from all the other emission lines in Figures~\ref{fig:3dvol_a} and \ref{fig:3dvol_b}.  

\subsection{3D peak/clump finding}
\label{sec:clumps}

In this Section we use the FellWalker clump-finding algorithm \citep{Berry2015} to quantify the 3D morphology of the different emission lines. A similar analysis was previously carried out on the MUSE data of SNR~0540 in \cite{Larsson2021}. The algorithm performs walks through the data cubes, following the path with the steepest gradient until a significant peak is identified. A clump is then defined as all the voxels (the individual elements of a cube) located on paths that reach a given peak. There are no restrictions on the shapes of the clumps, and all voxels in a cube that have an intensity above a given threshold will be assigned to a unique clump. As in \cite{Larsson2021}, we use a Python wrapper\footnote{\url{https://starlink-pywrapper.readthedocs.io/en/latest/index.html}} to run the Fellwalker algorithm implemented in the Starlink CUPID package \citep{Berry2007}.\footnote{https://starlink.eao.hawaii.edu/starlink/CUPID}  

We run the algorithm on the data cubes of all the lines listed in Table~\ref{tab:lines} and set the main input parameters based on the standard deviation ($\sigma$) measured in background regions of each cube. Specifically, we set the threshold intensity for assigning voxels to a clump to  $3\sigma$ (the \texttt{noise} parameter) and the smallest dip between two peaks that are considered to belong to different clumps to $2\sigma$ (the \texttt{MinDip} parameter). We further set the minimum number of voxels in a clump such that no clumps smaller than the resolution are included in the results. 

The number of clumps identified and the sizes of the clumps is strongly dependent on the resolution, which varies considerably with wavelength (Table~\ref{tab:lines}), as well as the FOV and S/N of a given line. We therefore only present result based on the positions (in velocity space) of the peaks of the clumps. We also investigated the peak intensity of the clumps, but found no notable patterns not already clear from the results presented above. The number of clumps identified ranges from 5 for [\ion{Fe}{1}]~17.936~$\mu$m and 9 for [\ion{S}{3}]~18.7130~$\mu$m at the longest wavelengths where the resolution is lowest, to 20--22 for [\ion{Fe}{2}]~$1.6440$~~$\mu$m, [\ion{Fe}{2}]~$5.3402$~~$\mu$m, [\ion{Ar}{2}]~$6.9853$~~$\mu$m, and [\ion{Ar}{3}]~$8.9914$~~$\mu$m. The remaining lines in Table~\ref{tab:lines} have 12--19 identified clumps.     

\begin{figure*}[t]
\centering
\includegraphics[width=\hsize]{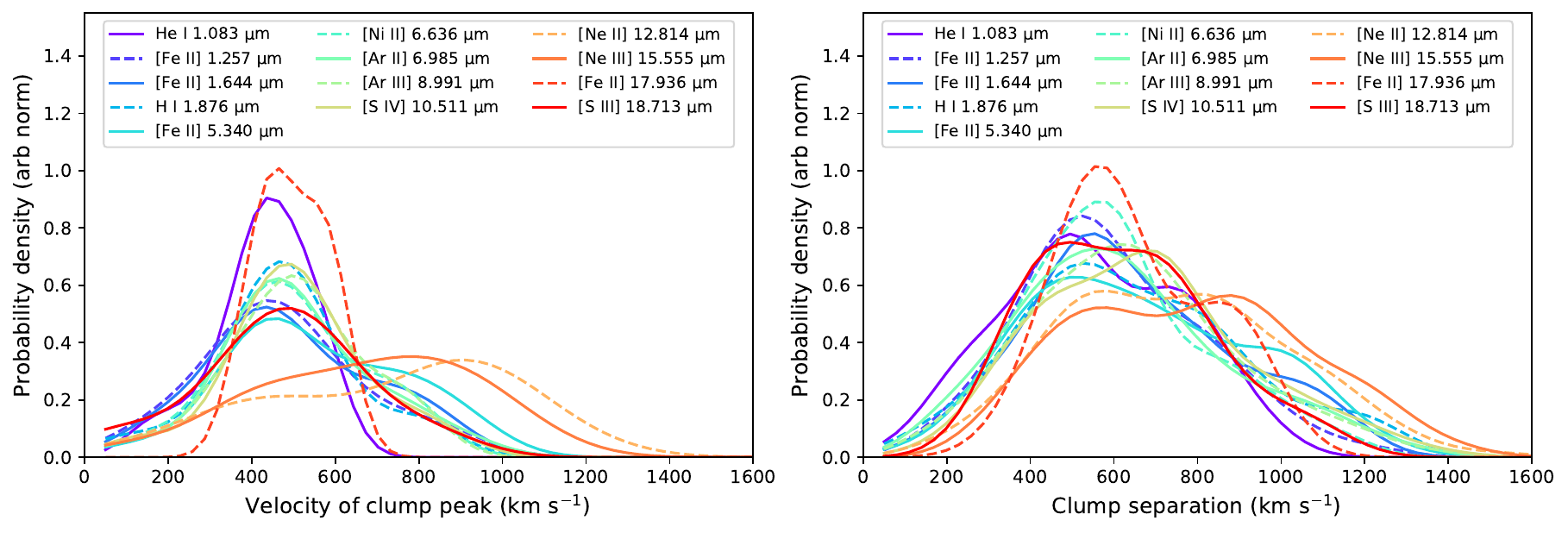}
\caption{Distributions of clumps for the different emission lines, shown as Gaussian kernel density estimates. Left: space velocities of the clump peaks. Right: separations in velocity between all pairs of clumps.}
\label{fig:clumpvel}
\end{figure*}
\begin{figure*}[t]
\centering
\includegraphics[width=\hsize]{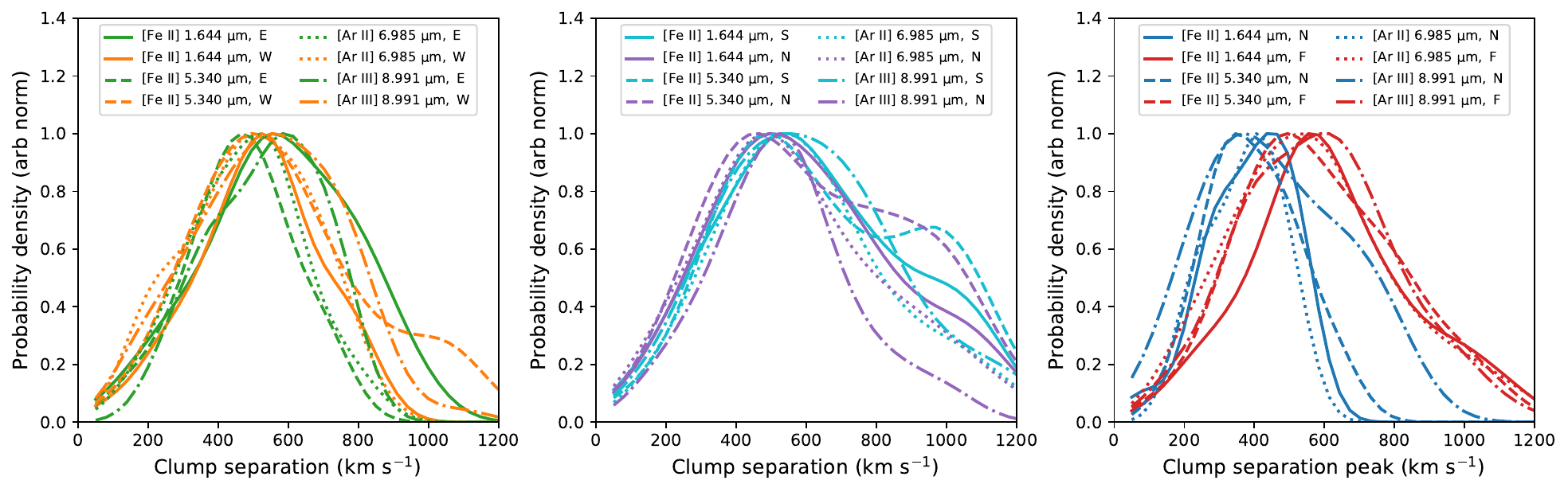}
\caption{Distributions of clump separations in different parts of the remnant. The left, middle, and right panels show the separations in the east/west, north/south, and near/far parts, respectively. All the Gaussian kernel density estimates have been normalized to 1 to facilitate comparison between the positions of the peaks.}
\label{fig:clumpsep}
\end{figure*}

The left panel of Figure~\ref{fig:clumpvel} shows the distribution of 3D space velocities of the peaks of the clumps. The distributions are shown as Gaussian kernel density estimates, which provide an unbinned representation of the probability distributions. The majority of the lines have distributions that peak in the range $\sim$430--500~\kms, with tails extending to $\sim$ 1000~\kms. The main outliers are the Ne lines, which show broader distributions with peaks at $\sim$ 800--900~\kms\ in line with their different observed morphologies discussed above (Figures~\ref{fig:3dvol_a} and \ref{fig:3dvol_b}). We also note that the \ion{He}{1}~1.083~$\mu$m and [\ion{Fe}{2}]~17.936~$\mu$m lines show narrower distributions with more pronounced peaks. In the case of [\ion{Fe}{2}] this is clearly related to the relatively low resolution and small number of clumps identified, while the \ion{He}{1} distribution is based on 14 clumps and is more likely to reflect differences in physical conditions within the ejecta (Section~\ref{sec:disc-mixing}).   

The right panel of Figure~\ref{fig:clumpvel} shows the distribution of distances between all pairs of clumps for all lines. This reveals a range of substructures in the ejecta with diameters between $\sim $300--800~\kms, with some differences for the Ne lines and the [\ion{Fe}{2}]~17.936~$\mu$m line,  as expected from the velocity distributions discussed above. Given the two main lobes of ejecta seen in the 3D maps, it is interesting to investigate the distribution of distances in different parts of the ejecta. This is illustrated in Figure~\ref{fig:clumpsep}, where we have calculated the distance distributions separately for the east/west, north/south, and near/far sections of the ejecta, respectively. We only include the emission lines that have at least 20 clumps in this analysis. The east/west and north/south distributions all show peaks in the range 470--580~\kms, with no systematic differences between the sides. This  shows that the two lobes are approximately the same size and symmetric around the pulsar in the sky plane. Inspection of the 3D maps further shows that the observed emission morphology can be approximated by two spheres with diameters of $\sim 600$~\kms\ (Figure~\ref{fig:cartoon}). The fact that these diameters are slightly larger than the peaks of the clump-separation distributions reflect the fragmented nature of the emission, with many clumps being located close together.

All the 3D maps also show that the symmetry point between the two lobes is offset to the redshifted side from the assumed center of explosion. This is quantified by the distributions of clump separations on the near/far sides in the right panel of Figure~\ref{fig:clumpsep}, which show a systematic difference with smaller separations on the near sides in all emission lines, with the peaks being offset by 120--240~\kms. These results have implications for the pulsar kick and/or center of explosion, as discussed further in Section~\ref{sec:disc-kick}.

\subsection{Line ratios in 3D}
\label{sec:ratios}

The spectrum of SNR~0540 enables the study of many line ratios that probe the physical conditions in the ejecta. Here we consider the ratios of lines that are bright enough to be investigated in 3D, while a full analysis of relevant line ratios will be presented in Tegkelidis et al. (in preparation). Specifically, we consider the ratios of the [\ion{Fe}{2}] lines 1.2570/1.6440~$\mu$m and 5.3402/1.6440~$\mu$m, where the former gives information about dust extinction  and the latter is sensitive to temperature and density. We note that the [\ion{Fe}{2}] 17.9360/1.6440~$\mu$m ratio also provides a density diagnostic, but this ratio is not useful to consider in 3D due to the large differences in spatial resolution between the two lines. 

To take line ratios in 3D we first convolve the data cube of the [\ion{Fe}{2}]~1.6440~$\mu$m line with Gaussians to match the somewhat lower resolution of the two other lines (see Table~\ref{tab:lines}, only the spectral dimension is adjusted when comparing to the 1.2570~$\mu$m line). We also convert the units of the data cubes from flux densities to fluxes, and limit the size of the cubes in the 5.3402/1.6440~$\mu$m comparison to only include the overlapping region, excluding the uncertain edges of the FOV. Finally, we only take the ratios of voxels where the flux of both lines are above a threshold value corresponding to 3$\sigma$ fluctuations of the background. 

\begin{figure*}[t]
\centering
\includegraphics[width=0.75 \hsize]{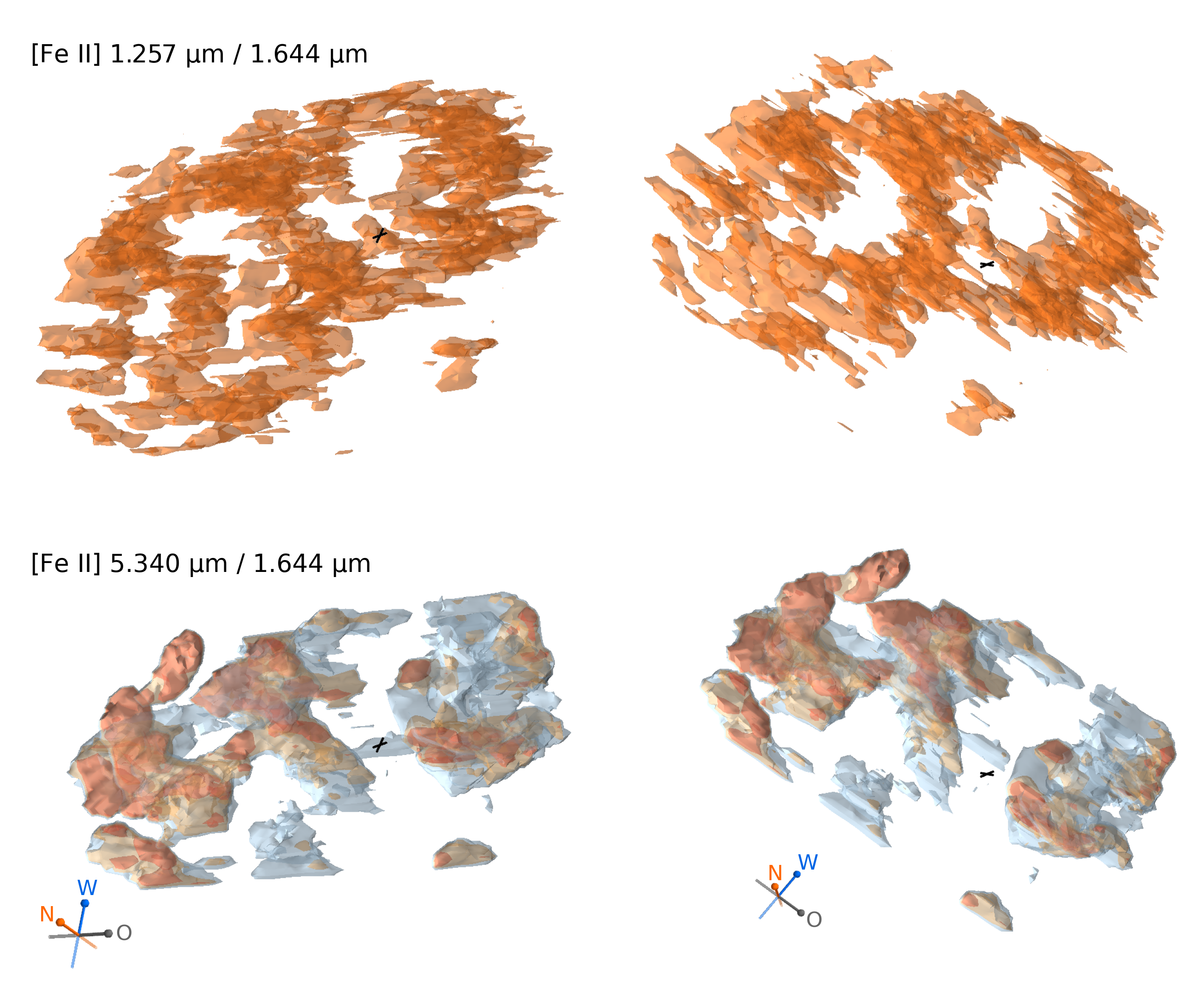}
\caption{Iso-surfaces of line ratios between [\ion{Fe}{2}] 1.2570/1.6440~$\mu$m (top row) and 5.3402/1.6440~$\mu$m (bottom row). The former ratio does not exhibit any significant spatial variations, so only one surface is shown (corresponding to a ratio of 1.11), while we show surfaces for three ratios in the latter case, corresponding to 0.85 (blue), 1.3 (orange), and 1.7 (red). All iso-surfaces are semi-transparent. The viewing angle in the left column is the same as in Figure~\ref{fig:3dvol_b}, while the one in the right column has been adjusted to best highlight the spatial variations. The direction to the north is out of the page in both cases, and the length of each compass axis is 200~\kms. An animated (rotating) version of this figure is available. The video shows one rotation of the panels in the bottom row.} 
\label{fig:3dratios}
\end{figure*}

The [\ion{Fe}{2}] 1.2570~$\mu$m and 1.6440~$\mu$m lines originate from the same upper level, which means that the intrinsic ratio is fixed and the observed value will depend on the amount of extinction. The intrinsic ratio has considerable uncertainties, however, where a flux ratio in the range between 1.18 (from \citealt{Quinet1996} Hartree-Fock) and 1.36 \citep{Nussbaumer1988} is generally favored, though an even wider range has been reported in the literature (see \citealt{Gianni2015,Koo2015} and references therein). Because of this, we are primarily interested in possible spatial variations of the ratio, which would indicate significant dust within the ejecta filaments. The very similar full line profiles in Figure~\ref{fig:profiles} (top left) already suggests that there is little variation within the remnant, and we find this to be the case also when considering the line ratios in 3D, the results of which are shown in Figure~\ref{fig:3dratios} (top row). 

The flux ratio from the full profiles is $F_{1.257~\mu \rm m}/F_{1.644~\mu \rm m} = 1.111 \pm 0.007$, where the small uncertainty reflects the excellent signal in the lines and we have excluded the uncertain region between $\pm 100$~\kms. With this observed flux ratio, the inferred $E(B-V)$ is $0.21 \pm 0.01$ and $0.73 \pm 0.01$, using the intrinsic ratios from \cite{Quinet1996} Hartree-Fock and \citep{Nussbaumer1988}, respectively. The former is close to $E(B-V)=0.20$, which was favored by \cite{Serafimovich2004}, and also consistent with $E(B-V)=0.27 \pm 0.07$ estimated from the H$\alpha$/H$\beta$ ratio in the ISM around SNR~0540 \citep{Tenhu2024}. By contrast, an $E(B-V)=0.73$ is much higher than any value that has previously been inferred for SNR~0540 and, if correct, would indicate substantial amounts of dust within the ejecta filaments. We consider this unlikely, however, given the lack of spatial variations of the ratio. The distribution of line ratios in 3D  has a symmetric spread about the peak value of 1.11 (the iso-surface plotted in Figure~\ref{fig:3dratios}), consistent with statistical fluctuations. This suggests that the 3D maps we observe are not strongly affected by dust, neither within individual filaments (which would give ratios that vary along the line of sight) or between different regions of the remnant. These results thus disfavor the high intrinsic ratio from \cite{Nussbaumer1988} and point to a lower ratio in line with \cite{Quinet1996} Hartree-Fock, which is also consistent with the observationally derived ratio by \cite{Gianni2015}.

The  results for the 5.3402/1.6440~$\mu$m ratio are very different, as is immediately clear from the differences in the integrated line profiles in Figure~\ref{fig:profiles} (bottom right). The full distribution of ratios in 3D peaks at $\sim 0.85$, but exhibits an asymmetric distribution extending to higher values of $\gtrsim 2$. To show this in 3D, we plot iso-surfaces corresponding to three different ratios in  Figure~\ref{fig:3dratios} (right column). It is clear that the highest ratios (corresponding to a lower temperature and/or electron density, see Section~\ref{sec:disc-mixing}) are found in a small faint region on the redshifted west side of the remnant (see also schematic in Figure~\ref{fig:cartoon}). In fact, it is below the lowest flux levels plotted in the 3D maps in Figures~\ref{fig:3dvol_a} and \ref{fig:3dvol_b}, while it can be seen faintly in Figure~\ref{fig:3dvol_ar2_neb} and in the collapsed images in Appendix~\ref{sec:app-sideviews}. There is also some evidence that the line ratios vary within the filaments, being higher further away from the center.

\section{Discussion}
\label{sec:discussion}

\subsection{Overall morphology and connection to the pulsar wind}
\label{sec:disc-morphology}

The large-scale morphology of SNR~0540 revealed by the 3D reconstruction of the IR emission lines is asymmetric. We discern two major fragmented lobes of similar size (diameters of $\sim 600$~\kms, equivalent to $\sim 0.7$~pc), which are largely empty inside, but with weak emission detected outside. The weak emission includes additional small ring-like structures on the redshifted side. Most lines show a similar morphology, with the exception of the Ne lines, which notably show a bright region on the northeast far side. 

The basic picture for how these structures are produced is that the relativistic particle wind from the pulsar sweeps up the ejecta into a thin shell, leaving an empty region in the interior.  The emission from the bright clumps and filaments is most likely powered by shock excitation, as evidenced from the many strong [\ion{Fe}{2}] lines observed in the spectrum, as well as previous models by \cite{Williams2008}. Photoionization by both the synchrotron emission from the pulsar wind and hot shocked gas is also expected to contribute, and most likely explains the weaker emission outside the main clumps (Figure~\ref{fig:3dvol_ar2_neb}), as well as the extended  [\ion{O}{3}] torus that is outside the FOV of the JWST observations \citep{Larsson2021}.

\begin{figure*}[t]
\centering
\includegraphics[width=\hsize]{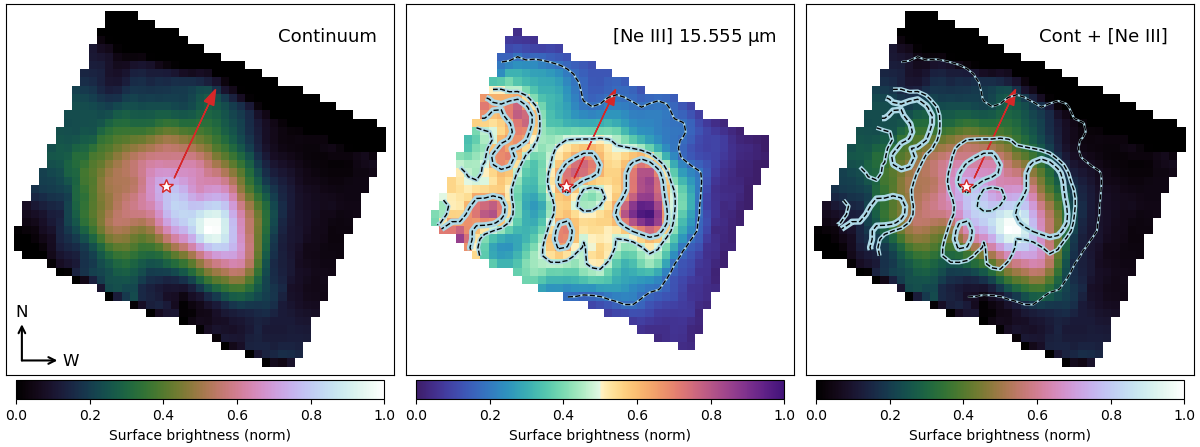}
\caption{Images of the synchrotron continuum compared to the [\ion{Ne}{3}]~15.5550~$\mu$m line. The left panel shows the continuum integrated over the 15.4--18.0~$\mu$m range (channel 3 LONG), the middle panel shows an image of the [\ion{Ne}{3}] line with its own contours, while the right panel shows the continuum image together with the  [\ion{Ne}{3}] contours. The star symbol and red arrow show the position of the pulsar and the proposed jet direction, respectively.} 
\label{fig:cont_ne3}
\end{figure*}

Given that the ejecta distribution is shaped by the pulsar wind, it is interesting to compare it with the morphology of the synchrotron continuum. Figure~\ref{fig:cont_ne3} (left) shows an image of the the continuum emission in the 15.4--18.0~$\mu$m range, which was produced by integrating over channel 3 LONG, after excluding prominent emission lines. This channel offers a good overview of the spatial distribution of the continuum due to its large FOV. The continuum spectrum in this channel is well described by a power law, $F_{\nu} \propto \nu^{\alpha}$, with $\alpha \sim -0.9$, consistent with expectations for synchrotron emission as well as previous measurements of the spectral index in the IR range \citep{Lundqvist2020,Tenhu2024}. 
 
The continuum emission shown in Figure~\ref{fig:cont_ne3}  is elongated along the northeast--southwest direction and is dominated by a bright blob located southwest of the pulsar, which has been noted in many previous optical and IR observations \citep{DeLuca2007,Lundqvist2011,Mignani2012,Tenhu2024}. The figure also shows a possible jet direction to the northwest, proposed based on the X-ray morphology \citep{Gotthelf2000}, polarization signatures \citep{Lundqvist2011}, optical spectral index variations \citep{Tenhu2024}, as well as the 3D morphology of the optical line emission, which shows an extension with a hole in the same direction (Figure~\ref{fig:3dvol_linecomp}, middle, \citealt{Sandin2013,Larsson2021,Lundqvist2022}). An alignment with this hole indicates a direction slightly away from the observer if the Doppler shift of the pulsar is close to zero, but toward the observer if the pulsar is on the redshifted side (Figures~\ref{fig:cartoon}, \ref{fig:3dvol_linecomp} Section~\ref{sec:disc-kick}). The evidence for a jet is not conclusive though and there is no evidence for a counter jet. 

With the proposed jet direction, the extension along the northeast--southwest direction is interpreted as the torus, formed by an enhanced injection of plasma in the equatorial plane of the spinning pulsar. This kind of torus-jet morphology is commonly observed in young PWNe, where most observations are in the X-ray range \citep{Kargaltsev2008}. Among the small number of objects with IR and optical observations, the Crab is by far the most well observed. A comparison between the Crab and SNR~0540 is especially relevant given the similar ages of $\sim 1000$~years and energetic pulsars, with spin periods of $\sim 30$~ms for the Crab \citep{Lyne2015} and $\sim 50$~ms for SNR~0540 \citep{Marshall2016}.  The most important difference between the systems is that the Crab has a low kinetic energy and was possibly an electron capture SN \citep{Nomoto1982,Yang2015}, while SNR~0540 likely stems from a more massive progenitor \citep{Serafimovich2005,Williams2008}. 

The Crab shows an overall elongation along the jet direction at all wavelengths, including recent IR images from JWST \citep{Temim2024}. A 3D reconstruction of the optically line emitting ejecta shows that the global morphology is not an ellipsoid, however, but heart-shaped with an indentation close to the plane of the torus \citep{Martin2021}. This was interpreted as evidence of restricted expansion due to the circumstellar medium (CSM). By contrast, the overall 3D structure of SNR~0540 is more irregular, where the two main fragmented lobes fall roughly along the  northeast--southwest direction of the likely torus. 

The Galactic PWN 3C\ 58 also shows some similarities with the Crab, including a large-scale elongation along the jet axis in X-rays \citep{Slane2004}, though 3D information from optical observations reveals a more complex picture that diverges from this simple geometry and raises questions about the age of the system \citep{Fesen2008}. From this small sample of PWNe with 3D information it is clear that they are not described by a simple common shape, making it probable that asymmetric ejecta distributions stemming from the explosions and/or CSM interaction play a major role in shaping them. For comparison, magneto-hydrodynamical (MHD) simulations of PWNe in 3D have been able to reproduce the jet-torus structure, but show a more spherical geometry on large scales, where the pressure is nearly uniform \citep{Porth2014b,Olmi2016}.  

MHD simulations of expanding PWNe also show a significant impact of Rayleigh-Taylor instabilities, which result in thin filaments extending inwards from the expanding shell of ejecta \citep{Porth2014b,Blondin2017}, as observed in the Crab \cite[e.g.,][]{Temim2024,Blair2026}. The greater distance of SNR~0540 compared to the Crab implies a difference of a factor $\sim 25$ in terms of spatial resolution, so we do not expect to resolve small-scale filamentary structures in SNR~0540, though it is evident from the 3D maps that the shell is highly fragmented. In this context it is interesting to note that the Crab shows synchrotron emission extending well outside the line-emitting region, where the pulsar wind has burst out through the shell of ejecta (see IR continuum and line emission in \citealt{Temim2024}). We see no clear sign of this in SNR~0540, as shown by the middle and right panels of Figure~\ref{fig:cont_ne3} , which show a comparison between the [\ion{Ne}{3}]~15.5550~$\mu$m line and the continuum (both in channel 3 LONG). Although the observed FOV does not fully cover the edges of the PWN in all directions, the extent of the line emission and continuum is similar in the eastern part, where we clearly see the edge. The same is true for the line emission and continuum in channel 4 (not shown), where the FOV captures the full extent of the PWN also to the south. 

The simulations of \cite{Blondin2017} show that the pulsar wind breaks out through the ejecta shell when the PWN is expanding through the outer part of the ejecta, where the density profile follows a power law, and only if the energy supplied by the PWN is high compared to the explosion energy. If these conditions are not met, the PWN is confined by the pressure of the ejecta. For SNR~0540, this implies that the PWN is in the inner flat part of the ejecta density profile and that the explosion had a typical energy, in line with expectations.  The fact that we see a highly fragmented ejecta structure with large holes instead points to a significant impact of asymmetries imprinted by the explosion itself. Such asymmetries are a ubiquitous feature of neutrino-driven explosions \cite[e.g.][]{Stockinger2020,Sandoval2021}. The so-called Ni-bubble (or bubbles), which are produced due to energy deposited by the radioactive decay of $^{56}$Ni, are expected to further add to the fragmentation \citep{Li1993,Basko1994,Gabler2021,Orlando2021}.   

\subsection{Implications for the pulsar kick}
\label{sec:disc-kick}

When creating the 3D maps we have assumed that the center of explosion is at the position of the pulsar in the sky plane and at the systemic velocity of the local ISM along the line of sight. The former is motivated by the upper limit on the proper motion of the pulsar, which corresponds to a transverse velocity of $<250$~\kms\ \citep{Mignani2010}, while the latter simply assumes that the progenitor had a velocity comparable to its local environment. It is naturally expected that the true explosion center will deviate somewhat from the assumed position. 

It is interesting to note that the two main lobes are symmetric about the pulsar position in the sky plane (seen e.g., in Figures~\ref{fig:sideviews_a}--\ref{fig:sideviews_d} and quantified by the analysis in Figure~\ref{fig:clumpsep}), but not along the line of sight, where the emission clearly extends further on the redshifted side (see line profiles in Figure~\ref{fig:profiles} and pseudo-images in Figures~\ref{fig:sideviews_a}--\ref{fig:sideviews_d}, as well as Figure 6 of \citealt{Lundqvist2022}). The centroid velocity of the middle peak of the total line profiles is $\sim 315 \pm 15$~\kms, which also corresponds to the approximate midpoint between the two lobes along the line of sight. This velocity and confidence interval represent the average and standard deviation from fitting the middle peak of all the lines in Figure~\ref{fig:profiles} with a Gaussian, excluding the \ion{He}{1}~1.083~$\mu$m line, which does not show a clear peak, as well as the lines in channel 4 where the resolution is lower. 
 
The offset of $\sim 300$~\kms\ may be caused by a high velocity of the progenitor, which implies that the assumed center of explosion is wrong. We find this scenario unlikely, however, as the  inferred velocity is above the escape speed of the LMC ($\sim 100-250$~\kms, \citealt{Boubert2017}) and such hyper-velocity stars are very rare \citep{Boubert2017,Lin2023}. The asymmetry is more readily explained by a pulsar kick of $\sim 300$~\kms\ directed away from the observer. This follows from the fact that the observed ejecta emission is powered by the pulsar wind, so it is  natural to assume that the pulsar is at the center of the two main lobes. 

The fact that the lobes are symmetric about the pulsar in the sky plane is consistent with a very low transverse component of the kick. A total kick velocity of $\sim 300$~\kms\ is consistent with the 3D kick distribution derived from the population of Galactic radio pulsars, which has a mean $\sim 400$~\kms\   \citep{Hobbs2005,Faucher2006}. We also note that if the pulsar is indeed located on the far side of the nebula, it implies that the putative jet discussed in Section~\ref{sec:disc-morphology} is directed toward the observer, which means that the absence of a counter jet could be explained by Doppler boosting.

An uncertainty in this estimate of the pulsar kick is that it only considers the peak of the emission along the line of sight and does not explain the weaker, more extended emission on the redshifted side. Furthermore, the observed ejecta distribution will also be affected by asymmetries in the explosion itself, as discussed above. We note, however, that the extended torus of [\ion{O}{3}] emission seen in the MUSE observations is symmetric around the systemic velocity \citep{Larsson2021}, which argues against an intrinsic global asymmetry of the ejected matter to the redshifted side. The symmetric [\ion{O}{3}] torus also disfavors the scenario with a high-velocity progenitor.

\subsection{Small-scale structure and mixing}
\label{sec:disc-mixing}

The observed differences in the 3D emissivities of different emission lines is expected to be due to a combination of spatial variations of the physical conditions as well as differences in chemical composition, just as for the Crab \cite[e.g.,][]{MacAlpine2008}. Disentangling these effects requires spatially resolved modeling of the spectra, which will be the subject of future work. Here we limit ourselves to some general conclusions that can be drawn from a comparison of the different 3D maps. 

First, we note that while the present analysis does not rule out the presence of dust within the filaments of SNR~0540, the 3D maps of the line emission are unlikely to be strongly affected by dust. This is seen both from the lack of spatial variations of the [\ion{Fe}{2}] 1.2570/1.6440~$\mu$m ratio, as well as the fact that there is no evidence of optical “shadows" due to dust absorption of background synchrotron emission, as has been seen in the Crab \citep{Fesen1990,Blair2026}. Previous reports of dust  in SNR~0540 are based on Spitzer and AKARI observations which do not resolve the PWN \citep{Williams2008,Lundqvist2020}. These studies suggest a warm dust component with  $\sim 55$~K and a mass of $\sim 2.5 \times 10^{-3}$~\msun\ that gives rise to an excess at wavelengths $> 20\ \mu$m. The presence of such a component in the JWST data will be discussed in a separate paper on the continuum emission.

Numerical simulations of neutrino-driven SNe have shown that the different abundance zones of the progenitor stars are broken up and mixed in the explosions \citep{Wongwathanarat2017,Utrobin2019,Giudici2025}. This results in inward mixing of hydrogen from the outer layers down to the core, as well as outward mixing of freshly synthesized $^{56}$Ni out to high velocities. The details of the mixing are a sensitive diagnostic of the explosion properties, and the observational constraints on mixing obtained from SNRs provide an important testbed for the models. In the case of SNR~0540, it is clear that the spatial distribution of elements has been modified by the pulsar wind, but this only mixes ejecta in the outward direction, so the observed distribution of elements still retains information about mixing in the explosion. Importantly, the detection of H in the innermost ejecta (through \ion{H}{1}~1.8756~$\mu$m) confirms previous findings and the classification of the SN as a Type II \citep{Serafimovich2005,Morse2006,Larsson2021}. The JWST observations also demonstrate that H is mixed down to very low velocities of $<400$~\kms, consistent with model predictions of mixing in neutrino-driven explosions \citep{Utrobin2019} and similar to the lowest velocities of H observed in the innermost ejecta of SN~1987A \citep{Kozma1998,Larsson2019}. 

While most regions show emission from all the analyzed lines, the different 3D morphology of the [\ion{Ne}{2}] and [\ion{Ne}{3}] lines suggest abundance variations within the ejecta. Specifically, the Ne emission shows a more spatially extended distribution than all the other lines in the JWST observations, including a bright small ring of ejecta on the far side, which is particularly prominent in the [\ion{Ne}{2}] maps (Figures~\ref{fig:3dvol_a} and \ref{fig:3dvol_b}). The comparison with [\ion{S}{3}]~18.713~$\mu$m in MRS channel 4 is especially interesting as the FOV fully covers the the Ne emission observed in channel 3. The same is true for the MUSE observations of [\ion{S}{3}]~0.9069~$\mu$m, which has a similar morphology as [\ion{S}{2}]~0.6717, 0.6731~$\mu$m and [\ion{Ar}{3}]~0.7136~$\mu$m \citep{Larsson2021}, while [\ion{O}{3}]~0.5007~$\mu$m is more similar to Ne (Figure~\ref{fig:3dvol_linecomp}). In terms of the structure of the progenitor star, Ne, along with O,C and Mg, is located outside the Si/S/Ar-rich zone, so the fact that Ne and S/Ar are observed in partly different locations shows that some of this structure is retained, supporting the scenario of macroscopic (rather than microscopic) mixing of ejecta, as has also been found in observations of Cas~A \citep{Ennis2006}. 

The clearest indication of differences in the physical conditions within the ejecta is provided by the 3D map of the [\ion{Fe}{2}] 5.3402/1.6440~$\mu$m ratio (Figure~\ref{fig:3dratios}). This ratio is sensitive to both the  temperature, $T_{\rm e}$, and electron density, $n_{\rm e}$. To investigate the dependence, we calculated the ratio using an \ion{Fe}{2} model atom with 750 levels, including recombination from \ion{Fe}{3}, similar to the one used in \cite{Kavanagh2026}.  Line transfer was handled by the Sobolev approximation. For atomic data and details about uncertainties, we also refer to \cite{Kavanagh2026}.  

\ion{Fe}{1} is likely to be nearly completely ionized, as seen from the absence of \ion{Si}{1} lines in the spectrum.\footnote{
This is clear from the excellent agreement between the [\ion{Fe}{2}]~1.2570 and 1.6440~$\mu$m lines (Figure~\ref{fig:profiles}). If neutral Si had been present, a line from [\ion{Si}{1}]~1.646~$\mu$m would be expected (as observed in SN~1987A, \citealt{Larsson2023,Larsson2025}), which would produce a broader $\sim 1.64~\mu$m feature due to the $\sim 360$~\kms\ offset between the [\ion{Si}{1}] and [\ion{Fe}{2}] lines. 
}
The \ion{Fe}{3}  fraction is, however, important as a source of recombination to some of the \ion{Fe}{2} lines. We can get an estimate of the \ion{Fe}{3}  fraction from the [\ion{Fe}{3}]~22.93~$\mu$m line relative to the [\ion{Fe}{2}]~25.99~$\mu$m line. 
If electron collisions are the only important excitation process, the ratio of the line fluxes is given by \citep{Kavanagh2026}
\begin{equation}
\frac{X({\rm Fe~ III})}{X({\rm Fe ~II})} = 2.0 \ e^{77/T_{\rm e}} \frac{F(\lambda 22.93)}{F(\lambda 25.99)}
\label{eq:fe3_fe2_2}
\end{equation}
The ratio is therefore insensitive to temperature.
From the integrated flux from the filaments (Tegkelidis, C., in prep.) we find $F({\rm \lambda 22.93)} /{F(\lambda 25.99)} \approx 0.11$, and X(Fe~III)/X(Fe~II) $ \approx 0.2$, which we use for the calculation. 
While the optical depth of the [\ion{Fe}{2}]~25.99~$\mu$m line can be large \citep{Kavanagh2026}, the optical depth of the [Fe~{\sc ii}]~1.6440 and 5.3402~$\mu$m lines are small ($\tau \propto \lambda^3$), and the ratio of these lines is therefore not sensitive to the Fe abundance.

\begin{figure}[t]
\centering
\includegraphics[width=\hsize]{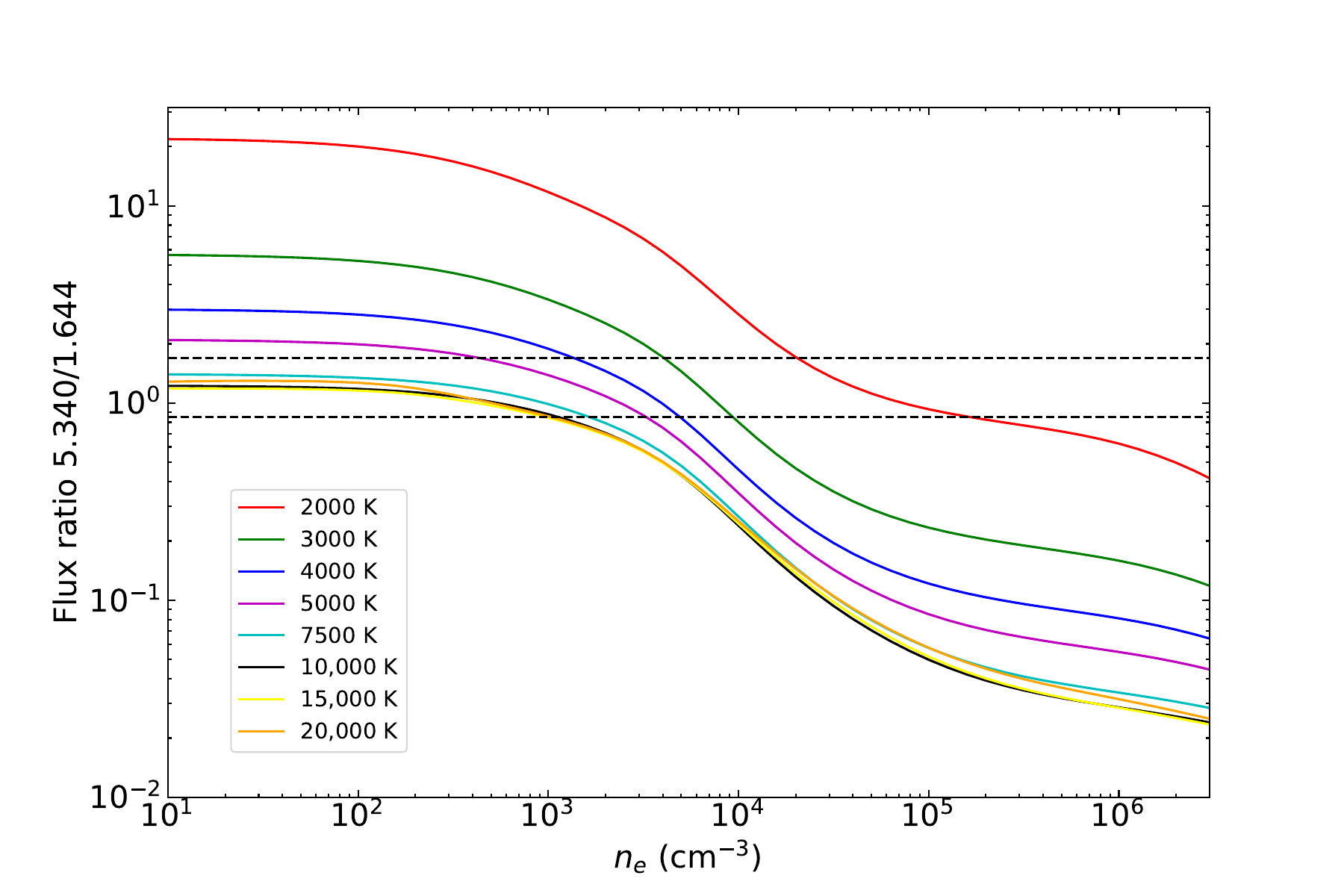}
\caption{Ratio of the [\ion{Fe}{2}] 5.3402/1.6440~$\mu$m lines as a function of electron density, $n_{\rm e}$, for a range of temperatures between 2,000--20,0000~K. The two dashed black horizontal lines correspond to the lowest (0.85) and highest (1.7) ratios, shown by the iso-surfaces in the bottom row Figure~\ref{fig:3dratios}.} 
\label{fig:fe2rat_curves}
\end{figure}

In Figure~\ref{fig:fe2rat_curves} we show the resulting [\ion{Fe}{2}] 5.3402/1.6440~$\mu$m ratio for different temperatures as function of $n_{\rm e}$. This shows that higher values of the ratio correspond to lower temperatures and densities, though the temperature dependence is very weak for $T_{\rm e} \gtrsim 7000$~K. The 3D maps in Figure~\ref{fig:3dratios} show a higher ratio of these lines in a small ring of ejecta outside the main lobes, as well as some evidence of a varying ratio within the main filaments. For regions in SNR~0540 with the lowest line ratios, these results suggest $n_{\rm e} \gtrsim 1.1 \times 10^3$~cm$^{-3}$, independent of $T_{\rm e}$. Regions with the highest line ratios, $\gtrsim 1.7$, are instead likely to  have low temperatures $T_{\rm e} \lesssim 5000$ K, while the lower limit for $n_{\rm e}$ is slightly reduced. 

The \ion{He}{1}~1.083~$\mu$m line provides further evidence of spatial variations of the physical conditions. Helium originates both in the inner Fe/He zone and in the outer layers of the progenitor, so in the scenario of macroscopic mixing it is expected to have a similar distribution as Fe and H. However, a comparison with [\ion{Fe}{2}]~1.644~$\mu$m and \ion{H}{1}~1.876~$\mu$m show that \ion{He}{1}~1.083~$\mu$m is significantly fainter on the redshifted side (top left panel of Figure~\ref{fig:profiles} and Figure~\ref{fig:sideviews_d}). As is well known \cite[e.g.,][]{Clegg1987}, the 1.083~$\mu$m line is sensitive to collisional excitation from the metastable $2 {}^3$S level to upper triplet states, which can enhance the 1.083~$\mu$m line by a large factor, depending on the electron density. A lower electron density on the redshifted side may therefore explain the fainter 1.083~$\mu$m line on this side.

\section{Summary and conclusions}
\label{sec:summary}

We have presented the first results from JWST NIRSpec and MRS IFU observations of the central region of SNR 0540.  We used the observations to reconstruct the 3D morphology of the strongest emission lines in the innermost ejecta, which are powered by a PWN. The results reveal the distribution of emission from \ion{H}{1},  \ion{He}{1},  [\ion{Ne}{2}--\small{III}], [\ion{S}{3}--\small{IV}],  [\ion{Ar}{2}--\small{III}], [\ion{Fe}{2}] and  [\ion{Ni}{2}].  Our main conclusions can be summarized as follows: 

\begin{itemize}

\item{The 3D line maps reveal a fragmented distribution of ejecta, including two major fragmented lobes of similar size (diameters of $\sim 600$~\kms, equivalent to $\sim 0.7$~pc), which are largely empty inside. There is additional weak emission outside the lobes, including small ring-like structures on the redshifted side. The inferred 3D velocities peak at $\sim 450$~\kms, with a tail extending to $\gtrsim 1000$~\kms.}

\item{The two lobes are symmetric about the pulsar in the plane of the sky, but centered around a velocity of $\sim 300$~\kms\ along the line of sight. Since the line-emitting ejecta are powered by the pulsar wind, it is natural to assume that the pulsar would be at the center of these structures. This implies a pulsar kick velocity of  $\sim 300$~\kms\ away from the observer. }

\item{Emission from \ion{H}{1}~1.8756~$\mu$m is clearly detected in the inner ejecta, showing a similar 3D morphology as the other emission lines. This confirms that SNR~0540 stems from a Type II SN and shows that H has been mixed down to low velocities of $< 400$~\kms\ in the explosion.}

\item{The  bright [\ion{Fe}{2}]~1.2570~$\mu$m  and 1.6440~$\mu$m lines have a flux ratio of $1.111 \pm 0.007$, consistent with the extinction along lines of sight to the ISM surrounding SNR~0540. A 3D map of the line ratio does not reveal any spatial variations across the remnant or within individual filaments, which shows that the 3D line maps are not significantly affected by dust.}

\item{There are differences between the 3D morphologies of different emission lines due to a combination of varying abundances and physical conditions. The clearest sign of abundance variations is a bright region of enhanced Ne emission, which provides evidence for macroscopic mixing in the explosion. The clearest sign of varying physical conditions is provided by the 3D map of the [\ion{Fe}{2}] 5.3402/1.6440~$\mu$m ratio, which shows significant variations that trace temperature and density.}

\item{A comparison between SNR~0540 and the Crab reveals a number of differences, despite their similar ages and energetic pulsars. Both remnants show a highly fragmented ejecta structure, but the IR synchrotron emission has a comparable extent as the ejecta in SNR~0540, while it has burst through the ejecta shell in the Crab. The overall 3D morphologies are also different, with the Crab showing a clear elongation along the jet axis, which is not seen in SNR~0540.  This suggests that asymmetries inherent to the explosions  play a major role in shaping the PWNe.}

\end{itemize}

These results highlight the value of IFU data for characterizing the complex morphology of SNRs. Further analysis and modelling of the spatially resolved line- and continuum-spectra from these observations will provide new insights about the progenitor, explosion and pulsar in SNR~0540.

\begin{acknowledgments}
This work is based on observations made with the NASA/ESA/CSA James Webb Space Telescope. The data were obtained from the Mikulski Archive for Space Telescopes at the Space Telescope ScienceInstitute, which is operated by the Association of Universities for Research in Astronomy, Inc., under NASA contract NAS 5-03127 for JWST. These observations are associated with program \#4712. The specific observations analyzed can be accessed via doi \dataset[DOI: 110.17909/zjsn-ws92]{ https://doi.org/10.17909/zjsn-ws92}. The animations in Figures~\ref{fig:3dvol_a}, \ref{fig:3dvol_b}, \ref{fig:3dvol_ar2_neb}, \ref{fig:3dvol_linecomp}, and \ref{fig:3dratios} are available on Zenodo: \url{https://doi.org/10.5281/zenodo.19453672}

J. L. acknowledges support from the Knut and Alice Wallenberg foundation and a STIAS Fellowship during the 2nd semester of 2025. 

\end{acknowledgments}





\facilities{JWST (NIRSpec and MRS), VLT (MUSE)}

\software{astropy \citep{Astropy2022},
 CUPID \citep{Berry2007},
 Matplotlib \citep{Hunter2007}, 
 Mayavi \citep{Ramachandran2011}.
          }


\appendix

\section{Collapsed images from different directions}
\label{sec:app-sideviews}

\begin{figure*}[t]
\centering
\includegraphics[width=\hsize]{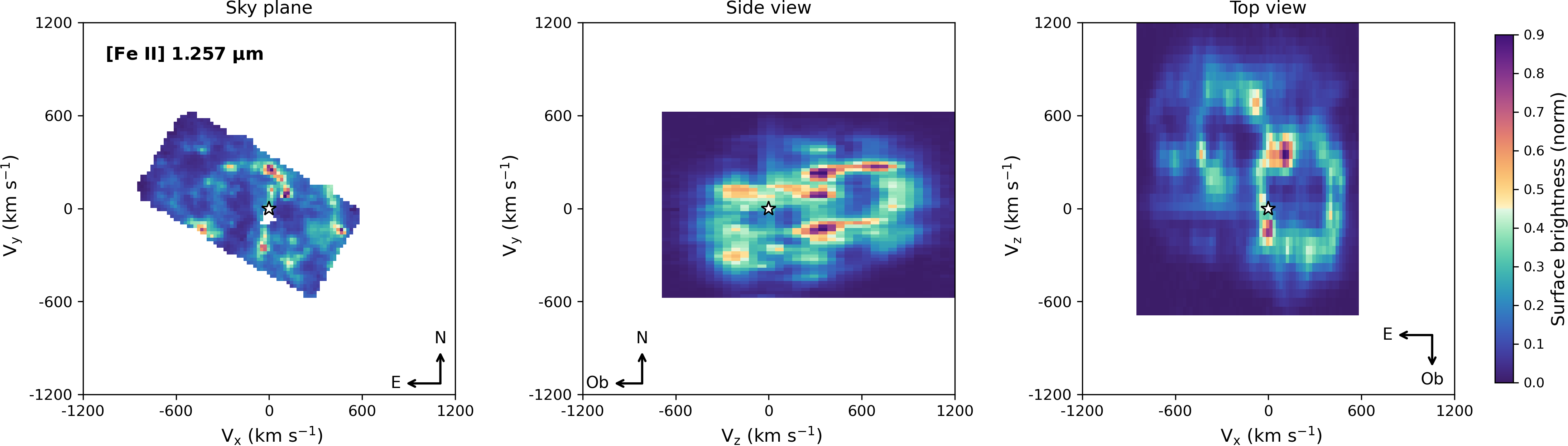}
\includegraphics[width=\hsize]{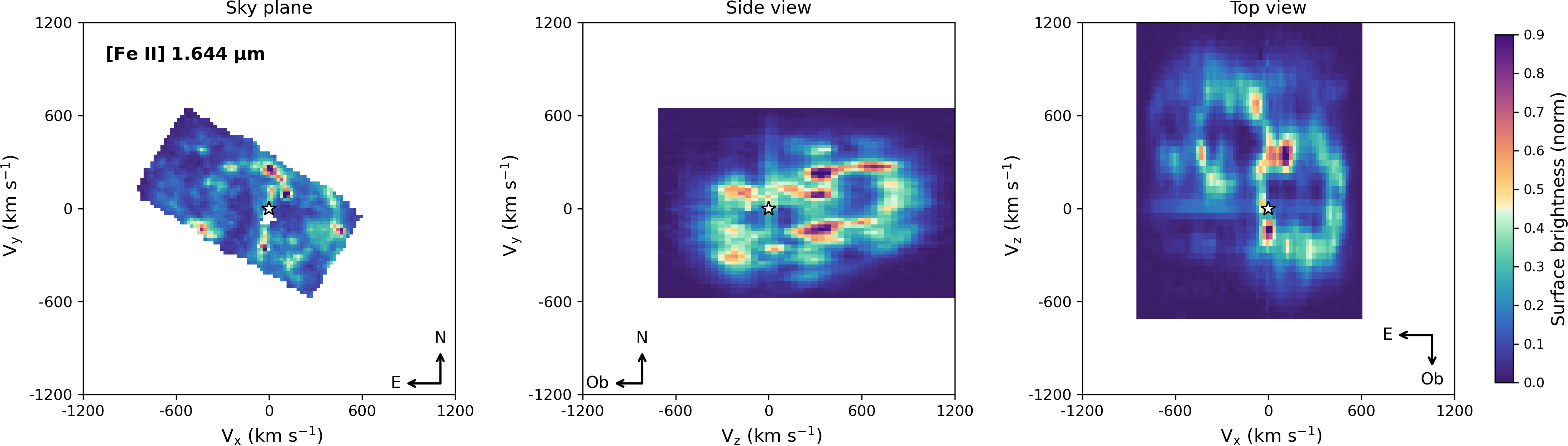}
\includegraphics[width=\hsize]{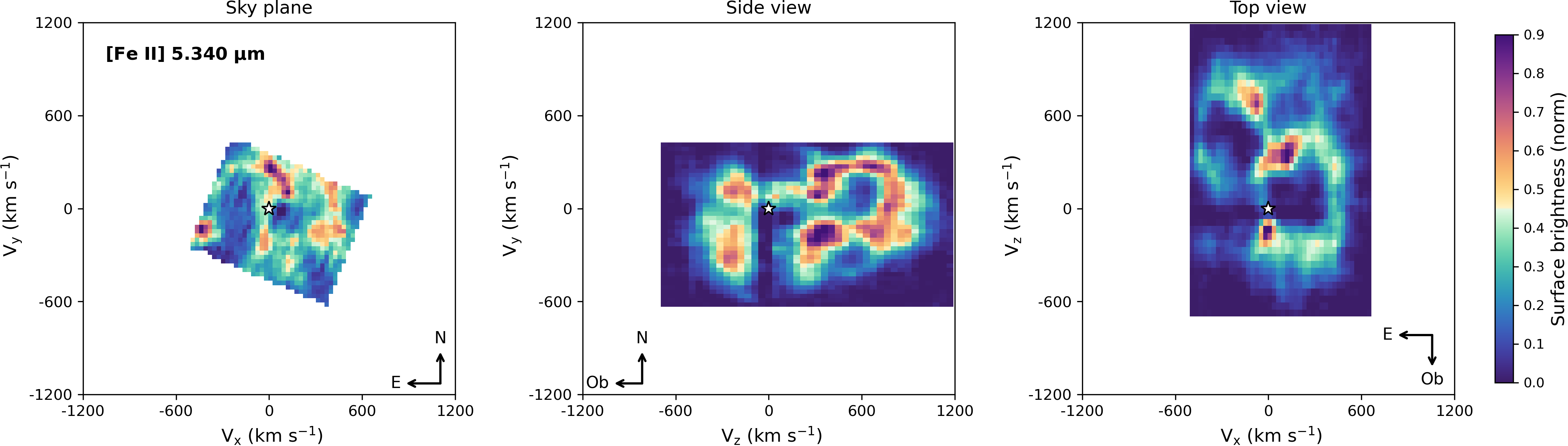}
\includegraphics[width=\hsize]{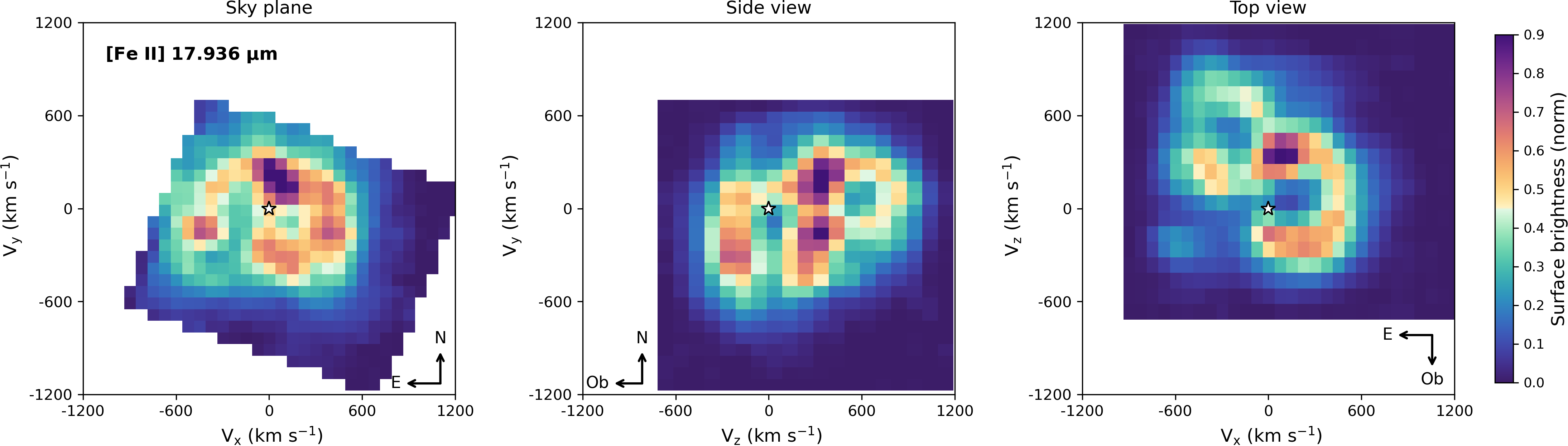}
\caption{“Images" of the [\ion{Fe}{2}] lines from different directions. Left: the regular image in the sky plane. Middle: image created by integrated over the east--west direction. The observer is located to the left, while the y-axis is the same as in the left panel with north pointing up. Right: image created by integrating over the north--south direction. The observer is located at the bottom of the page, while the x-axis is the same as in the left panel with east to the left. The white region just south of the pulsar in the left panels of the top two rows was masked out due to contamination by a star in the NIRSpec data.} 
\label{fig:sideviews_a}
\end{figure*}

\begin{figure*}[t]
\centering
\includegraphics[width=\hsize]{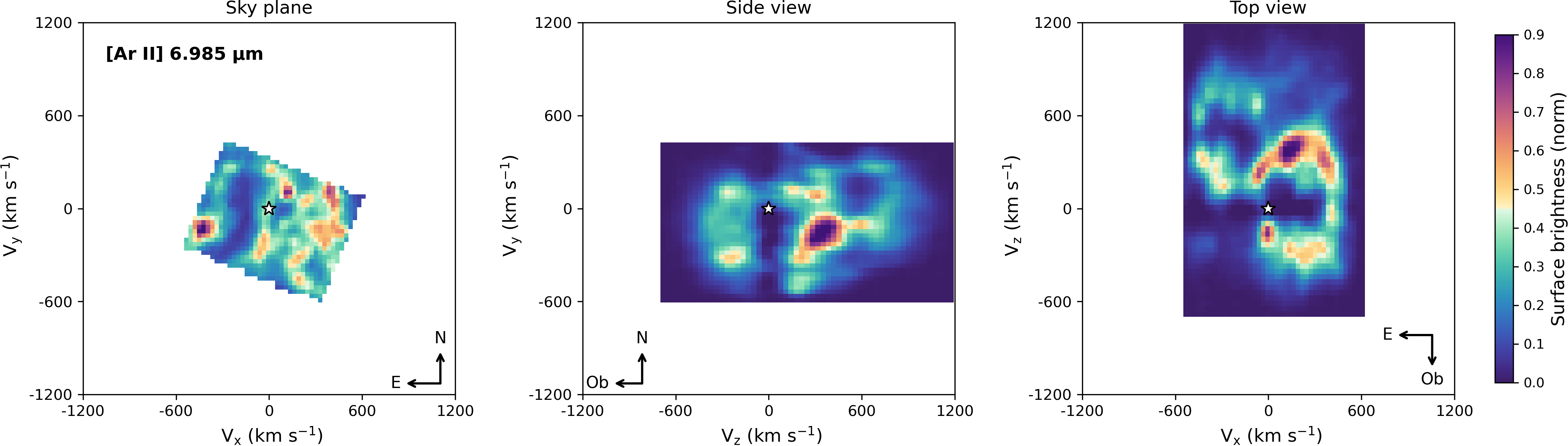}
\includegraphics[width=\hsize]{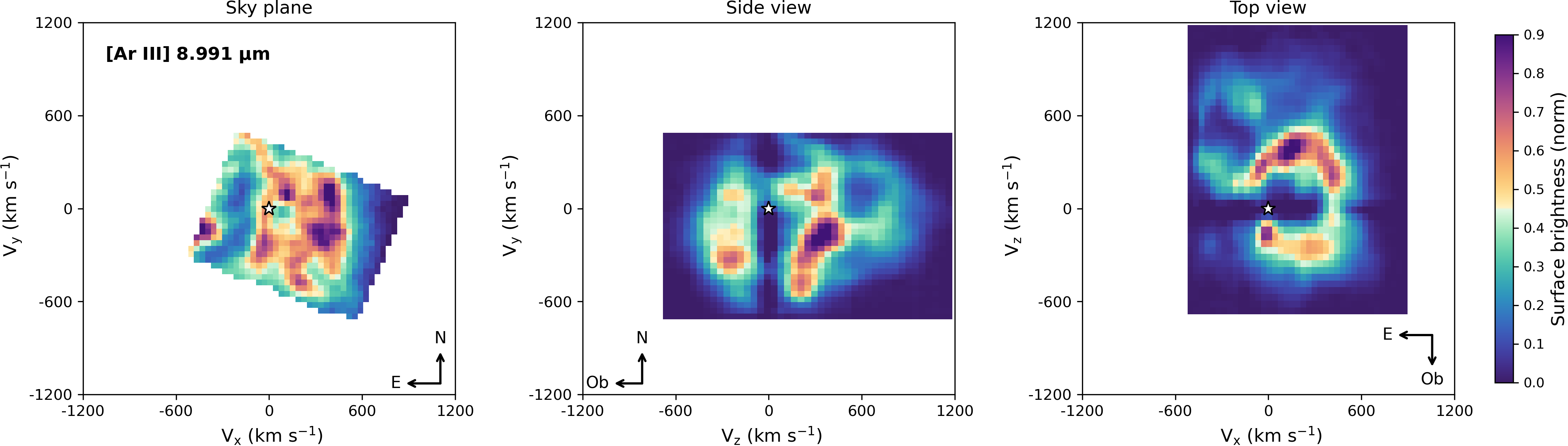}
\includegraphics[width=\hsize]{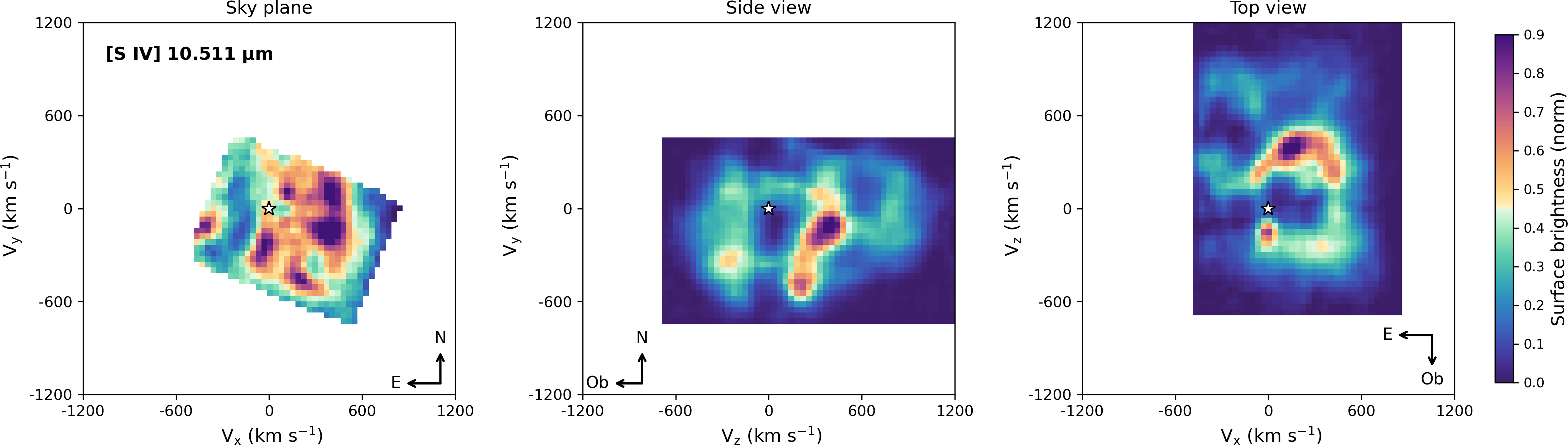}
\includegraphics[width=\hsize]{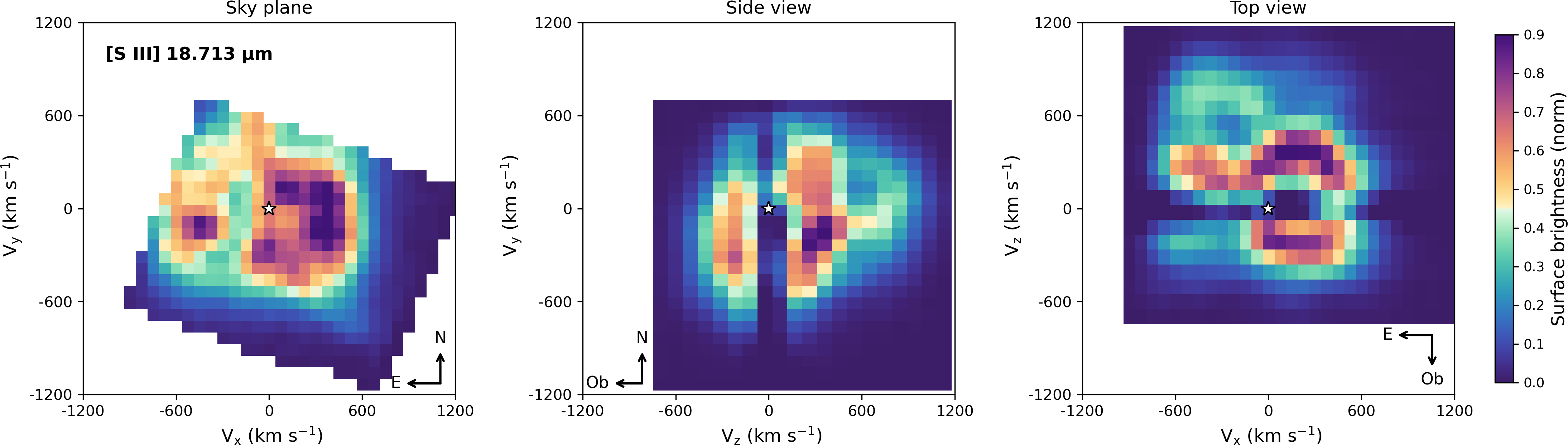}
\caption{Same as Figure~\ref{fig:sideviews_a}, but showing Ar and S lines.} 
\label{fig:sideviews_b}
\end{figure*}

\begin{figure*}[t]
\centering
\includegraphics[width=\hsize]{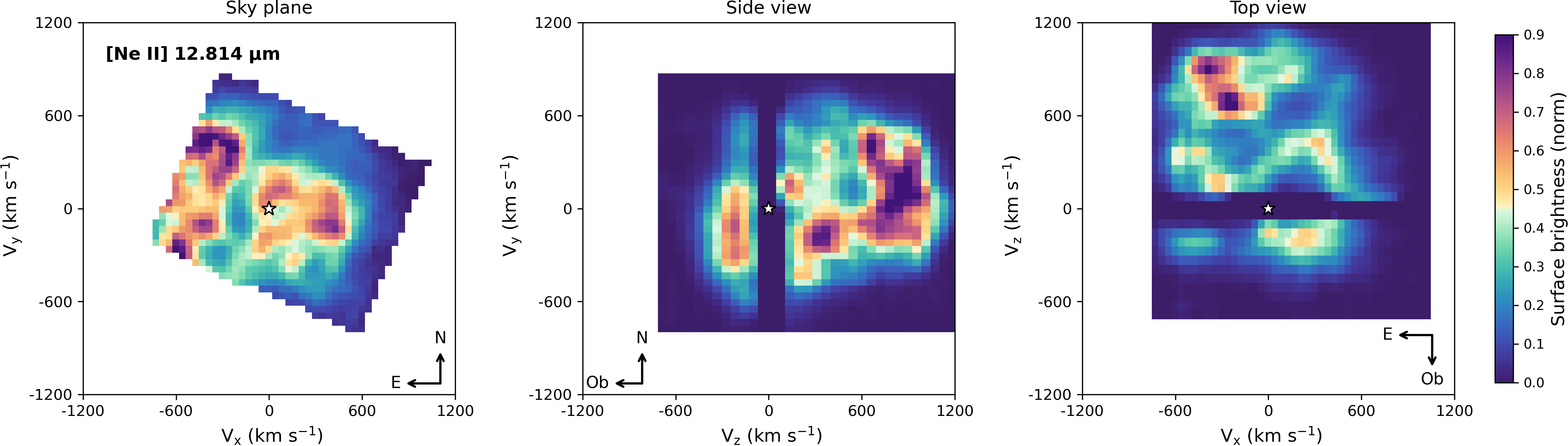}
\includegraphics[width=\hsize]{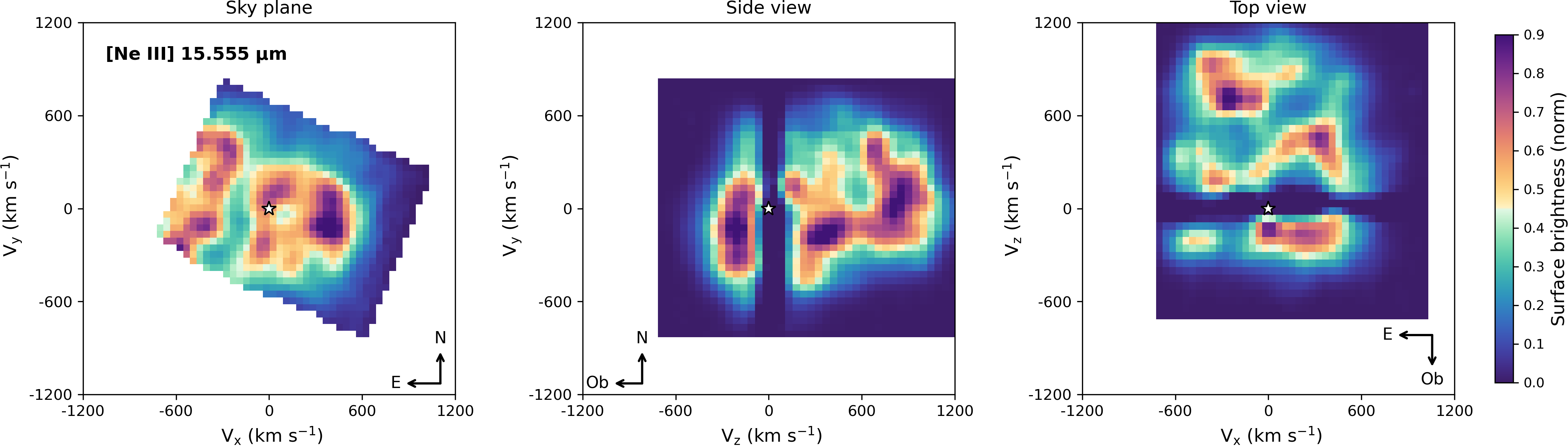}
\includegraphics[width=\hsize]{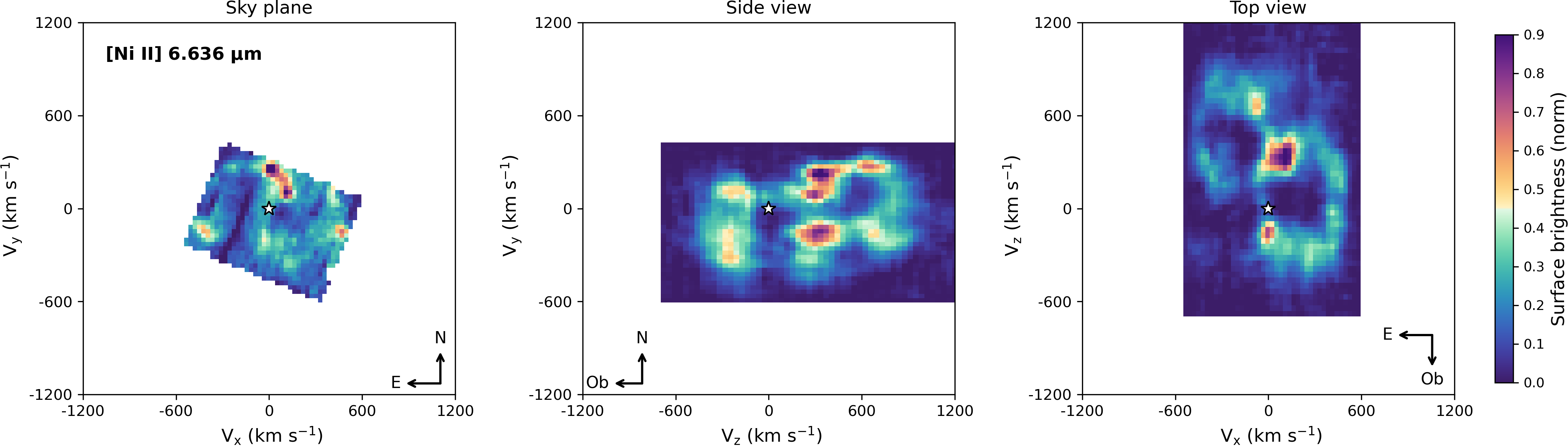}
\caption{Same as Figure~\ref{fig:sideviews_a}, but showing Ne and Ni lines.} 
\label{fig:sideviews_c}
\end{figure*}

\begin{figure*}[t]
\centering
\includegraphics[width=\hsize]{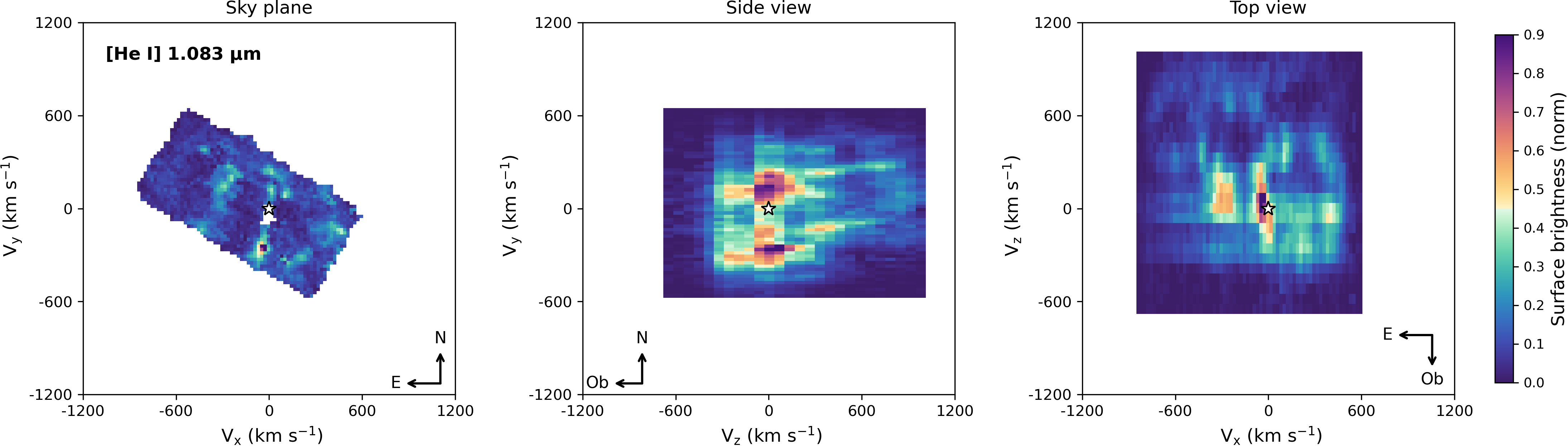}
\includegraphics[width=\hsize]{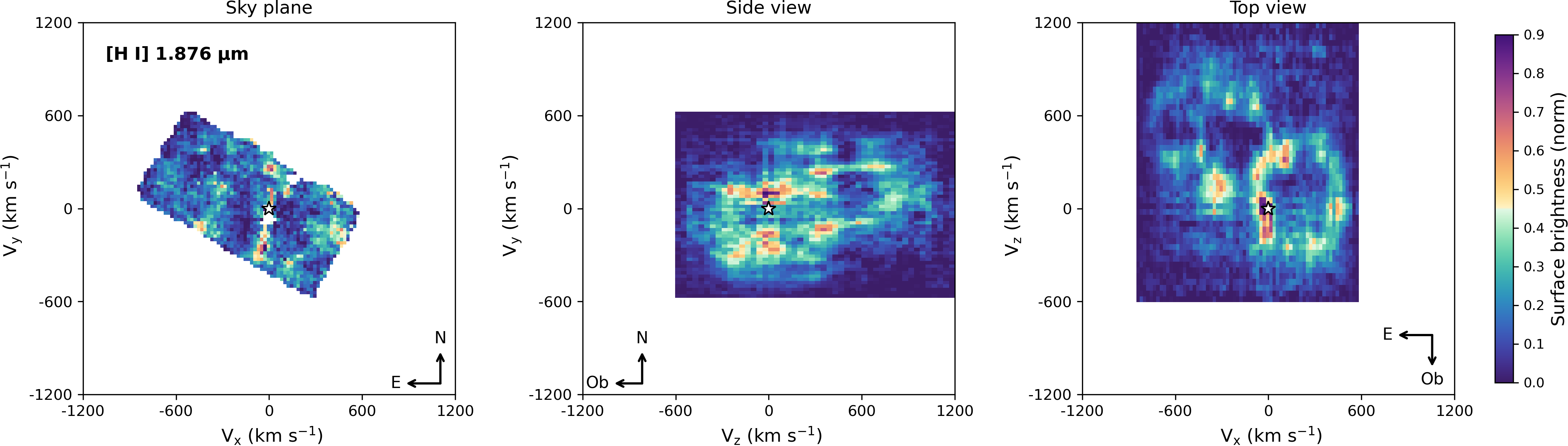}
\includegraphics[width=\hsize]{fe2_1p64_1300kms_3view.png}
\caption{Same as Figure~\ref{fig:sideviews_a}, showing the He and H lines. [\ion{Fe}{2}]~1.6640~$\mu$m is repeated from Figure~\ref{fig:sideviews_a} for comparison. The white region just south of the pulsar in the left column was masked out due to contamination by a star in the NIRSpec data.} 
\label{fig:sideviews_d}
\end{figure*}

Figures~\ref{fig:sideviews_a}--\ref{fig:sideviews_d} show images of the emission lines from Table~\ref{tab:lines} from different viewing angles. They were produced by integrating the cubes along the spectral dimension (left columns), the east--west direction (middle columns) and the north--south direction (right columns). The total area of all the images corresponds to $\pm 1200$~\kms, which is the greatest extent observed in any direction (highest redshift and greatest extent to the west and south for MRS channel 4). The observed emission covers different parts of this space for different lines, reflecting the differences in the FOV (Figure~\ref{fig:fov}). 





\clearpage
\newpage

\bibliography{snr0540_refs}{}

@ARTICLE{Larsson2025,
       author = {{Larsson}, J. and {Fransson}, C. and {Kavanagh}, P.~J. and {Sargent}, B. and {Barlow}, M.~J. and {Matsuura}, M. and {Gall}, C. and {Gehrz}, R.~D. and {Habel}, N. and {Hirschauer}, A.~S. and {Jones}, O.~C. and {Kirshner}, R.~P. and {Meixner}, M. and {Rosu}, S. and {Temim}, T.},
        title = "{The Compact Object and Innermost Ejecta of SN 1987A}",
      journal = {\apj},
         year = 2025,
        month = oct,
       volume = {991},
       number = {2},
          eid = {130},
        pages = {130},
          doi = {10.3847/1538-4357/adf741},
archivePrefix = {arXiv},
       eprint = {2508.03395},
 primaryClass = {astro-ph.HE},
       adsurl = {https://ui.adsabs.harvard.edu/abs/2025ApJ...991..130L}
}

@ARTICLE{Larsson2023,
       author = {{Larsson}, J. and {Fransson}, C. and {Sargent}, B. and {Jones}, O.~C. and {Barlow}, M.~J. and {Bouchet}, P. and {Meixner}, M. and {Blommaert}, J.~A.~D.~L. and {Coulais}, A. and {Fox}, O.~D. and {Gastaud}, R. and {Glasse}, A. and {Habel}, N. and {Hirschauer}, A.~S. and {Hjorth}, J. and {Jaspers}, J. and {Kavanagh}, P.~J. and {Krause}, O. and {Lau}, R.~M. and {Lenki{\'c}}, L. and {Nayak}, O. and {Rest}, A. and {Temim}, T. and {Tikkanen}, T. and {Wesson}, R. and {Wright}, G.~S.},
        title = "{JWST NIRSpec Observations of Supernova 1987A-From the Inner Ejecta to the Reverse Shock}",
      journal = {\apjl},
         year = 2023,
        month = jun,
       volume = {949},
       number = {2},
          eid = {L27},
        pages = {L27},
          doi = {10.3847/2041-8213/acd555},
archivePrefix = {arXiv},
       eprint = {2302.03576},
 primaryClass = {astro-ph.HE},
       adsurl = {https://ui.adsabs.harvard.edu/abs/2023ApJ...949L..27L}
}

@ARTICLE{Kavanagh2026,
       author = {{Kavanagh}, P.~J. and {Barlow}, M.~J. and {Fransson}, C. and {Larsson}, J. and {Matsuura}, M. and {Sargent}, B. and {Jones}, O.~C. and {Meixner} and  {Wesson}, R. and {Blommaert}, J. A. D. L. and {Bouchet}, P. and {Coulais}, A: and {Gastaud}, R. and {Gehrz}, R.~D. and {Habel}, N. and {Hirschauer}, A.~S. and {Jaspers}, 6 J. and {Kirshner}, R.~P. and {Lenki}, L. and {Nayak}, O. and {Rosu}, S. and {Temim}, T.},       
        title = "{The evolution of the mid-infrared spectrum of SN 1987A observed with the JWST/MIRI-MRS}",
      journal = {\apj},
         year = 2026, 
        month = xx,
       volume = {submitted},
       number = {},
          eid = {xx},
        pages = {},
          doi = {},
archivePrefix = {arXiv},
       eprint = {},
 primaryClass = {astro-ph.HE},
       adsurl = {https://ui.adsabs.harvard.edu/abs/2026xx}
}

@ARTICLE{Fesen1990,
       author = {{Fesen}, Robert and {Blair}, William P.},
        title = "{Optical Identification of Dust within the Crab Nebula's Filaments}",
      journal = {\apjl},
         year = 1990,
        month = mar,
       volume = {351},
        pages = {L45},
          doi = {10.1086/185676},
       adsurl = {https://ui.adsabs.harvard.edu/abs/1990ApJ...351L..45F}
}

@ARTICLE{Clegg1987,
       author = {{Clegg}, R.~E.~S.},
        title = "{Collisional effects in He I lines and helium abundances in planetary nebulae.}",
      journal = {\mnras},
         year = 1987,
        month = nov,
       volume = {229},
        pages = {31P-39},
          doi = {10.1093/mnras/229.1.31P},
       adsurl = {https://ui.adsabs.harvard.edu/abs/1987MNRAS.229P..31C}
}

@ARTICLE{Serafimovich2004,
       author = {{Serafimovich}, N.~I. and {Shibanov}, Yu. A. and {Lundqvist}, P. and {Sollerman}, J.},
        title = "{The young pulsar PSR B0540-69.3 and its synchrotron nebula in the optical and X-rays}",
      journal = {\aap},
         year = 2004,
        month = oct,
       volume = {425},
        pages = {1041-1060},
          doi = {10.1051/0004-6361:20040499},
archivePrefix = {arXiv},
       eprint = {astro-ph/0407226},
 primaryClass = {astro-ph},
       adsurl = {https://ui.adsabs.harvard.edu/abs/2004A&A...425.1041S}
}

@ARTICLE{Woosley2010,
       author = {{Woosley}, S.~E.},
        title = "{Bright Supernovae from Magnetar Birth}",
      journal = {\apjl},
         year = 2010,
        month = aug,
       volume = {719},
       number = {2},
        pages = {L204-L207},
          doi = {10.1088/2041-8205/719/2/L204},
archivePrefix = {arXiv},
       eprint = {0911.0698},
 primaryClass = {astro-ph.HE},
       adsurl = {https://ui.adsabs.harvard.edu/abs/2010ApJ...719L.204W}
}

@ARTICLE{Gaensler2006,
       author = {{Gaensler}, Bryan M. and {Slane}, Patrick O.},
        title = "{The Evolution and Structure of Pulsar Wind Nebulae}",
      journal = {\araa},
         year = 2006,
        month = sep,
       volume = {44},
       number = {1},
        pages = {17-47},
          doi = {10.1146/annurev.astro.44.051905.092528},
archivePrefix = {arXiv},
       eprint = {astro-ph/0601081},
 primaryClass = {astro-ph},
       adsurl = {https://ui.adsabs.harvard.edu/abs/2006ARA&A..44...17G}
}

@ARTICLE{Astropy2022,
       author = {{Astropy Collaboration} and {Price-Whelan}, Adrian M. and {Lim}, Pey Lian and {Earl}, Nicholas and {Starkman}, Nathaniel and {Bradley}, Larry and {Shupe}, David L. and {Patil}, Aarya A. and {Corrales}, Lia and {Brasseur}, C.~E. and {N{\"o}the}, Maximilian and {Donath}, Axel and {Tollerud}, Erik and {Morris}, Brett M. and {Ginsburg}, Adam and {Vaher}, Eero and {Weaver}, Benjamin A. and {Tocknell}, James and {Jamieson}, William and {van Kerkwijk}, Marten H. and {Robitaille}, Thomas P. and {Merry}, Bruce and {Bachetti}, Matteo and {G{\"u}nther}, H. Moritz and {Aldcroft}, Thomas L. and {Alvarado-Montes}, Jaime A. and {Archibald}, Anne M. and {B{\'o}di}, Attila and {Bapat}, Shreyas and {Barentsen}, Geert and {Baz{\'a}n}, Juanjo and {Biswas}, Manish and {Boquien}, M{\'e}d{\'e}ric and {Burke}, D.~J. and {Cara}, Daria and {Cara}, Mihai and {Conroy}, Kyle E. and {Conseil}, Simon and {Craig}, Matthew W. and {Cross}, Robert M. and {Cruz}, Kelle L. and {D'Eugenio}, Francesco and {Dencheva}, Nadia and {Devillepoix}, Hadrien A.~R. and {Dietrich}, J{\"o}rg P. and {Eigenbrot}, Arthur Davis and {Erben}, Thomas and {Ferreira}, Leonardo and {Foreman-Mackey}, Daniel and {Fox}, Ryan and {Freij}, Nabil and {Garg}, Suyog and {Geda}, Robel and {Glattly}, Lauren and {Gondhalekar}, Yash and {Gordon}, Karl D. and {Grant}, David and {Greenfield}, Perry and {Groener}, Austen M. and {Guest}, Steve and {Gurovich}, Sebastian and {Handberg}, Rasmus and {Hart}, Akeem and {Hatfield-Dodds}, Zac and {Homeier}, Derek and {Hosseinzadeh}, Griffin and {Jenness}, Tim and {Jones}, Craig K. and {Joseph}, Prajwel and {Kalmbach}, J. Bryce and {Karamehmetoglu}, Emir and {Ka{\l}uszy{\'n}ski}, Miko{\l}aj and {Kelley}, Michael S.~P. and {Kern}, Nicholas and {Kerzendorf}, Wolfgang E. and {Koch}, Eric W. and {Kulumani}, Shankar and {Lee}, Antony and {Ly}, Chun and {Ma}, Zhiyuan and {MacBride}, Conor and {Maljaars}, Jakob M. and {Muna}, Demitri and {Murphy}, N.~A. and {Norman}, Henrik and {O'Steen}, Richard and {Oman}, Kyle A. and {Pacifici}, Camilla and {Pascual}, Sergio and {Pascual-Granado}, J. and {Patil}, Rohit R. and {Perren}, Gabriel I. and {Pickering}, Timothy E. and {Rastogi}, Tanuj and {Roulston}, Benjamin R. and {Ryan}, Daniel F. and {Rykoff}, Eli S. and {Sabater}, Jose and {Sakurikar}, Parikshit and {Salgado}, Jes{\'u}s and {Sanghi}, Aniket and {Saunders}, Nicholas and {Savchenko}, Volodymyr and {Schwardt}, Ludwig and {Seifert-Eckert}, Michael and {Shih}, Albert Y. and {Jain}, Anany Shrey and {Shukla}, Gyanendra and {Sick}, Jonathan and {Simpson}, Chris and {Singanamalla}, Sudheesh and {Singer}, Leo P. and {Singhal}, Jaladh and {Sinha}, Manodeep and {Sip{\H{o}}cz}, Brigitta M. and {Spitler}, Lee R. and {Stansby}, David and {Streicher}, Ole and {{\v{S}}umak}, Jani and {Swinbank}, John D. and {Taranu}, Dan S. and {Tewary}, Nikita and {Tremblay}, Grant R. and {de Val-Borro}, Miguel and {Van Kooten}, Samuel J. and {Vasovi{\'c}}, Zlatan and {Verma}, Shresth and {de Miranda Cardoso}, Jos{\'e} Vin{\'\i}cius and {Williams}, Peter K.~G. and {Wilson}, Tom J. and {Winkel}, Benjamin and {Wood-Vasey}, W.~M. and {Xue}, Rui and {Yoachim}, Peter and {Zhang}, Chen and {Zonca}, Andrea and {Astropy Project Contributors}},
        title = "{The Astropy Project: Sustaining and Growing a Community-oriented Open-source Project and the Latest Major Release (v5.0) of the Core Package}",
      journal = {\apj},
         year = 2022,
        month = aug,
       volume = {935},
       number = {2},
          eid = {167},
        pages = {167},
          doi = {10.3847/1538-4357/ac7c74},
archivePrefix = {arXiv},
       eprint = {2206.14220},
 primaryClass = {astro-ph.IM},
       adsurl = {https://ui.adsabs.harvard.edu/abs/2022ApJ...935..167A}
}

@ARTICLE{Hester2008,
       author = {{Hester}, J.~J.},
        title = "{The Crab Nebula : an astrophysical chimera.}",
      journal = {\araa},
         year = 2008,
        month = sep,
       volume = {46},
        pages = {127-155},
          doi = {10.1146/annurev.astro.45.051806.110608},
       adsurl = {https://ui.adsabs.harvard.edu/abs/2008ARA&A..46..127H}
}

@ARTICLE{Olmi2023,
       author = {{Olmi}, Barbara and {Bucciantini}, Niccol{\`o}},
        title = "{The Dawes Review 11: From young to old: The evolutionary path of Pulsar Wind Nebulae}",
      journal = {\pasa},
         year = 2023,
        month = mar,
       volume = {40},
          eid = {e007},
        pages = {e007},
          doi = {10.1017/pasa.2023.5},
archivePrefix = {arXiv},
       eprint = {2301.12903},
 primaryClass = {astro-ph.HE},
       adsurl = {https://ui.adsabs.harvard.edu/abs/2023PASA...40....7O}
}

@ARTICLE{Rodriguez2024,
       author = {{Rodr{\'\i}guez}, {\'O}smar and {Nakar}, Ehud and {Maoz}, Dan},
        title = "{Stripped-envelope supernova light curves argue for central engine activity}",
      journal = {\nat},
         year = 2024,
        month = apr,
       volume = {628},
       number = {8009},
        pages = {733-735},
          doi = {10.1038/s41586-024-07262-x},
archivePrefix = {arXiv},
       eprint = {2404.10846},
 primaryClass = {astro-ph.HE},
       adsurl = {https://ui.adsabs.harvard.edu/abs/2024Natur.628..733R}
}

@ARTICLE{Metzger2015,
       author = {{Metzger}, Brian D. and {Margalit}, Ben and {Kasen}, Daniel and {Quataert}, Eliot},
        title = "{The diversity of transients from magnetar birth in core collapse supernovae}",
      journal = {\mnras},
         year = 2015,
        month = dec,
       volume = {454},
       number = {3},
        pages = {3311-3316},
          doi = {10.1093/mnras/stv2224},
archivePrefix = {arXiv},
       eprint = {1508.02712},
 primaryClass = {astro-ph.HE},
       adsurl = {https://ui.adsabs.harvard.edu/abs/2015MNRAS.454.3311M}
}

@ARTICLE{Buhler2014,
       author = {{B{\"u}hler}, R. and {Blandford}, R.},
        title = "{The surprising Crab pulsar and its nebula: a review}",
      journal = {Reports on Progress in Physics},
         year = 2014,
        month = jun,
       volume = {77},
       number = {6},
          eid = {066901},
        pages = {066901},
          doi = {10.1088/0034-4885/77/6/066901},
archivePrefix = {arXiv},
       eprint = {1309.7046},
 primaryClass = {astro-ph.HE},
       adsurl = {https://ui.adsabs.harvard.edu/abs/2014RPPh...77f6901B}
}

@ARTICLE{Petre2007,
       author = {{Petre}, R. and {Hwang}, U. and {Holt}, S.~S. and {Safi-Harb}, S. and {Williams}, R.~M.},
        title = "{The X-Ray Structure and Spectrum of the Pulsar Wind Nebula Surrounding PSR B0540-69.3}",
      journal = {\apj},
         year = 2007,
        month = jun,
       volume = {662},
       number = {2},
        pages = {988-997},
          doi = {10.1086/518019},
       adsurl = {https://ui.adsabs.harvard.edu/abs/2007ApJ...662..988P}
}

@ARTICLE{Mathewson1980,
       author = {{Mathewson}, D.~S. and {Dopita}, M.~A. and {Tuohy}, I.~R. and {Ford}, V.~L.},
        title = "{A new oxygen-rich supernova remnant in the Large Magellanic Cloud}",
      journal = {\apjl},
         year = 1980,
        month = dec,
       volume = {242},
        pages = {L73-L76},
          doi = {10.1086/183406},
       adsurl = {https://ui.adsabs.harvard.edu/abs/1980ApJ...242L..73M}
}

@ARTICLE{Giudici2025,
       author = {{Giudici}, Beatrice and {Gabler}, Michael and {Janka}, Hans-Thomas},
        title = "{Hydrodynamic instabilities in long-term three-dimensional simulations of neutrino-driven supernovae of 13 red supergiant progenitors}",
      journal = {arXiv e-prints},
         year = 2025,
        month = nov,
          eid = {arXiv:2511.11796},
        pages = {arXiv:2511.11796},
          doi = {10.48550/arXiv.2511.11796},
archivePrefix = {arXiv},
       eprint = {2511.11796},
 primaryClass = {astro-ph.HE},
       adsurl = {https://ui.adsabs.harvard.edu/abs/2025arXiv251111796G}
}

@ARTICLE{Ennis2006,
       author = {{Ennis}, Jessica A. and {Rudnick}, Lawrence and {Reach}, William T. and {Smith}, J.~D. and {Rho}, Jeonghee and {DeLaney}, Tracey and {Gomez}, Haley and {Kozasa}, Takashi},
        title = "{Spitzer IRAC Images and Sample Spectra of Cassiopeia A's Explosion}",
      journal = {\apj},
         year = 2006,
        month = nov,
       volume = {652},
       number = {1},
        pages = {376-386},
          doi = {10.1086/508142},
archivePrefix = {arXiv},
       eprint = {astro-ph/0610838},
 primaryClass = {astro-ph},
       adsurl = {https://ui.adsabs.harvard.edu/abs/2006ApJ...652..376E}
}

@ARTICLE{MacAlpine2008,
       author = {{MacAlpine}, Gordon M. and {Satterfield}, Timothy J.},
        title = "{The Crab Nebula's Composition and Precursor Star Mass}",
      journal = {\aj},
         year = 2008,
        month = nov,
       volume = {136},
       number = {5},
        pages = {2152-2157},
          doi = {10.1088/0004-6256/136/5/2152},
archivePrefix = {arXiv},
       eprint = {0806.1342},
 primaryClass = {astro-ph},
       adsurl = {https://ui.adsabs.harvard.edu/abs/2008AJ....136.2152M}
}

@ARTICLE{Yang2015,
       author = {{Yang}, Haifeng and {Chevalier}, Roger A.},
        title = "{Evolution of the Crab Nebula in a Low Energy Supernova}",
      journal = {\apj},
         year = 2015,
        month = jun,
       volume = {806},
       number = {2},
          eid = {153},
        pages = {153},
          doi = {10.1088/0004-637X/806/2/153},
archivePrefix = {arXiv},
       eprint = {1505.03211},
 primaryClass = {astro-ph.HE},
       adsurl = {https://ui.adsabs.harvard.edu/abs/2015ApJ...806..153Y}
}

@ARTICLE{Nomoto1982,
       author = {{Nomoto}, Ken'ichi and {Sparks}, Warren M. and {Fesen}, Robert A. and {Gull}, Theodore R. and {Miyaji}, S. and {Sugimoto}, D.},
        title = "{The Crab Nebula's progenitor}",
      journal = {\nat},
         year = 1982,
        month = oct,
       volume = {299},
       number = {5886},
        pages = {803-805},
          doi = {10.1038/299803a0},
       adsurl = {https://ui.adsabs.harvard.edu/abs/1982Natur.299..803N}
}

@ARTICLE{Lyne2015,
       author = {{Lyne}, A.~G. and {Jordan}, C.~A. and {Graham-Smith}, F. and {Espinoza}, C.~M. and {Stappers}, B.~W. and {Weltevrede}, P.},
        title = "{45 years of rotation of the Crab pulsar}",
      journal = {\mnras},
         year = 2015,
        month = jan,
       volume = {446},
       number = {1},
        pages = {857-864},
          doi = {10.1093/mnras/stu2118},
archivePrefix = {arXiv},
       eprint = {1410.0886},
 primaryClass = {astro-ph.HE},
       adsurl = {https://ui.adsabs.harvard.edu/abs/2015MNRAS.446..857L}
}

@ARTICLE{Marshall2016,
       author = {{Marshall}, F.~E. and {Guillemot}, L. and {Harding}, A.~K. and {Martin}, P. and {Smith}, D.~A.},
        title = "{A New, Low Braking Index for the LMC Pulsar B0540-69}",
      journal = {\apjl},
         year = 2016,
        month = aug,
       volume = {827},
       number = {2},
          eid = {L39},
        pages = {L39},
          doi = {10.3847/2041-8205/827/2/L39},
archivePrefix = {arXiv},
       eprint = {1608.01901},
 primaryClass = {astro-ph.HE},
       adsurl = {https://ui.adsabs.harvard.edu/abs/2016ApJ...827L..39M}
}

@ARTICLE{Olmi2016,
       author = {{Olmi}, B. and {Del Zanna}, L. and {Amato}, E. and {Bucciantini}, N. and {Mignone}, A.},
        title = "{Multi-D magnetohydrodynamic modelling of pulsar wind nebulae: recent progress and open questions}",
      journal = {Journal of Plasma Physics},
         year = 2016,
        month = dec,
       volume = {82},
       number = {6},
          eid = {635820601},
        pages = {635820601},
          doi = {10.1017/S0022377816000957},
archivePrefix = {arXiv},
       eprint = {1610.07956},
 primaryClass = {astro-ph.HE},
       adsurl = {https://ui.adsabs.harvard.edu/abs/2016JPlPh..82f6301O}
}

@ARTICLE{Porth2014b,
       author = {{Porth}, Oliver and {Komissarov}, Serguei S. and {Keppens}, Rony},
        title = "{Rayleigh-Taylor instability in magnetohydrodynamic simulations of the Crab nebula}",
      journal = {\mnras},
         year = 2014,
        month = sep,
       volume = {443},
       number = {1},
        pages = {547-558},
          doi = {10.1093/mnras/stu1082},
archivePrefix = {arXiv},
       eprint = {1405.4029},
 primaryClass = {astro-ph.HE},
       adsurl = {https://ui.adsabs.harvard.edu/abs/2014MNRAS.443..547P}
}

@ARTICLE{Blondin2017,
       author = {{Blondin}, John M. and {Chevalier}, Roger A.},
        title = "{Pulsar Wind Bubble Blowout from a Supernova}",
      journal = {\apj},
         year = 2017,
        month = aug,
       volume = {845},
       number = {2},
          eid = {139},
        pages = {139},
          doi = {10.3847/1538-4357/aa8267},
archivePrefix = {arXiv},
       eprint = {1707.07021},
 primaryClass = {astro-ph.HE},
       adsurl = {https://ui.adsabs.harvard.edu/abs/2017ApJ...845..139B}
}

@ARTICLE{Hobbs2005,
       author = {{Hobbs}, G. and {Lorimer}, D.~R. and {Lyne}, A.~G. and {Kramer}, M.},
        title = "{A statistical study of 233 pulsar proper motions}",
      journal = {\mnras},
         year = 2005,
        month = jul,
       volume = {360},
       number = {3},
        pages = {974-992},
          doi = {10.1111/j.1365-2966.2005.09087.x},
archivePrefix = {arXiv},
       eprint = {astro-ph/0504584},
 primaryClass = {astro-ph},
       adsurl = {https://ui.adsabs.harvard.edu/abs/2005MNRAS.360..974H}
}

@ARTICLE{Faucher2006,
       author = {{Faucher-Gigu{\`e}re}, Claude-Andr{\'e} and {Kaspi}, Victoria M.},
        title = "{Birth and Evolution of Isolated Radio Pulsars}",
      journal = {\apj},
         year = 2006,
        month = may,
       volume = {643},
       number = {1},
        pages = {332-355},
          doi = {10.1086/501516},
archivePrefix = {arXiv},
       eprint = {astro-ph/0512585},
 primaryClass = {astro-ph},
       adsurl = {https://ui.adsabs.harvard.edu/abs/2006ApJ...643..332F}
}

@ARTICLE{Boubert2017,
       author = {{Boubert}, D. and {Erkal}, D. and {Evans}, N.~W. and {Izzard}, R.~G.},
        title = "{Hypervelocity runaways from the Large Magellanic Cloud}",
      journal = {\mnras},
         year = 2017,
        month = aug,
       volume = {469},
       number = {2},
        pages = {2151-2162},
          doi = {10.1093/mnras/stx848},
archivePrefix = {arXiv},
       eprint = {1704.01373},
 primaryClass = {astro-ph.GA},
       adsurl = {https://ui.adsabs.harvard.edu/abs/2017MNRAS.469.2151B}
}

@ARTICLE{Lin2023,
       author = {{Lin}, Zehao and {Xu}, Ye and {Hao}, Chaojie and {Li}, Yingjie and {Liu}, Dejian and {Bian}, Shuaibo},
        title = "{Massive Hypervelocity Runaway Stars in the Large Magellanic Cloud}",
      journal = {\apj},
         year = 2023,
        month = jul,
       volume = {952},
       number = {1},
          eid = {64},
        pages = {64},
          doi = {10.3847/1538-4357/acd644},
       adsurl = {https://ui.adsabs.harvard.edu/abs/2023ApJ...952...64L}
}

@INPROCEEDINGS{Kargaltsev2008,
       author = {{Kargaltsev}, O. and {Pavlov}, G.~G.},
        title = "{Pulsar Wind Nebulae in the Chandra Era}",
    booktitle = {40 Years of Pulsars: Millisecond Pulsars, Magnetars and More},
         year = 2008,
       editor = {{Bassa}, C. and {Wang}, Z. and {Cumming}, A. and {Kaspi}, V.~M.},
       series = {American Institute of Physics Conference Series},
       volume = {983},
        month = feb,
    publisher = {AIP},
        pages = {171-185},
          doi = {10.1063/1.2900138},
archivePrefix = {arXiv},
       eprint = {0801.2602},
 primaryClass = {astro-ph},
       adsurl = {https://ui.adsabs.harvard.edu/abs/2008AIPC..983..171K}
}

@ARTICLE{Fesen2008,
       author = {{Fesen}, Robert and {Rudie}, Gwen and {Hurford}, Alan and {Soto}, Aljeandro},
        title = "{Optical Imaging and Spectroscopy of the Galactic Supernova Remnant 3C 58 (G130.7+3.1)}",
      journal = {\apjs},
         year = 2008,
        month = feb,
       volume = {174},
       number = {2},
        pages = {379-395},
          doi = {10.1086/522781},
       adsurl = {https://ui.adsabs.harvard.edu/abs/2008ApJS..174..379F}
}

@ARTICLE{Slane2004,
       author = {{Slane}, Patrick and {Helfand}, David J. and {van der Swaluw}, Eric and {Murray}, Stephen S.},
        title = "{New Constraints on the Structure and Evolution of the Pulsar Wind Nebula 3C 58}",
      journal = {\apj},
         year = 2004,
        month = nov,
       volume = {616},
       number = {1},
        pages = {403-413},
          doi = {10.1086/424814},
archivePrefix = {arXiv},
       eprint = {astro-ph/0405380},
 primaryClass = {astro-ph},
       adsurl = {https://ui.adsabs.harvard.edu/abs/2004ApJ...616..403S}
}

@ARTICLE{Temim2024,
       author = {{Temim}, Tea and {Laming}, J. Martin and {Kavanagh}, P.~J. and {Smith}, Nathan and {Slane}, Patrick and {Blair}, William P. and {De Looze}, Ilse and {Bucciantini}, Niccol{\`o} and {Jerkstrand}, Anders and {Gountanis}, Nicole Marcelina and {Sankrit}, Ravi and {Milisavljevic}, Dan and {Rest}, Armin and {Lyutikov}, Maxim and {DePasquale}, Joseph and {Martin}, Thomas and {Drissen}, Laurent and {Raymond}, John and {Fox}, Ori D. and {Modjaz}, Maryam and {Spitkovsky}, Anatoly and {Strolger}, Louis-Gregory},
        title = "{Dissecting the Crab Nebula with JWST: Pulsar Wind, Dusty Filaments, and Ni/Fe Abundance Constraints on the Explosion Mechanism}",
      journal = {\apjl},
         year = 2024,
        month = jun,
       volume = {968},
       number = {2},
          eid = {L18},
        pages = {L18},
          doi = {10.3847/2041-8213/ad50d1},
archivePrefix = {arXiv},
       eprint = {2406.00172},
 primaryClass = {astro-ph.HE},
       adsurl = {https://ui.adsabs.harvard.edu/abs/2024ApJ...968L..18T}
}

@ARTICLE{Blair2026,
       author = {{Blair}, William P. and {Sankrit}, Ravi and {Milisavljevic}, Dan and {Temim}, Tea and {Laming}, J. Martin and {Slane}, Patrick and {Ding}, Ziwei and {Martin}, Thomas},
        title = "{The Crab Nebula Revisited Using HST/WFC3}",
      journal = {\apj},
         year = 2026,
        month = jan,
       volume = {997},
       number = {1},
          eid = {81},
        pages = {81},
          doi = {10.3847/1538-4357/ae2adc},
archivePrefix = {arXiv},
       eprint = {2512.11103},
 primaryClass = {astro-ph.SR},
       adsurl = {https://ui.adsabs.harvard.edu/abs/2026ApJ...997...81B}
}

@ARTICLE{Quinet1996,
       author = {{Quinet}, P. and {Le Dourneuf}, M. and {Zeippen}, C.~J.},
        title = "{Atomic data from the IRON Project. XIX. Radiative transition probabilities for forbidden lines in Fe II.}",
      journal = {\aaps},
         year = 1996,
        month = dec,
       volume = {120},
        pages = {361-371},
       adsurl = {https://ui.adsabs.harvard.edu/abs/1996A&AS..120..361Q}
}

@ARTICLE{Nussbaumer1988,
       author = {{Nussbaumer}, H. and {Storey}, P.~J.},
        title = "{Transition probabilities for Fe II infrared lines.}",
      journal = {\aap},
         year = 1988,
        month = mar,
       volume = {193},
        pages = {327-333},
       adsurl = {https://ui.adsabs.harvard.edu/abs/1988A&A...193..327N}
}

@ARTICLE{Gianni2015,
       author = {{Giannini}, T. and {Antoniucci}, S. and {Nisini}, B. and {Lorenzetti}, D. and {Alcal{\'a}}, J.~M. and {Bacciotti}, F. and {Bonito}, R. and {Podio}, L. and {Stelzer}, B.},
        title = "{Empirical Determination of Einstein A-coefficient Ratios of Bright [Fe II] Lines}",
      journal = {\apj},
         year = 2015,
        month = jan,
       volume = {798},
       number = {1},
          eid = {33},
        pages = {33},
          doi = {10.1088/0004-637X/798/1/33},
archivePrefix = {arXiv},
       eprint = {1410.8304},
 primaryClass = {astro-ph.SR},
       adsurl = {https://ui.adsabs.harvard.edu/abs/2015ApJ...798...33G}
}

@ARTICLE{Koo2015,
       author = {{Koo}, Bon-Chul and {Lee}, Yong-Hyun},
        title = "{Near-Infrared Spectroscopy of Young Galactic Supernova Remnants}",
      journal = {Publication of Korean Astronomical Society},
         year = 2015,
        month = sep,
       volume = {30},
       number = {2},
        pages = {145-148},
          doi = {10.5303/PKAS.2015.30.2.145},
archivePrefix = {arXiv},
       eprint = {1502.00048},
 primaryClass = {astro-ph.GA},
       adsurl = {https://ui.adsabs.harvard.edu/abs/2015PKAS...30..145K}
}

@ARTICLE{Tenhu2025,
       author = {{Tenhu}, L. and {Larsson}, J. and {Lundqvist}, P. and {Saathoff}, I. and {Lyman}, J.~D. and {Sollerman}, J.},
        title = "{MUSE observations reveal optical coronal iron lines from shock emission in supernova remnant 0540‑69.3}",
      journal = {\mnras},
         year = 2025,
        month = oct,
       volume = {542},
       number = {4},
        pages = {2830-2856},
          doi = {10.1093/mnras/staf1390},
archivePrefix = {arXiv},
       eprint = {2508.14570},
 primaryClass = {astro-ph.HE},
       adsurl = {https://ui.adsabs.harvard.edu/abs/2025MNRAS.542.2830T}
}

@ARTICLE{Tenhu2024,
       author = {{Tenhu}, L. and {Larsson}, J. and {Sollerman}, J. and {Lundqvist}, P. and {Spyromilio}, J. and {Lyman}, J.~D. and {Olofsson}, G.},
        title = "{Spatial Variations and Breaks in the Optical{\textendash}Near-infrared Spectra of the Pulsar and Pulsar Wind Nebula in Supernova Remnant 0540{\textendash}69.3}",
      journal = {\apj},
         year = 2024,
        month = may,
       volume = {966},
       number = {1},
          eid = {125},
        pages = {125},
          doi = {10.3847/1538-4357/ad3214},
archivePrefix = {arXiv},
       eprint = {2403.05206},
 primaryClass = {astro-ph.HE},
       adsurl = {https://ui.adsabs.harvard.edu/abs/2024ApJ...966..125T}
}

@ARTICLE{Argyriou2023,
       author = {{Argyriou}, Ioannis and {Glasse}, Alistair and {Law}, David R. and {Labiano}, Alvaro and {{\'A}lvarez-M{\'a}rquez}, Javier and {Patapis}, Polychronis and {Kavanagh}, Patrick J. and {Gasman}, Danny and {Mueller}, Michael and {Larson}, Kirsten and {Vandenbussche}, Bart and {Glauser}, Adrian M. and {Royer}, Pierre and {Dicken}, Daniel and {Harkett}, Jake and {Sargent}, Beth A. and {Engesser}, Michael and {Jones}, Olivia C. and {Kendrew}, Sarah and {Noriega-Crespo}, Alberto and {Brandl}, Bernhard and {Rieke}, George H. and {Wright}, Gillian S. and {Lee}, David and {Wells}, Martyn},
        title = "{JWST MIRI flight performance: The Medium-Resolution Spectrometer}",
      journal = {\aap},
         year = 2023,
        month = jul,
       volume = {675},
          eid = {A111},
        pages = {A111},
          doi = {10.1051/0004-6361/202346489},
archivePrefix = {arXiv},
       eprint = {2303.13469},
 primaryClass = {astro-ph.IM},
       adsurl = {https://ui.adsabs.harvard.edu/abs/2023A&A...675A.111A}
}

@ARTICLE{Larsson2021,
       author = {{Larsson}, J. and {Sollerman}, J. and {Lyman}, J.~D. and {Spyromilio}, J. and {Tenhu}, L. and {Fransson}, C. and {Lundqvist}, P.},
        title = "{Clumps and Rings of Ejecta in SNR 0540-69.3 as Seen in 3D}",
      journal = {\apj},
         year = 2021,
        month = dec,
       volume = {922},
       number = {2},
          eid = {265},
        pages = {265},
          doi = {10.3847/1538-4357/ac2a41},
archivePrefix = {arXiv},
       eprint = {2109.03683},
 primaryClass = {astro-ph.HE},
       adsurl = {https://ui.adsabs.harvard.edu/abs/2021ApJ...922..265L}
}

@ARTICLE{Pontoppidan2024,
       author = {{Pontoppidan}, Klaus M. and {Salyk}, Colette and {Banzatti}, Andrea and {Zhang}, Ke and {Pascucci}, Ilaria and {{\"O}berg}, Karin I. and {Long}, Feng and {Romero-Mirza}, Carlos E. and {Carr}, John and {Najita}, Joan and {Blake}, Geoffrey A. and {Arulanantham}, Nicole and {Andrews}, Sean and {Ballering}, Nicholas P. and {Bergin}, Edwin and {Calahan}, Jenny and {Cobb}, Douglas and {Colmenares}, Maria Jose and {Dickson-Vandervelde}, Annie and {Dignan}, Anna and {Green}, Joel and {Heretz}, Phoebe and {Herczeg}, Gregory and {Kalyaan}, Anusha and {Krijt}, Sebastiaan and {Pauly}, Tyler and {Pinilla}, Paola and {Trapman}, Leon and {Xie}, Chengyan},
        title = "{High-contrast JWST-MIRI Spectroscopy of Planet-forming Disks for the JDISC Survey}",
      journal = {\apj},
         year = 2024,
        month = mar,
       volume = {963},
       number = {2},
          eid = {158},
        pages = {158},
          doi = {10.3847/1538-4357/ad20f0},
archivePrefix = {arXiv},
       eprint = {2311.17020},
 primaryClass = {astro-ph.EP},
       adsurl = {https://ui.adsabs.harvard.edu/abs/2024ApJ...963..158P}
}

@ARTICLE{Law2023,
       author = {{Law}, David R. and {E. Morrison}, Jane and {Argyriou}, Ioannis and {Patapis}, Polychronis and {{\'A}lvarez-M{\'a}rquez}, J. and {Labiano}, Alvaro and {Vandenbussche}, Bart},
        title = "{A 3D Drizzle Algorithm for JWST and Practical Application to the MIRI Medium Resolution Spectrometer}",
      journal = {\aj},
         year = 2023,
        month = aug,
       volume = {166},
       number = {2},
          eid = {45},
        pages = {45},
          doi = {10.3847/1538-3881/acdddc},
archivePrefix = {arXiv},
       eprint = {2306.05520},
 primaryClass = {astro-ph.IM},
       adsurl = {https://ui.adsabs.harvard.edu/abs/2023AJ....166...45L}
}

@ARTICLE{Dumont2025,
       author = {{Dumont}, Antoine and {Neumayer}, Nadine and {Seth}, Anil C. and {B{\"o}ker}, Torsten and {Eracleous}, Michael and {Goold}, Kameron and {Greene}, Jenny E. and {G{\"u}ltekin}, Kayhan and {Ho}, Luis C. and {Walsh}, Jonelle L. and {L{\"u}tzgendorf}, Nora},
        title = "{WIggle Corrector Kit for NIRSpEc Data: WICKED}",
      journal = {arXiv e-prints},
         year = 2025,
        month = mar,
          eid = {arXiv:2503.09697},
        pages = {arXiv:2503.09697},
          doi = {10.48550/arXiv.2503.09697},
archivePrefix = {arXiv},
       eprint = {2503.09697},
 primaryClass = {astro-ph.IM},
       adsurl = {https://ui.adsabs.harvard.edu/abs/2025arXiv250309697D}
}

@software{Bushouse2023,
       author = {{Bushouse}, Howard and {Eisenhamer}, Jonathan and {Dencheva}, Nadia and {Davies}, James and {Greenfield}, Perry and {Morrison}, Jane and {Hodge}, Phil and {Simon}, Bernie and {Grumm}, David and {Droettboom}, Michael and {Slavich}, Edward and {Sosey}, Megan and {Pauly}, Tyler and {Miller}, Todd and {Jedrzejewski}, Robert and {Hack}, Warren and {Davis}, David and {Crawford}, Steven and {Law}, David and {Gordon}, Karl and {Regan}, Michael and {Cara}, Mihai and {MacDonald}, Ken and {Bradley}, Larry and {Shanahan}, Clare and {Jamieson}, William and {Teodoro}, Mairan and {Williams}, Thomas and {Pena-Guerrero}, Maria},
        title = "{JWST Calibration Pipeline}",
         year = 2023,
        month = oct,
          eid = {10.5281/zenodo.10022973},
          doi = {10.5281/zenodo.10022973},
      version = {1.12.5},
    publisher = {Zenodo},
       adsurl = {https://ui.adsabs.harvard.edu/abs/2023zndo..10022973B}
}

@ARTICLE{Wells2015,
       author = {{Wells}, Martyn and {Pel}, J. -W. and {Glasse}, Alistair and {Wright}, G.~S. and {Aitink-Kroes}, Gabby and {Azzollini}, Ruym{\'a}n and {Beard}, Steven and {Brandl}, B.~R. and {Gallie}, Angus and {Geers}, V.~C. and {Glauser}, A.~M. and {Hastings}, Peter and {Henning}, Th. and {Jager}, Rieks and {Justtanont}, K. and {Kruizinga}, Bob and {Lahuis}, Fred and {Lee}, David and {Martinez-Delgado}, I. and {Mart{\'\i}nez-Galarza}, J.~R. and {Meijers}, M. and {Morrison}, Jane E. and {M{\"u}ller}, Friedrich and {Nakos}, Thodori and {O'Sullivan}, Brian and {Oudenhuysen}, Ad and {Parr-Burman}, P. and {Pauwels}, Evert and {Rohloff}, R. -R. and {Schmalzl}, Eva and {Sykes}, Jon and {Thelen}, M.~P. and {van Dishoeck}, E.~F. and {Vandenbussche}, Bart and {Venema}, Lars B. and {Visser}, Huib and {Waters}, L.~B.~F.~M. and {Wright}, David},
        title = "{The Mid-Infrared Instrument for the James Webb Space Telescope, VI: The Medium Resolution Spectrometer}",
      journal = {\pasp},
         year = 2015,
        month = jul,
       volume = {127},
       number = {953},
        pages = {646},
          doi = {10.1086/682281},
archivePrefix = {arXiv},
       eprint = {1508.03070},
 primaryClass = {astro-ph.IM},
       adsurl = {https://ui.adsabs.harvard.edu/abs/2015PASP..127..646W}
}

@ARTICLE{Jakobsen2022,
       author = {{Jakobsen}, P. and {Ferruit}, P. and {Alves de Oliveira}, C. and {Arribas}, S. and {Bagnasco}, G. and {Barho}, R. and {Beck}, T.~L. and {Birkmann}, S. and {B{\"o}ker}, T. and {Bunker}, A.~J. and {Charlot}, S. and {de Jong}, P. and {de Marchi}, G. and {Ehrenwinkler}, R. and {Falcolini}, M. and {Fels}, R. and {Franx}, M. and {Franz}, D. and {Funke}, M. and {Giardino}, G. and {Gnata}, X. and {Holota}, W. and {Honnen}, K. and {Jensen}, P.~L. and {Jentsch}, M. and {Johnson}, T. and {Jollet}, D. and {Karl}, H. and {Kling}, G. and {K{\"o}hler}, J. and {Kolm}, M. -G. and {Kumari}, N. and {Lander}, M.~E. and {Lemke}, R. and {L{\'o}pez-Caniego}, M. and {L{\"u}tzgendorf}, N. and {Maiolino}, R. and {Manjavacas}, E. and {Marston}, A. and {Maschmann}, M. and {Maurer}, R. and {Messerschmidt}, B. and {Moseley}, S.~H. and {Mosner}, P. and {Mott}, D.~B. and {Muzerolle}, J. and {Pirzkal}, N. and {Pittet}, J. -F. and {Plitzke}, A. and {Posselt}, W. and {Rapp}, B. and {Rauscher}, B.~J. and {Rawle}, T. and {Rix}, H. -W. and {R{\"o}del}, A. and {Rumler}, P. and {Sabbi}, E. and {Salvignol}, J. -C. and {Schmid}, T. and {Sirianni}, M. and {Smith}, C. and {Strada}, P. and {te Plate}, M. and {Valenti}, J. and {Wettemann}, T. and {Wiehe}, T. and {Wiesmayer}, M. and {Willott}, C.~J. and {Wright}, R. and {Zeidler}, P. and {Zincke}, C.},
        title = "{The Near-Infrared Spectrograph (NIRSpec) on the James Webb Space Telescope. I. Overview of the instrument and its capabilities}",
      journal = {\aap},
         year = 2022,
        month = may,
       volume = {661},
          eid = {A80},
        pages = {A80},
          doi = {10.1051/0004-6361/202142663},
archivePrefix = {arXiv},
       eprint = {2202.03305},
 primaryClass = {astro-ph.IM},
       adsurl = {https://ui.adsabs.harvard.edu/abs/2022A&A...661A..80J}
}

@ARTICLE{Basko1994,
       author = {{Basko}, Mikhail},
        title = "{Nickel Bubble Instability and Mixing in SN 1987A}",
      journal = {\apj},
         year = 1994,
        month = apr,
       volume = {425},
        pages = {264},
          doi = {10.1086/173983},
       adsurl = {https://ui.adsabs.harvard.edu/abs/1994ApJ...425..264B}
}

@INPROCEEDINGS{Berry2007,
       author = {{Berry}, D.~S. and {Reinhold}, K. and {Jenness}, T. and {Economou}, F.},
        title = "{CUPID: A Clump Identification and Analysis Package}",
    booktitle = {Astronomical Data Analysis Software and Systems XVI},
         year = 2007,
       editor = {{Shaw}, R.~A. and {Hill}, F. and {Bell}, D.~J.},
       series = {Astronomical Society of the Pacific Conference Series},
       volume = {376},
        month = oct,
        pages = {425},
       adsurl = {https://ui.adsabs.harvard.edu/abs/2007ASPC..376..425B}
}

@ARTICLE{Berry2015,
       author = {{Berry}, D.~S.},
        title = "{FellWalker-A clump identification algorithm}",
      journal = {Astronomy and Computing},
         year = 2015,
        month = apr,
       volume = {10},
        pages = {22-31},
          doi = {10.1016/j.ascom.2014.11.004},
archivePrefix = {arXiv},
       eprint = {1411.6267},
 primaryClass = {astro-ph.IM},
       adsurl = {https://ui.adsabs.harvard.edu/abs/2015A&C....10...22B}
}

@ARTICLE{Brantseg2014,
       author = {{Brantseg}, T. and {McEntaffer}, R.~L. and {Bozzetto}, L.~M. and {Filipovic}, M. and {Grieves}, N.},
        title = "{A Multi-wavelength Look at the Young Plerionic Supernova Remnant 0540-69.3}",
      journal = {\apj},
         year = 2014,
        month = jan,
       volume = {780},
       number = {1},
          eid = {50},
        pages = {50},
          doi = {10.1088/0004-637X/780/1/50},
archivePrefix = {arXiv},
       eprint = {1312.4790},
 primaryClass = {astro-ph.HE},
       adsurl = {https://ui.adsabs.harvard.edu/abs/2014ApJ...780...50B}
}

@ARTICLE{DeLuca2007,
       author = {{De Luca}, A. and {Mignani}, R.~P. and {Caraveo}, P.~A. and {Bignami}, G.~F.},
        title = "{Hubble Space Telescope Multiepoch Imaging of the PSR B0540-69 System Unveils a Highly Dynamic Synchrotron Nebula}",
      journal = {\apjl},
         year = 2007,
        month = sep,
       volume = {667},
       number = {1},
        pages = {L77-L80},
          doi = {10.1086/522033},
archivePrefix = {arXiv},
       eprint = {0708.0290},
 primaryClass = {astro-ph},
       adsurl = {https://ui.adsabs.harvard.edu/abs/2007ApJ...667L..77D}
}

@ARTICLE{Gabler2021,
       author = {{Gabler}, Michael and {Wongwathanarat}, Annop and {Janka}, Hans-Thomas},
        title = "{The infancy of core-collapse supernova remnants}",
      journal = {\mnras},
         year = 2021,
        month = apr,
       volume = {502},
       number = {3},
        pages = {3264-3293},
          doi = {10.1093/mnras/stab116},
archivePrefix = {arXiv},
       eprint = {2008.01763},
 primaryClass = {astro-ph.SR},
       adsurl = {https://ui.adsabs.harvard.edu/abs/2021MNRAS.502.3264G}
}

@ARTICLE{Gotthelf2000,
       author = {{Gotthelf}, E.~V. and {Wang}, Q. Daniel},
        title = "{A Spatially Resolved Plerionic X-Ray Nebula around PSR B0540-69}",
      journal = {\apjl},
         year = 2000,
        month = apr,
       volume = {532},
       number = {2},
        pages = {L117-L120},
          doi = {10.1086/312568},
archivePrefix = {arXiv},
       eprint = {astro-ph/0002166},
 primaryClass = {astro-ph},
       adsurl = {https://ui.adsabs.harvard.edu/abs/2000ApJ...532L.117G}
}

@ARTICLE{Hunter2007,
   author = {{Hunter}, J.~D.},
    title = "{Matplotlib: A 2D Graphics Environment}",
  journal = {Computing in Science and Engineering},
     year = 2007,
    month = may,
   volume = 9,
    pages = {90-95},
      doi = {10.1109/MCSE.2007.55},
   adsurl = {http://adsabs.harvard.edu/abs/2007CSE.....9...90H}
}

@ARTICLE{Hwang2001,
       author = {{Hwang}, Una and {Petre}, Robert and {Holt}, Stephen S. and {Szymkowiak}, Andrew E.},
        title = "{The Thermal X-Ray-emitting Shell of Large Magellanic Cloud Supernova Remnant 0540-69.3}",
      journal = {\apj},
         year = 2001,
        month = oct,
       volume = {560},
       number = {2},
        pages = {742-748},
          doi = {10.1086/322962},
archivePrefix = {arXiv},
       eprint = {astro-ph/0106415},
 primaryClass = {astro-ph},
       adsurl = {https://ui.adsabs.harvard.edu/abs/2001ApJ...560..742H}
}

@ARTICLE{Kirshner1989,
       author = {{Kirshner}, Robert P. and {Morse}, Jon A. and {Winkler}, P. Frank and {Blair}, William P.},
        title = "{The Penultimate Supernova in the Large Magellanic Cloud: SNR 0540-69.3}",
      journal = {\apj},
         year = 1989,
        month = jul,
       volume = {342},
        pages = {260},
          doi = {10.1086/167590},
       adsurl = {https://ui.adsabs.harvard.edu/abs/1989ApJ...342..260K}
}

@ARTICLE{Kozma1998,
       author = {{Kozma}, Cecilia and {Fransson}, Claes},
        title = "{Late Spectral Evolution of SN 1987A. II. Line Emission}",
      journal = {\apj},
         year = 1998,
        month = apr,
       volume = {497},
       number = {1},
        pages = {431-457},
          doi = {10.1086/305452},
archivePrefix = {arXiv},
       eprint = {astro-ph/9712224},
 primaryClass = {astro-ph},
       adsurl = {https://ui.adsabs.harvard.edu/abs/1998ApJ...497..431K}
}

@ARTICLE{Larsson2019,
       author = {{Larsson}, J. and {Spyromilio}, J. and {Fransson}, C. and {Indebetouw}, R. and {Matsuura}, M. and {Abell{\'a}n}, F.~J. and {Cigan}, P. and {Gomez}, H. and {Leibundgut}, B.},
        title = "{A Three-dimensional View of Molecular Hydrogen in SN 1987A}",
      journal = {\apj},
         year = 2019,
        month = mar,
       volume = {873},
       number = {1},
          eid = {15},
        pages = {15},
          doi = {10.3847/1538-4357/ab03d1},
archivePrefix = {arXiv},
       eprint = {1901.11235},
 primaryClass = {astro-ph.HE},
       adsurl = {https://ui.adsabs.harvard.edu/abs/2019ApJ...873...15L}
}

@ARTICLE{Li1993,
       author = {{Li}, Hongwei and {McCray}, Richard and {Sunyaev}, Rashid A.},
        title = "{Iron, Cobalt, and Nickel in SN 1987A}",
      journal = {\apj},
         year = 1993,
        month = dec,
       volume = {419},
        pages = {824},
          doi = {10.1086/173534},
       adsurl = {https://ui.adsabs.harvard.edu/abs/1993ApJ...419..824L}
}

@ARTICLE{Lundqvist2011,
       author = {{Lundqvist}, N. and {Lundqvist}, P. and {Bj{\"o}rnsson}, C. -I. and {Olofsson}, G. and {Pires}, S. and {Shibanov}, Yu. A. and {Zyuzin}, D.~A.},
        title = "{Spectral evolution and polarization of variable structures in the pulsar wind nebula of PSR B0540-69.3}",
      journal = {\mnras},
         year = 2011,
        month = may,
       volume = {413},
       number = {1},
        pages = {611-627},
          doi = {10.1111/j.1365-2966.2010.18159.x},
archivePrefix = {arXiv},
       eprint = {1008.3750},
 primaryClass = {astro-ph.HE},
       adsurl = {https://ui.adsabs.harvard.edu/abs/2011MNRAS.413..611L}
}

@ARTICLE{Lundqvist2020,
       author = {{Lundqvist}, P. and {Lundqvist}, N. and {Vlahakis}, C. and {Bj{\"o}rnsson}, C. -I. and {Dickel}, J.~R. and {Matsuura}, M. and {Shibanov}, Yu A. and {Zyuzin}, D.~A. and {Olofsson}, G.},
        title = "{Atacama Compact Array observations of the pulsar-wind nebula of SNR 0540-69.3}",
      journal = {\mnras},
         year = 2020,
        month = aug,
       volume = {496},
       number = {2},
        pages = {1834-1844},
          doi = {10.1093/mnras/staa1675},
archivePrefix = {arXiv},
       eprint = {2006.05222},
 primaryClass = {astro-ph.HE},
       adsurl = {https://ui.adsabs.harvard.edu/abs/2020MNRAS.496.1834L}
}

@ARTICLE{Lundqvist2022,
       author = {{Lundqvist}, P. and {Lundqvist}, N. and {Shibanov}, Yu. A.},
        title = "{Kinematics, structure and abundances of supernova remnant 0540-69.3}",
      journal = {\aap},
         year = 2022,
        month = feb,
       volume = {658},
          eid = {A30},
        pages = {A30},
          doi = {10.1051/0004-6361/202141931},
archivePrefix = {arXiv},
       eprint = {2109.03287},
 primaryClass = {astro-ph.HE},
       adsurl = {https://ui.adsabs.harvard.edu/abs/2022A&A...658A..30L}
}

@ARTICLE{Martin2021,
       author = {{Martin}, T. and {Milisavljevic}, D. and {Drissen}, L.},
        title = "{3D mapping of the Crab Nebula with SITELLE - I. Deconvolution and kinematic reconstruction}",
      journal = {\mnras},
         year = 2021,
        month = apr,
       volume = {502},
       number = {2},
        pages = {1864-1881},
          doi = {10.1093/mnras/staa4046},
archivePrefix = {arXiv},
       eprint = {2101.02709},
 primaryClass = {astro-ph.HE},
       adsurl = {https://ui.adsabs.harvard.edu/abs/2021MNRAS.502.1864M}
}

@ARTICLE{Mignani2012,
       author = {{Mignani}, R.~P. and {De Luca}, A. and {Hummel}, W. and {Zajczyk}, A. and {Rudak}, B. and {Kanbach}, G. and {S{\l}owikowska}, A.},
        title = "{The near-infrared detection of PSR B0540-69 and its nebula}",
      journal = {\aap},
         year = 2012,
        month = aug,
       volume = {544},
          eid = {A100},
        pages = {A100},
          doi = {10.1051/0004-6361/201219177},
archivePrefix = {arXiv},
       eprint = {1204.6655},
 primaryClass = {astro-ph.HE},
       adsurl = {https://ui.adsabs.harvard.edu/abs/2012A&A...544A.100M}
}

@ARTICLE{Mignani2010,
       author = {{Mignani}, R.~P. and {Sartori}, A. and {de Luca}, A. and {Rudak}, B. and {S{\l}owikowska}, A. and {Kanbach}, G. and {Caraveo}, P.~A.},
        title = "{HST/WFPC2 observations of the LMC pulsar PSR B0540-69}",
      journal = {\aap},
         year = 2010,
        month = jun,
       volume = {515},
          eid = {A110},
        pages = {A110},
          doi = {10.1051/0004-6361/200913870},
archivePrefix = {arXiv},
       eprint = {1003.0786},
 primaryClass = {astro-ph.SR},
       adsurl = {https://ui.adsabs.harvard.edu/abs/2010A&A...515A.110M}
}

@ARTICLE{Morse2006,
       author = {{Morse}, Jon A. and {Smith}, Nathan and {Blair}, William P. and {Kirshner}, Robert P. and {Winkler}, P. Frank and {Hughes}, John P.},
        title = "{Hubble Space Telescope Observations of Oxygen-rich Supernova Remnants in the Magellanic Clouds. III. WFPC2 Imaging of the Young, Crab-like Supernova Remnant SNR 0540-69.3}",
      journal = {\apj},
         year = 2006,
        month = jun,
       volume = {644},
       number = {1},
        pages = {188-197},
          doi = {10.1086/503313},
       adsurl = {https://ui.adsabs.harvard.edu/abs/2006ApJ...644..188M}
}

@ARTICLE{Orlando2021,
       author = {{Orlando}, S. and {Wongwathanarat}, A. and {Janka}, H. -T. and {Miceli}, M. and {Ono}, M. and {Nagataki}, S. and {Bocchino}, F. and {Peres}, G.},
        title = "{The fully developed remnant of a neutrino-driven supernova. Evolution of ejecta structure and asymmetries in SNR Cassiopeia A}",
      journal = {\aap},
         year = 2021,
        month = jan,
       volume = {645},
          eid = {A66},
        pages = {A66},
          doi = {10.1051/0004-6361/202039335},
archivePrefix = {arXiv},
       eprint = {2009.01789},
 primaryClass = {astro-ph.HE},
       adsurl = {https://ui.adsabs.harvard.edu/abs/2021A&A...645A..66O}
}

@ARTICLE{Pietrzynski2019,
       author = {{Pietrzy{\'n}ski}, G. and {Graczyk}, D. and {Gallenne}, A. and {Gieren}, W. and {Thompson}, I.~B. and {Pilecki}, B. and {Karczmarek}, P. and {G{\'o}rski}, M. and {Suchomska}, K. and {Taormina}, M. and {Zgirski}, B. and {Wielg{\'o}rski}, P. and {Ko{\l}aczkowski}, Z. and {Konorski}, P. and {Villanova}, S. and {Nardetto}, N. and {Kervella}, P. and {Bresolin}, F. and {Kudritzki}, R.~P. and {Storm}, J. and {Smolec}, R. and {Narloch}, W.},
        title = "{A distance to the Large Magellanic Cloud that is precise to one per cent}",
      journal = {\nat},
         year = 2019,
        month = mar,
       volume = {567},
       number = {7747},
        pages = {200-203},
          doi = {10.1038/s41586-019-0999-4},
archivePrefix = {arXiv},
       eprint = {1903.08096},
 primaryClass = {astro-ph.GA},
       adsurl = {https://ui.adsabs.harvard.edu/abs/2019Natur.567..200P}
}

@ARTICLE{Ramachandran2011,
   author = {{Ramachandran}, P. and {Varoquaux}, G.},
    title = "{Mayavi: 3D Visualization of Scientific Data}",
  journal = {Computing in Science and Engineering},
archivePrefix = "arXiv",
   eprint = {1010.4891},
 primaryClass = "cs.SE",
     year = 2011,
    month = mar,
   volume = 13,
   number = 2,
    pages = {40-51},
      doi = {10.1109/MCSE.2011.35},
   adsurl = {http://adsabs.harvard.edu/abs/2011CSE....13b..40R}
}

@ARTICLE{Sandin2013,
       author = {{Sandin}, C. and {Lundqvist}, P. and {Lundqvist}, N. and {Bj{\"o}rnsson}, C. -I. and {Olofsson}, G. and {Shibanov}, Yu. A.},
        title = "{Properties of the three-dimensional structure in the central region of the supernova remnant SNR 0540-69.3}",
      journal = {\mnras},
         year = 2013,
        month = jul,
       volume = {432},
       number = {4},
        pages = {2854-2868},
          doi = {10.1093/mnras/stt641},
archivePrefix = {arXiv},
       eprint = {1304.4052},
 primaryClass = {astro-ph.GA},
       adsurl = {https://ui.adsabs.harvard.edu/abs/2013MNRAS.432.2854S}
}

@ARTICLE{Sandoval2021,
       author = {{Sandoval}, Michael A. and {Hix}, W. Raphael and {Messer}, O.~E. Bronson and {Lentz}, Eric J. and {Harris}, J. Austin},
        title = "{Three-dimensional Core-collapse Supernova Simulations with 160 Isotopic Species Evolved to Shock Breakout}",
      journal = {\apj},
         year = 2021,
        month = nov,
       volume = {921},
       number = {2},
          eid = {113},
        pages = {113},
          doi = {10.3847/1538-4357/ac1d49},
archivePrefix = {arXiv},
       eprint = {2106.01389},
 primaryClass = {astro-ph.HE},
       adsurl = {https://ui.adsabs.harvard.edu/abs/2021ApJ...921..113S}
}

@ARTICLE{Serafimovich2005,
       author = {{Serafimovich}, N.~I. and {Lundqvist}, P. and {Shibanov}, Yu. A. and {Sollerman}, J.},
        title = "{Optical observations of the young supernova remnant SNR 0540-69.3 and its pulsar}",
      journal = {Advances in Space Research},
         year = 2005,
        month = jan,
       volume = {35},
       number = {6},
        pages = {1106-1111},
          doi = {10.1016/j.asr.2005.01.071},
archivePrefix = {arXiv},
       eprint = {astro-ph/0501523},
 primaryClass = {astro-ph},
       adsurl = {https://ui.adsabs.harvard.edu/abs/2005AdSpR..35.1106S}
}

@ARTICLE{Stockinger2020,
       author = {{Stockinger}, G. and {Janka}, H. -T. and {Kresse}, D. and {Melson}, T. and {Ertl}, T. and {Gabler}, M. and {Gessner}, A. and {Wongwathanarat}, A. and {Tolstov}, A. and {Leung}, S. -C. and {Nomoto}, K. and {Heger}, A.},
        title = "{Three-dimensional models of core-collapse supernovae from low-mass progenitors with implications for Crab}",
      journal = {\mnras},
         year = 2020,
        month = aug,
       volume = {496},
       number = {2},
        pages = {2039-2084},
          doi = {10.1093/mnras/staa1691},
archivePrefix = {arXiv},
       eprint = {2005.02420},
 primaryClass = {astro-ph.HE},
       adsurl = {https://ui.adsabs.harvard.edu/abs/2020MNRAS.496.2039S}
}

@ARTICLE{Utrobin2019,
       author = {{Utrobin}, V.~P. and {Wongwathanarat}, A. and {Janka}, H. -Th. and {M{\"u}ller}, E. and {Ertl}, T. and {Woosley}, S.~E.},
        title = "{Three-dimensional mixing and light curves: constraints on the progenitor of supernova 1987A}",
      journal = {\aap},
         year = 2019,
        month = apr,
       volume = {624},
          eid = {A116},
        pages = {A116},
          doi = {10.1051/0004-6361/201834976},
archivePrefix = {arXiv},
       eprint = {1812.11083},
 primaryClass = {astro-ph.HE},
       adsurl = {https://ui.adsabs.harvard.edu/abs/2019A&A...624A.116U}
}

@ARTICLE{Williams2008,
       author = {{Williams}, Brian J. and {Borkowski}, Kazimierz J. and {Reynolds}, Stephen P. and {Raymond}, John C. and {Long}, Knox S. and {Morse}, Jon and {Blair}, William P. and {Ghavamian}, Parviz and {Sankrit}, Ravi and {Hendrick}, Sean P. and {Smith}, R. Chris and {Points}, Sean and {Winkler}, P. Frank},
        title = "{Ejecta, Dust, and Synchrotron Radiation in SNR B0540-69.3: A More Crab-Like Remnant than the Crab}",
      journal = {\apj},
         year = 2008,
        month = nov,
       volume = {687},
       number = {2},
        pages = {1054-1069},
          doi = {10.1086/592139},
archivePrefix = {arXiv},
       eprint = {0807.4155},
 primaryClass = {astro-ph},
       adsurl = {https://ui.adsabs.harvard.edu/abs/2008ApJ...687.1054W}
}

@ARTICLE{Wongwathanarat2017,
       author = {{Wongwathanarat}, Annop and {Janka}, Hans-Thomas and {M{\"u}ller}, Ewald and {Pllumbi}, Else and {Wanajo}, Shinya},
        title = "{Production and Distribution of $^{44}$Ti and $^{56}$Ni in a Three-dimensional Supernova Model Resembling Cassiopeia A}",
      journal = {\apj},
         year = 2017,
        month = jun,
       volume = {842},
       number = {1},
          eid = {13},
        pages = {13},
          doi = {10.3847/1538-4357/aa72de},
archivePrefix = {arXiv},
       eprint = {1610.05643},
 primaryClass = {astro-ph.HE},
       adsurl = {https://ui.adsabs.harvard.edu/abs/2017ApJ...842...13W}
}

@ARTICLE{Patapis2024,
       author = {{Patapis}, Polychronis and {Argyriou}, Ioannis and {Law}, David R. and {Glauser}, Adrian M. and {Glasse}, Alistair and {Labiano}, Alvaro and {{\'A}lvarez-M{\'a}rquez}, Javier and {Kavanagh}, Patrick J. and {Gasman}, Danny and {Mueller}, Michael and {Larson}, Kirsten and {Vandenbussche}, Bart and {Lee}, David and {Klaassen}, Pamela and {Guillard}, Pierre and {Wright}, Gillian S.},
        title = "{Geometric distortion and astrometric calibration of the JWST MIRI Medium Resolution Spectrometer}",
      journal = {\aap},
         year = 2024,
        month = feb,
       volume = {682},
          eid = {A53},
        pages = {A53},
          doi = {10.1051/0004-6361/202347339},
archivePrefix = {arXiv},
       eprint = {2307.01025},
 primaryClass = {astro-ph.IM},
       adsurl = {https://ui.adsabs.harvard.edu/abs/2024A&A...682A..53P}
}

@ARTICLE{GaiaCollaboration2016,
       author = {{Gaia Collaboration} and {Prusti}, T. and {de Bruijne}, J.~H.~J. and {Brown}, A.~G.~A. and {Vallenari}, A. and {Babusiaux}, C. and {Bailer-Jones}, C.~A.~L. and {Bastian}, U. and {Biermann}, M. and {Evans}, D.~W. and {Eyer}, L. and {Jansen}, F. and {Jordi}, C. and {Klioner}, S.~A. and {Lammers}, U. and {Lindegren}, L. and {Luri}, X. and {Mignard}, F. and {Milligan}, D.~J. and {Panem}, C. and {Poinsignon}, V. and {Pourbaix}, D. and {Randich}, S. and {Sarri}, G. and {Sartoretti}, P. and {Siddiqui}, H.~I. and {Soubiran}, C. and {Valette}, V. and {van Leeuwen}, F. and {Walton}, N.~A. and {Aerts}, C. and {Arenou}, F. and {Cropper}, M. and {Drimmel}, R. and {H{\o}g}, E. and {Katz}, D. and {Lattanzi}, M.~G. and {O'Mullane}, W. and {Grebel}, E.~K. and {Holland}, A.~D. and {Huc}, C. and {Passot}, X. and {Bramante}, L. and {Cacciari}, C. and {Casta{\~n}eda}, J. and {Chaoul}, L. and {Cheek}, N. and {De Angeli}, F. and {Fabricius}, C. and {Guerra}, R. and {Hern{\'a}ndez}, J. and {Jean-Antoine-Piccolo}, A. and {Masana}, E. and {Messineo}, R. and {Mowlavi}, N. and {Nienartowicz}, K. and {Ord{\'o}{\~n}ez-Blanco}, D. and {Panuzzo}, P. and {Portell}, J. and {Richards}, P.~J. and {Riello}, M. and {Seabroke}, G.~M. and {Tanga}, P. and {Th{\'e}venin}, F. and {Torra}, J. and {Els}, S.~G. and {Gracia-Abril}, G. and {Comoretto}, G. and {Garcia-Reinaldos}, M. and {Lock}, T. and {Mercier}, E. and {Altmann}, M. and {Andrae}, R. and {Astraatmadja}, T.~L. and {Bellas-Velidis}, I. and {Benson}, K. and {Berthier}, J. and {Blomme}, R. and {Busso}, G. and {Carry}, B. and {Cellino}, A. and {Clementini}, G. and {Cowell}, S. and {Creevey}, O. and {Cuypers}, J. and {Davidson}, M. and {De Ridder}, J. and {de Torres}, A. and {Delchambre}, L. and {Dell'Oro}, A. and {Ducourant}, C. and {Fr{\'e}mat}, Y. and {Garc{\'\i}a-Torres}, M. and {Gosset}, E. and {Halbwachs}, J. -L. and {Hambly}, N.~C. and {Harrison}, D.~L. and {Hauser}, M. and {Hestroffer}, D. and {Hodgkin}, S.~T. and {Huckle}, H.~E. and {Hutton}, A. and {Jasniewicz}, G. and {Jordan}, S. and {Kontizas}, M. and {Korn}, A.~J. and {Lanzafame}, A.~C. and {Manteiga}, M. and {Moitinho}, A. and {Muinonen}, K. and {Osinde}, J. and {Pancino}, E. and {Pauwels}, T. and {Petit}, J. -M. and {Recio-Blanco}, A. and {Robin}, A.~C. and {Sarro}, L.~M. and {Siopis}, C. and {Smith}, M. and {Smith}, K.~W. and {Sozzetti}, A. and {Thuillot}, W. and {van Reeven}, W. and {Viala}, Y. and {Abbas}, U. and {Abreu Aramburu}, A. and {Accart}, S. and {Aguado}, J.~J. and {Allan}, P.~M. and {Allasia}, W. and {Altavilla}, G. and {{\'A}lvarez}, M.~A. and {Alves}, J. and {Anderson}, R.~I. and {Andrei}, A.~H. and {Anglada Varela}, E. and {Antiche}, E. and {Antoja}, T. and {Ant{\'o}n}, S. and {Arcay}, B. and {Atzei}, A. and {Ayache}, L. and {Bach}, N. and {Baker}, S.~G. and {Balaguer-N{\'u}{\~n}ez}, L. and {Barache}, C. and {Barata}, C. and {Barbier}, A. and {Barblan}, F. and {Baroni}, M. and {Barrado y Navascu{\'e}s}, D. and {Barros}, M. and {Barstow}, M.~A. and {Becciani}, U. and {Bellazzini}, M. and {Bellei}, G. and {Bello Garc{\'\i}a}, A. and {Belokurov}, V. and {Bendjoya}, P. and {Berihuete}, A. and {Bianchi}, L. and {Bienaym{\'e}}, O. and {Billebaud}, F. and {Blagorodnova}, N. and {Blanco-Cuaresma}, S. and {Boch}, T. and {Bombrun}, A. and {Borrachero}, R. and {Bouquillon}, S. and {Bourda}, G. and {Bouy}, H. and {Bragaglia}, A. and {Breddels}, M.~A. and {Brouillet}, N. and {Br{\"u}semeister}, T. and {Bucciarelli}, B. and {Budnik}, F. and {Burgess}, P. and {Burgon}, R. and {Burlacu}, A. and {Busonero}, D. and {Buzzi}, R. and {Caffau}, E. and {Cambras}, J. and {Campbell}, H. and {Cancelliere}, R. and {Cantat-Gaudin}, T. and {Carlucci}, T. and {Carrasco}, J.~M. and {Castellani}, M. and {Charlot}, P. and {Charnas}, J. and {Charvet}, P. and {Chassat}, F. and {Chiavassa}, A. and {Clotet}, M. and {Cocozza}, G. and {Collins}, R.~S. and {Collins}, P. and {Costigan}, G.},
        title = "{The Gaia mission}",
      journal = {\aap},
         year = 2016,
        month = nov,
       volume = {595},
          eid = {A1},
        pages = {A1},
          doi = {10.1051/0004-6361/201629272},
archivePrefix = {arXiv},
       eprint = {1609.04153},
 primaryClass = {astro-ph.IM},
       adsurl = {https://ui.adsabs.harvard.edu/abs/2016A&A...595A...1G}
}

@ARTICLE{GaiaCollaboration2023,
       author = {{Gaia Collaboration} and {Vallenari}, A. and {Brown}, A.~G.~A. and {Prusti}, T. and {de Bruijne}, J.~H.~J. and {Arenou}, F. and {Babusiaux}, C. and {Biermann}, M. and {Creevey}, O.~L. and {Ducourant}, C. and {Evans}, D.~W. and {Eyer}, L. and {Guerra}, R. and {Hutton}, A. and {Jordi}, C. and {Klioner}, S.~A. and {Lammers}, U.~L. and {Lindegren}, L. and {Luri}, X. and {Mignard}, F. and {Panem}, C. and {Pourbaix}, D. and {Randich}, S. and {Sartoretti}, P. and {Soubiran}, C. and {Tanga}, P. and {Walton}, N.~A. and {Bailer-Jones}, C.~A.~L. and {Bastian}, U. and {Drimmel}, R. and {Jansen}, F. and {Katz}, D. and {Lattanzi}, M.~G. and {van Leeuwen}, F. and {Bakker}, J. and {Cacciari}, C. and {Casta{\~n}eda}, J. and {De Angeli}, F. and {Fabricius}, C. and {Fouesneau}, M. and {Fr{\'e}mat}, Y. and {Galluccio}, L. and {Guerrier}, A. and {Heiter}, U. and {Masana}, E. and {Messineo}, R. and {Mowlavi}, N. and {Nicolas}, C. and {Nienartowicz}, K. and {Pailler}, F. and {Panuzzo}, P. and {Riclet}, F. and {Roux}, W. and {Seabroke}, G.~M. and {Sordo}, R. and {Th{\'e}venin}, F. and {Gracia-Abril}, G. and {Portell}, J. and {Teyssier}, D. and {Altmann}, M. and {Andrae}, R. and {Audard}, M. and {Bellas-Velidis}, I. and {Benson}, K. and {Berthier}, J. and {Blomme}, R. and {Burgess}, P.~W. and {Busonero}, D. and {Busso}, G. and {C{\'a}novas}, H. and {Carry}, B. and {Cellino}, A. and {Cheek}, N. and {Clementini}, G. and {Damerdji}, Y. and {Davidson}, M. and {de Teodoro}, P. and {Nu{\~n}ez Campos}, M. and {Delchambre}, L. and {Dell'Oro}, A. and {Esquej}, P. and {Fern{\'a}ndez-Hern{\'a}ndez}, J. and {Fraile}, E. and {Garabato}, D. and {Garc{\'\i}a-Lario}, P. and {Gosset}, E. and {Haigron}, R. and {Halbwachs}, J. -L. and {Hambly}, N.~C. and {Harrison}, D.~L. and {Hern{\'a}ndez}, J. and {Hestroffer}, D. and {Hodgkin}, S.~T. and {Holl}, B. and {Jan{\ss}en}, K. and {Jevardat de Fombelle}, G. and {Jordan}, S. and {Krone-Martins}, A. and {Lanzafame}, A.~C. and {L{\"o}ffler}, W. and {Marchal}, O. and {Marrese}, P.~M. and {Moitinho}, A. and {Muinonen}, K. and {Osborne}, P. and {Pancino}, E. and {Pauwels}, T. and {Recio-Blanco}, A. and {Reyl{\'e}}, C. and {Riello}, M. and {Rimoldini}, L. and {Roegiers}, T. and {Rybizki}, J. and {Sarro}, L.~M. and {Siopis}, C. and {Smith}, M. and {Sozzetti}, A. and {Utrilla}, E. and {van Leeuwen}, M. and {Abbas}, U. and {{\'A}brah{\'a}m}, P. and {Abreu Aramburu}, A. and {Aerts}, C. and {Aguado}, J.~J. and {Ajaj}, M. and {Aldea-Montero}, F. and {Altavilla}, G. and {{\'A}lvarez}, M.~A. and {Alves}, J. and {Anders}, F. and {Anderson}, R.~I. and {Anglada Varela}, E. and {Antoja}, T. and {Baines}, D. and {Baker}, S.~G. and {Balaguer-N{\'u}{\~n}ez}, L. and {Balbinot}, E. and {Balog}, Z. and {Barache}, C. and {Barbato}, D. and {Barros}, M. and {Barstow}, M.~A. and {Bartolom{\'e}}, S. and {Bassilana}, J. -L. and {Bauchet}, N. and {Becciani}, U. and {Bellazzini}, M. and {Berihuete}, A. and {Bernet}, M. and {Bertone}, S. and {Bianchi}, L. and {Binnenfeld}, A. and {Blanco-Cuaresma}, S. and {Blazere}, A. and {Boch}, T. and {Bombrun}, A. and {Bossini}, D. and {Bouquillon}, S. and {Bragaglia}, A. and {Bramante}, L. and {Breedt}, E. and {Bressan}, A. and {Brouillet}, N. and {Brugaletta}, E. and {Bucciarelli}, B. and {Burlacu}, A. and {Butkevich}, A.~G. and {Buzzi}, R. and {Caffau}, E. and {Cancelliere}, R. and {Cantat-Gaudin}, T. and {Carballo}, R. and {Carlucci}, T. and {Carnerero}, M.~I. and {Carrasco}, J.~M. and {Casamiquela}, L. and {Castellani}, M. and {Castro-Ginard}, A. and {Chaoul}, L. and {Charlot}, P. and {Chemin}, L. and {Chiaramida}, V. and {Chiavassa}, A. and {Chornay}, N. and {Comoretto}, G. and {Contursi}, G. and {Cooper}, W.~J. and {Cornez}, T. and {Cowell}, S. and {Crifo}, F. and {Cropper}, M. and {Crosta}, M. and {Crowley}, C. and {Dafonte}, C. and {Dapergolas}, A. and {David}, M. and {David}, P. and {de Laverny}, P. and {De Luise}, F. and {De March}, R.},
        title = "{Gaia Data Release 3. Summary of the content and survey properties}",
      journal = {\aap},
         year = 2023,
        month = jun,
       volume = {674},
          eid = {A1},
        pages = {A1},
          doi = {10.1051/0004-6361/202243940},
archivePrefix = {arXiv},
       eprint = {2208.00211},
 primaryClass = {astro-ph.GA},
       adsurl = {https://ui.adsabs.harvard.edu/abs/2023A&A...674A...1G}
}

@ARTICLE{Bentz2025,
       author = {{Bentz}, Misty C.},
        title = "{The NIRSpec IFU Point Spread Function}",
      journal = {Research Notes of the American Astronomical Society},
         year = 2025,
        month = may,
       volume = {9},
       number = {5},
          eid = {128},
        pages = {128},
          doi = {10.3847/2515-5172/adddac},
       adsurl = {https://ui.adsabs.harvard.edu/abs/2025RNAAS...9..128B}
}
\bibliographystyle{aasjournalv7}


\end{document}